%% file: math_part/main.tex
\documentclass[parskip=half,bibliography=totoc,fontsize=10pt]{scrartcl}

\input{math_part/settings}
\author[1]{Q.~Bonnefoy}
\author[2]{V.~Cortés}
\author[3]{E.~Gendy} 
\author[4,5]{C.~Grojean}
\author[2]{K.~Ritter~von~Merkl} 
\author[2,4]{P.~N.~Pilatus}

\title{Geometry of effective field theory positivity cones}

\date{}

\affil[1]{Université de Strasbourg, CNRS, IPHC UMR7178, 23 rue du Loess, 67037 Strasbourg, France\smallskip}
\affil[2]{Fachbereich Mathematik, Universit\"at Hamburg, Bundesstra\ss e 55, 20146 Hamburg, Germany \smallskip}
\affil[3]{Technische Universität München, 
James-Franck-Strasse 1, 85748 Garching, Germany\smallskip}
\affil[4]{Deutsches Elektronen-Synchrotron DESY, Notkestr.\ 85, 22607 Hamburg, Germany\smallskip}
\affil[5]{Institut für Physik, Humboldt-Universit\"at zu Berlin, Newtonstraße 15, 12489 Berlin, Germany}

\begin{document}

\maketitle

\begin{abstract}
    Positivity bounds are theoretical constraints on the Wilson coefficients of an effective field theory. These bounds emerge from the requirement that a given effective field theory must be the low-energy limit of a relativistic quantum theory that satisfies the fundamental principles of unitarity, locality, and causality. The task of deriving these bounds can be reformulated as the geometric problem of finding the extremal representation of a closed convex cone~$\mathcal C_W$. More precisely, in the presence of multiple particle flavors, the forward-limit positivity cone $\mathcal C_W$ consists of all 
    positive semi-definite tensors in $W =\left\{ S \in \mathrm{Sym}^2 (\mathrm{Sym}^2\, V^*)\oplus \mathrm{Sym}^2 \left({\Lambda}^2 V^*\right)
:
\tau S = S \right\}
    \subset \mathrm{Sym}^2(V^*\otimes V^*)$, where $\tau$ denotes transposition in the second and fourth tensor factor and $V\cong\mathbb{R}^n$, where $n$ is the number of flavors.  In this work, we solve this question up to three flavors, i.e.~$n=3$, proving a full classification of all extremal elements in these cases. We furthermore study the implications of our findings, deriving the full positivity bounds for amplitudes with and without additional symmetries. In the cases with additional symmetries that we consider, we find that the so-called elastic bounds are sufficient to give rise to the full positivity bounds.
    \bigskip
    
\emph{Preprint numbers: DESY-25-119, HU-EP-25/31}

\medskip


\end{abstract}
\newpage
\tableofcontents
\section{Introduction}
Effective field theories (EFTs) are a way to approximate a quantum field theory (QFT) in terms of the field content at lower energies.
All the physical information of an EFT is captured in a set of parameters called \emph{Wilson coefficients} (together with a power counting rule). They can be probed by performing scattering experiments. Generalizing earlier results in Chiral Perturbation Theory ($\chi$PT) \cite{Pham:1985cr,Pennington:1994kc,Ananthanarayan:1994hf}, it was shown in Ref.~\cite{Adams06} that, given an EFT, requiring that it is the low energy limit of a unitary, local and causal QFT results in constraints on the Wilson coefficients of the EFT\footnote{Although we refer to QFTs in this paper, positivity can also be shown to hold for EFTs of some non-QFT relativistic quantum theories, such as string theory~\cite{Bellazzini:2015cra,Guerrieri:2021ivu,Bern:2022yes}.}. These constraints are commonly referred to as \emph{positivity bounds}. They are a powerful tool in order to reduce the landscape of physically meaningful EFTs: they have been applied in a great variety of physical contexts, such as the Standard Model Effective Field Theory~\cite{Bellazzini:2018paj,Bi:2019phv,Remmen:2019cyz,Remmen:2020vts,Fuks:2020ujk,Yamashita:2020gtt,Gu:2020ldn,Bonnefoy:2020yee,Henriksson:2021ymi,Zhang:2021eeo,Li:2022rag,Ghosh:2022qqq,Li:2022aby,Gu:2023emi,Chen:2023bhu,Davighi:2023acq} and related EFTs~\cite{Remmen:2024hry,Chakraborty:2024ciu}, gravitational theories~\cite{Cheung:2016yqr,Bellazzini:2017fep,deRham:2017xox,Alberte:2020jsk,Tokuda:2020mlf,Alberte:2020bdz,Caron-Huot:2021rmr,Alberte:2021dnj,Caron-Huot:2022ugt,Chiang:2022jep,deRham:2022gfe,Hong:2023zgm,Alviani:2024sxx} and their UV completion in consistent theories of quantum gravity~\cite{Cheung:2014ega,Bellazzini:2015cra,Andriolo:2018lvp,DeRham:2018bgz,Bellazzini:2019xts,Arkani-Hamed:2021ajd}, early-universe cosmology~\cite{Melville:2019wyy,Grall:2021xxm,deRham:2021fpu,Freytsis:2022aho,Xu:2023lpq,Kim:2023bbs}, pion and Goldstone boson physics~\cite{deRham:2017imi,Wang:2020jxr,Guerrieri:2020bto,Fernandez:2022kzi,Acanfora:2023axz}, renormalization~\cite{Chala:2023jyx,Liao:2025npz,Chala:2023xjy}, massive spinning particles~\cite{Bonifacio:2018vzv,deRham:2018qqo,Bellazzini:2019bzh,Alberte:2019xfh,Wang:2020xlt,deRham:2022sdl,Bellazzini:2023nqj,Aoki:2023khq} and supersymmetry breaking~\cite{Melville:2019tdc,Bonnefoy:2022rcw}.

State-of-the-art positivity bounds are associated with a specific kind of scattering, known as two-to-two scattering\footnote{See Refs.~\cite{Chandrasekaran:2018qmx,Arkani-Hamed:2023jwn,Bresciani:2025toe,Cheung:2025krg} for recent developments beyond two-to-two scattering.}: two incoming particles interact and yield two outgoing particles. Additionally,
the forward limit --namely, identical
ingoing and
outgoing
kinematics-- is
often invoked,
and we will stick
to this assumption
in what follows\footnote{Furthermore, we only consider amplitudes computed at tree level, we restrict to the positivity of the part of these amplitudes quartic in momenta, we focus on non-truncated cones, and we do not include the complete constraints arising from crossing symmetry. See respectively Refs.~\cite{deRham:2017avq,Davighi:2021osh,Beadle:2024hqg}, Refs.~\cite{Chala:2021wpj,Bellazzini:2021oaj,Hamada:2023cyt,Caron-Huot:2024tsk,Ye:2024rzr,Peng:2025klv,Chang:2025cxc,Beadle:2025cdx}, Refs.~\cite{Bellazzini:2020cot,Arkani-Hamed:2020blm,Chiang:2021ziz}, Refs.~\cite{Caron-Huot:2020cmc,Chiang:2022ltp,EliasMiro:2022xaa,Riembau:2022yse,Wan:2024eto} and Refs.~\cite{Tolley:2020gtv,Chowdhury:2021ynh,Du:2021byy} for progress on these fronts.}. The formulation of such positivity bounds involves a real four-tensor, which is related to the amplitude (see Appendix~\ref{app: tensor M} for details on how to extract this four-tensor from the amplitude). More precisely, it was argued in Ref.~\cite{zhang20} (see also Refs.~\cite{Bellazzini:2014waa,Trott:2020ebl,Yang:2023ncf}) that positivity bounds translate to the requirement that the aforementioned four-tensor lies in the closed convex cone $\mathcal{C}$ that is defined by
\begin{align*}
    \mathcal{C}=\conv\{ Q^{\otimes 2}+\tau Q^{\otimes 2}: Q\in \Lambda^2V\text{ or }Q\in\Sym^2 V\},
\end{align*}
where $V$ is a real vector space whose dimension agrees with the number of flavors and where $\tau S(x,y,z,w)=S(x,w,z,y)$ for $S\in V^{\otimes 4}$. The symmetries of the elements of $\mathcal{C}$ are related to the different symmetries of the amplitude: for instance, the symmetry under $\tau$ is related to crossing symmetry.

In Ref.~\cite{Li_et_al21} it was suggested to derive the positivity bounds via the dual cone $\mathcal{C}^*$ of $\mathcal{C}$, which is of the form
\[ \mathcal{C}^*=\mathcal{C}_W:= \{ S\in W : S\ge 0\},\] 
where 
\[ W :=\left\{ S \in \mathrm{Sym}^2 (\mathrm{Sym}^2\, V^*)\oplus \mathrm{Sym}^2 \left({\Lambda}^2 V^*\right)
:
\tau S = S \right\}\subset {V^*}^{\otimes 4},\]
with $\tau$ as above and $S\geq 0$ as an element in $\Sym^2(V^*\otimes V^*)$. As it was argued in Ref.~\cite{Li_et_al21}, every extremal ray of the cone $\mathcal{C}_W$ gives rise to an inequality constraining the cone $\mathcal{C}$. The system of all these inequalities completely determines the cone $\mathcal{C}$ and hence gives rise to the full set of positivity bounds. This set of bounds can be further split into two categories: \emph{elastic} and  \emph{inelastic} positivity bounds. The former, which have received the most attention in the literature, arise from scattering processes where the initial and final states of the scattering process are identical. However, in Ref.~\cite{zhang20} it was pointed out that in general the description of ${\cal C}$ will involve additional constraints, the inelastic positivity bounds, which cannot be deduced from the elastic bounds.

It is worth noting that the complexity of the problem of finding all extremal rays of the cone $\mathcal{C}_W$ grows fast with  $n:=\dim V$. This is due to $\dim W=n(n+1)(n(n+1)+2)/8$ growing fast with $n$. The problem of determining the extremal representation of $\mathcal{C}_W$ for $\dim V=2$ was already solved in Ref.~\cite{Li_et_al21}, where it was shown that in this case every extremal ray is elastic. In contrast, if $\dim V>2$ there are extremal rays that give rise to constraints that cannot be deduced from the elastic positivity bounds. In this work, we establish a full classification of all extremal rays of the cone $\mathcal{C}_W$ in the case where $\dim V=3$. Our main result is the following theorem, see Theorem~\ref{thm: ers in 3d} (as well as Section~\ref{sec:notation} for our notations). 
\begin{thm*}
Suppose $\dim V=3$. Then $S\in \mathcal{C}_W$ is extremal if and only if it is of one of the following types: 
\begin{enumerate}
    \item For some $\alpha\in V^*\setminus \{0\}$, $S=\alpha^4$.
    \item For linearly independent $\alpha_1,\alpha_2\in V^*$, $S=(\alpha_1\vee \alpha_2)^{\otimes 2}+(\alpha_1\wedge \alpha_2)^{\otimes 2}$.
    \item For a basis $\{\alpha_1,\alpha_2,\alpha_3\}$ of $V^*$ and parameters $g,d,h$ such that $g^2>1-d^2+dh$ we have
    \begin{align*}
        S=S_{\mathrm{tot}}+2\,(g^2+d^2-1-dh)\left[(\alpha_2\otimes \alpha_3)^{\otimes 2}+(\alpha_3\otimes \alpha_2)^{\otimes 2}\right],
    \end{align*}
    where
    \begin{align*}
    S_{\mathrm{tot}}=&\,\alpha_1^4+6\,\alpha_1^2\alpha_2^2+6\,\alpha_1^2\alpha_3^2+12d\,\alpha_1\alpha_2^2\alpha_3+12g\,\alpha_1\alpha_2\alpha_3^2+4h\,\alpha_1\alpha_3^3+(1+d^2)\,\alpha_2^4\\&+4dg\,\alpha_2^3\alpha_3+6(1+dh)\,\alpha_2^2\alpha_3^2+4g(d+h)\,\alpha_2\alpha_3^3+(1+g^2+h^2)\,\alpha_3^4.    
    \end{align*}
\end{enumerate}
\end{thm*}
This theorem gives rise to all positivity bounds for scattering amplitudes corresponding to processes involving up to three flavors. Along the way, we prove several general properties of extremal elements of the cone $\mathcal{C}_W$ in arbitrary dimensions. We also study in more detail the corresponding positivity bounds for amplitudes involving three flavors and respecting an additional~$\O(3)$, $\Z_2^{\,3}$ or $\SO(2)$ symmetry. We prove that in these cases the elastic bounds are sufficient to describe the full set of positivity bounds, while in the case without additional symmetries there are inelastic positivity bounds, coming from the third class of the extremal rays in our classification result.

The structure of the paper is as follows. In Section~\ref{sect: prelim}, we fix notation and recall general notions and results from convex geometry. Moreover, we state some useful auxiliary results on symmetric bilinear forms. In the more specific  Section~\ref{sect: pos cone}, we then describe our general setup for the study of the cone $\mathcal{C}_W$. We study some properties of the space $W$ together with some natural linear maps defined on $W$ that will be useful to formulate and prove our classification result. We then investigate some general properties of the extremal elements of the cone $\mathcal{C}_W$. In the end of Section~\ref{sect: pos cone}, we recover the result of Ref.~\cite{Li_et_al21} that in the case where $\dim V=2$ the elastic bounds are sufficient to describe the entire cone. In Section~\ref{sect: ers in 3d}, we focus on the case where $\dim V=3$ and prove a full classification of all extremal elements of $\mathcal{C}_W$. Lastly, in Section~\ref{sect: pos bounds}, we study the positivity bounds generated by the extremal elements found in Section~\ref{sect: ers in 3d}. In particular, we consider the positivity bounds for amplitudes with additional $\O(3)$, $\Z_2^{\,3}$ and $\SO(2)$ symmetries, respectively, proving that in these cases the elastic bounds are sufficient. As a physical application we show that the elastic positivity bounds are sufficient to bound pion scattering amplitudes in chiral perturbation theory. In the Appendix, we describe in more detail the connection of the mathematical objects studied in this paper to physics. More precisely, in Appendix~\ref{app: tensor M}, we show how the space $W$ and the cone $\mathcal{C}_W^*$ arise in physics and how the elements thereof are related to amplitudes. In Appendix~\ref{app:elasticScattering}, we explicitly link elastic bounds and the two first kinds of extremal tensors of Theorem~\ref{thm: ers in 3d}. Finally, in Appendix~\ref{app: ChPT} we take a closer look at the two-to-two pion scattering amplitude in chiral perturbation theory, which, as mentioned above, serves as a first application of our results.

\section{Preliminaries}
\label{sect: prelim}
In this section, we will fix some notations and recall relevant definitions and results from convex geometry, as well as some general results on symmetric bilinear forms.
\subsection{Notation}\label{sec:notation}
For the symmetric and anti-symmetric product of vectors $u,v\in V$ in some finite-dimensional real vector space $V$, we use the convention
\[
u\vee v=\frac{1}{2}(u\otimes v+v\otimes u),\quad u\wedge v=\frac{1}{2}(u\otimes v-v\otimes u)
\]
so that
\[
u\otimes v=u\vee v+u\wedge v.
\]

Moreover, for some vector $v\in V$ and subspace $U\subset V$, we will denote by $v\vee U$ the space
\[
v\vee U:=\{ v\vee u\,:\, u\in U\}
\]
and similarly, we write
\[
v\wedge U:=\{ v\wedge u\,:\, u\in U\}.
\]

We will furthermore denote the totally symmetric product of one-forms $\beta_i$, $i=1,...,k$ by
\begin{align*}
    \beta_1\beta_2\cdots\beta_k:=\frac{1}{k!}\sum_{\sigma\in S_k}\beta_{\sigma(1)}\otimes\beta_{\sigma(2)}\otimes\cdots\otimes\beta_{\sigma(k)}.
\end{align*}
When $\beta_1=\beta=\beta_2=\cdots\beta_k$, we will also write $\beta^k=\beta\beta\cdots\beta$.

Given a symmetric or anti-symmetric bilinear form $b$ on some vector space $V$, we will denote by
\begin{align*}
    \ker b:=\ker\left(b\colon V\rightarrow V^*\right)
\end{align*}
the \emph{kernel} of $b$ and by
\begin{align*}
    \rk b := \dim V-\dim(\ker b)
\end{align*}
the \emph{rank} of $b$.

\subsection{Extremal representation of spectrahedra}

In this section we will briefly recall some definitions and review some classical results in convex geometry relevant to this work. 

We start recalling the following topological notions as well as the notion of a \emph{face} of a non-empty convex set:
\begin{mydef}
    Let $\mathcal{C}\subset \R^n$ be a non-empty convex set.
    \begin{itemize}
        \item The \emph{affine hull} of $\mathcal{C}$ is the smallest affine set $\aff (\mathcal{C})$ containing $\mathcal{C}$.
        \item The \emph{dimension} $\dim \mathcal{C}$ of $\mathcal{C}$ is the dimension of its affine hull $\aff(\mathcal{C})$.
        \item The \emph{relative interior} $\ri (\mathcal{C})$ of $\mathcal{C}$ is defined as the interior of $\mathcal{C}$ with respect to the induced topology of $\aff(\mathcal{C})$.
        \item The \emph{relative boundary} of $\mathcal{C}$ is defined as $\rb \mathcal{C}:=\cl(\mathcal{C})\setminus \ri (\mathcal{C})$, where $\cl({\cal C})$ refers to the closure of $\cal C$.
        \item A \emph{face} of 
        $\mathcal{C}$ is a convex subset $\mathcal{C}'$ of $\mathcal{C}$ such that every (closed) line segment in $\mathcal{C}$ with a relative interior point in $\mathcal{C}'$ has both endpoints in $\mathcal{C}'$. We will consider the whole convex set $\mathcal{C}$ a face of itself.
    \end{itemize}
  
\end{mydef}

The faces of a convex set give rise to a partition thereof:
\begin{thm}[Theorem 18.2 in \cite{rockafellar}]
    Let $\mathcal{C}$ be a non-empty convex set, and let $\mathcal{U}$ be the collection of all relative interiors of non-empty faces of $\mathcal{C}$. Then $\mathcal{U}$ is a partition of $\mathcal{C}$, i.e. the sets in $\mathcal{U}$ are disjoint and their union is $\mathcal{C}$.
\end{thm}

In particular, this means that every point in $\mathcal{C}$ is contained in the relative interior of precisely one face of $\mathcal{C}$.

\begin{mydef}
    For $x\in\mathcal{C}$ the unique face $F_{\mathcal{C}}(x)$ such that $x\in\ri F_{\mathcal{C}}(x)$ is called the \emph{minimal face} of $x\in\mathcal{C}$.
\end{mydef}

We will now give a definition of \emph{extremal rays} of a convex cone in terms of faces. 
\begin{mydef}
An \emph{extremal ray} of a convex cone $\mathcal{C}$ is a face of $\mathcal{C}$ which is of the form $\R_{\geq 0}\, x\cap \mathcal{C}$ for some $x\in\mathcal{C}\setminus \{0\}$.
An element $x\in \mathcal{C}\setminus \{0\}$ generating an extremal ray is called \emph{extremal}.
\end{mydef}

\begin{rem}
    Note that an element $x\in \mathcal{C}$ of a line-free convex cone $\mathcal{C}$ is extremal if and only if $\dim (F_{\mathcal{C}}(x))=1$. 
\end{rem}
We also have the following equivalent characterization of extremal elements:
\begin{prop}
\label{prop: alternative extremal} 
    Let $\mathcal{C}$ be a convex cone. Then $x\in\mathcal{C}\setminus\{0\}$ is extremal if and only if for every $x_0,x_1\in\mathcal{C}$ such that $x=x_0+x_1$ we must have $x_0,x_1\in\R_{\geq0}\, x$.
\end{prop}
\begin{myproof}
    If $x\in\mathcal{C}\setminus\{0\}$ generates an extremal ray, every closed line segment 
    \[\{(1-t)\, y_1+t\, y_2\,:\, t\in [0,1]\}\subset\mathcal{C}\]
    starting at some $y_1\in\mathcal{C}$, $y_1\neq x$, ending at some $y_2\in\mathcal{C}$, $y_2\neq x$, and  passing through $x$ must be contained in $\R_{\geq0}x$, which is only possible if $y_1, y_2\in\R_{\geq0}\, x$. 

    Conversely, if $x$ is not extremal, there is a line segment 
    \[L:=\{(1-t)\, y_1+t\, y_2\,:\,t\in[0,1]\}\subset\mathcal{C},\] whose interior intersects $\R_{\geq0}x$ in precisely one point. Then the vectors $\{x,y_1,y_2\}\subset \mathcal{C}$ are pairwise linearly independent. We may furthermore assume that, up to translating and rescaling $L$, we have $L\cap\R_{\geq0}x=\{x\}$ and conclude that $x=(1-s)\, y_1+s\, y_2$ for some $s\in(0,1)$.
\end{myproof}

A non-trivial closed convex cone containing no lines is completely determined by its extremal rays.
\begin{thm}[\cite{klee1957extremal}, see also Corollary 18.5.2 in \cite{rockafellar}]
    Let $\mathcal{C}$ be a closed convex cone containing more than just the origin but no lines. Let $T$ be any set of vectors in $\mathcal{C}$ such that every extremal ray of $\mathcal{C}$ is generated by some $x \in T$. Then $\mathcal{C}$ is the convex cone generated by $T$.
\end{thm}
We will refer to this characterization of a closed, line-free, non-trivial convex cone $\mathcal{C}$ in terms of its extremal rays as the \emph{extremal representation} of $\mathcal{C}$. In combination with Carathéodory's Theorem (see e.g.~Theorem 17.1 in~\cite{rockafellar}) this implies the following:
\begin{thm}
\label{thm: finite sum of extremals}
    Every non-zero element of a closed convex cone $\mathcal{C}$ containing more than just the origin but no lines can be written as a finite positive linear combination of extremal elements in~$\mathcal{C}$.
\end{thm}
The following is a consequence of Theorem~\ref{thm: finite sum of extremals} and will be useful for us. 
\begin{prop}
\label{prop: er proj to er general}
Let $\mathcal{C}\subset \R^n$ be a closed convex cone, $U\subset \R^n$ a linear subspace and $p\colon \R^n\to U$ a linear projection to $U$. We denote by $\tilde{\mathcal{C}}:=p(\mathcal{C})$ the projection of $\mathcal{C}$, which is itself a closed convex cone. Then every extremal ray of $\tilde{\mathcal{C}}$ is the projection of some extremal ray in $\mathcal{C}$.
\end{prop}
\begin{proof}
    The case where $\tilde{\mathcal{C}}=0$ is trivial. We may therefore assume that $\tilde{\mathcal{C}}\neq 0$. Let $x\in \tilde{\mathcal{C}}\setminus\{0\}$ be an extremal element. Let $y\in\mathcal{C}$ such that $p(y)=x$. Suppose that $y=y_0+y_1$ for some $y_0,y_1\in\mathcal{C}$. Then $x=p(y_0)+p(y_1)$. We may furthermore assume that $p(y_0)\neq 0$. Since by assumption $x$ is extremal, it follows from Proposition~\ref{prop: alternative extremal} that $p(y_0)=\lambda x$ for some $\lambda>0$. In particular, $1/\lambda\,y_0$ projects onto $x$. Since by Theorem~\ref{thm: finite sum of extremals} every element in $\mathcal{C}$ is a finite positive linear combination of extremal elements, we may iterate the procedure described above (reducing in each step the number of terms in the linear combination) to obtain an extremal element projecting onto $x\in\tilde{\mathcal{C}}$.
\end{proof}

A particular class of closed convex sets are \emph{spectrahedra}. To state the definition of a  spectrahedron we denote by $\mathcal{S}_n$ the space of real symmetric $n\times n$-matrices and by $\mathcal{P}_n$ the closed convex cone of positive semi-definite real symmetric $n\times n$-matrices.

\begin{mydef}
    An intersection of $\mathcal{P}_n$ with an affine subspace of $\mathcal{S}_n$ is called a \emph{spectrahedron}. A spectrahedron is called \emph{linear} if it is the intersection of $\mathcal{P}_n$ with a linear subspace of $\mathcal{S}_n$.
\end{mydef}

The faces of a spectrahedron have been studied in \cite{ramana1995some}. The following characterization was obtained:

\begin{thm}[Theorem 1 in~\cite{ramana1995some}, see also Theorem~4.4 in~\cite{vinzant20}]
\label{thm: faces of spectrahedra}
    Let $\mathcal{C}$ be a spectrahedron. Then the minimal face of $x\in\mathcal{C}$ is given by
        \[F_{\mathcal{C}}(x)=\{z\in\mathcal{C}\,:\, \ker z\supset \ker x\}.\]
\end{thm}

We also note the following description of the relative boundary of the minimal face of a point $x\in \mathcal{C}$:

\begin{cor}
\label{cor: bdr faces spectrahedron}
    Let $\mathcal{C}$ be a spectrahedron, $x\in\mathcal{C}$. Then 
    \[\rb F_{\mathcal{C}}(x)=\{z\in\mathcal{C}\,:\, \ker z\supsetneq \ker x\}.\]
\end{cor}
\begin{myproof}
    Let $x\in\mathcal{C}$, $y\in\rb F_{\mathcal{C}}(x)$. Since $\mathcal{C}$ is closed so are its faces and in particular $F_{\mathcal{C}}(x)$ is closed. Therefore, $y\in F_{\mathcal{C}}(x)$ and by Theorem \ref{thm: faces of spectrahedra}, it follows that $\ker y\supset \ker x$. If $\ker y= \ker x$, Theorem \ref{thm: faces of spectrahedra} implies that also $x\in F_{\mathcal{C}}(y)$, which is impossible since $y$ was assumed to be in the relative boundary of $F_{\mathcal{C}}(x)$ and therefore $\dim F_{\mathcal{C}}(y)<\dim F_{\mathcal{C}}(x)$. Conversely, if $\ker y\supsetneq \ker x$ for some $y\in \mathcal{C}$, then $y\in F_{\mathcal{C}}(x)\setminus \ri F_{\mathcal{C}}(x)=\rb F_{\mathcal{C}}(x)$. 
\end{myproof}

Linear spectrahedra are in particular closed line-free convex cones and it makes sense to ask for their extremal representation. The first property in Theorem \ref{thm: faces of spectrahedra} yields a criterion to determine whether an element $x\in\mathcal{C}$ is extremal in $\mathcal{C}$:

\begin{cor}[Corollary 4 in~\cite{ramana1995some}]
    \label{cor: extremal ramana}
    Let $\mathcal{C}$ be a linear spectrahedron. Then $x\in\mathcal{C}\setminus\{0\}$ is extremal if and only if for every $z\in\mathcal{C}$ the inclusion $\ker x\subset \ker z$ implies $z\in\R\, x$.
\end{cor}

\begin{cor}
\label{cor: nec extremal}
Let $\mathcal{C}$ be a linear spectrahedron. Then $x\in\mathcal{C}\setminus\{0\}$ is extremal if and only if for every $z\in\mathcal{C}$ the proper inclusion $\ker x\subsetneq \ker z$ implies $z= 0$.
\end{cor}
\begin{myproof}
    By Corollary \ref{cor: bdr faces spectrahedron}, the property that for every $z\in\mathcal{C}$ the inclusion $\ker x\subsetneq \ker z$ implies $z=0$ is equivalent to saying that the relative boundary of the minimal face $F_{\mathcal{C}}(x)$ of $x$ is the point $0\in\mathcal{C}$. We claim that this is equivalent to the statement that $\dim F_{\mathcal{C}}(x)=1$. Indeed,  suppose that $\dim F_{\mathcal{C}}(x)>1$ and $\rb F_{\mathcal{C}}(x)=\{0\}$. Then there is some $y\in \ri F_{\mathcal{C}}(x)$ linearly independent of $x$. We consider the line $L$ in $\mathcal{S}_n$ through $x$ and $y$. Since by assumption $x$ is linearly independent of $y$ this line does not intersect the origin and hence it does not intersect the boundary of $F_{\mathcal{C}}(x)$. Therefore, $L\subset \ri F_{\mathcal{C}}(x)$, which is impossible, since $\mathcal{P}_n$ is line-free and so must be $\mathcal{C}\subset \mathcal{P}_n$. Conversely, it is clear that if $\dim F_{\mathcal{C}}(x)=1$, we must have $\rb F_{\mathcal{C}}(x)=\{0\}$, since $\mathcal{C}$ is line-free.
\end{myproof}

\subsection{Some auxiliary results}

In the following, we will list some useful general results on symmetric bilinear forms: 
\begin{lem}
\label{lem: existence u2-v2}
Let $V$ be a real two-dimensional vector space. Then any subspace $U\subset\Sym^2 V$ such that $\dim U\geq 2$ contains an element of the form $u\vee v$ for $u,v\in V$.    
\end{lem}
\begin{myproof}
    Let $V$, $U$ as above. If $\dim U=3$, the statement is trivial. Suppose $\dim U=2$ and that $U$ contains no element of rank one (otherwise, the statement is also trivial). Let $\{u_1^{\otimes 2}+\varepsilon_1 v_1^{\otimes 2},u_2^{\otimes 2}+\varepsilon_2 v_2^{\otimes 2}\}$ be a basis for $U$, where $u_i,v_i\in V$ are linearly independent, $\varepsilon_i=\pm 1$ for $i\in\{1,2\}$. If $\varepsilon_i=-1$ for $i=1$ or $i=2$, we observe $u^{\otimes 2}-v^{\otimes 2}=(u+v)\vee (u-v)$ and the assertion follows. In the case where $\varepsilon_1=1=\varepsilon_2$, we note that $u_1^{\otimes 2}+v_1^{\otimes 2}$ defines a metric on $V^*$. Let $A\in O(V^*)$ with respect to that metric, such that, possibly after rescaling $u_2$ and $v_2$, we have $A^*(u_1)=u_2$. Since $A^*(u_1)^{\otimes 2}+A^*(v_1)^{\otimes 2}=u_1^{\otimes 2}+v_1^{\otimes 2}$, it follows that $v_2^{\otimes 2}-A^*(v_1)^{\otimes 2}\in U$.
\end{myproof}

\begin{lem}
\label{lem: existence of degenerate el in 3d}
    Let $V$ be a real vector space such that $\dim V=3$. Then any subspace $U\subset \Sym^2 V$ such that $\dim U\geq 2$ has an element $T\in U$ such that $\rk T<3$.
\end{lem}
\begin{myproof}
    Consider two linearly independent elements $T_1, T_2\in U$. Suppose that $\rk T_i=3$ for $i=1,2$. We consider the polynomial
    \begin{align*}
    \R\ni t\mapsto p(t):=&\,\det(T_1 +t\, T_2)\\
    =&\,\det(T_2)\, t^3+\mu_2(T_1,T_2)\, t^2+\mu_1(T_1,T_2)\, t+\det(T_1),    
    \end{align*}
    where the coefficients $\mu_i\in\R$ depend on $T_1,T_2$. Since by assumption $\det(T_2)\neq 0$, there is some $t_0\in \R$ such that $p(t_0)=0$. In particular, $T:=T_1+t_0\, T_2\in U$ has a non-trivial kernel.
\end{myproof}

We also recall the following well-known property of semi-definite bilinear forms:

\begin{lem}
\label{lem: ker of psd bilin form}
    Suppose $b$ is a positive semi-definite bilinear form on some vector space $V$. Then $v\in \ker(b\colon V\to V^*)$ if and only if $b(v,v)=0$.
\end{lem}
\section{The positivity cone}
\label{sect: pos cone}
In this section we will describe our general setup. We begin by introducing a vector space $W$ consisting of four-tensors on a finite-dimensional vector space $V$ with certain symmetries. We will construct two  natural linear maps, $R$ and $R'$, associated with 
$W$, which are defined via restriction to certain subspaces. Then, we will study the properties of the space $W$ and of the restriction maps $R$ and $R'$. Furthermore, we will define a cone $\mathcal{C}_W\subset W$ in $W$, which is the main object of interest in this work. Our aim is to determine the extremal elements of this cone. In Section~\ref{sect: ers} we will establish some properties of the extremal elements of $\mathcal{C}_W$ as well as a complete classification of all extremal rays in the case where $\dim V=2$.

\subsection{Relevant spaces and linear maps}
Let $V$ be a real $n$-dimensional vector space. We denote by $W$ the vector space
\[ W :=\left\{ S \in \mathrm{Sym}^2 (\mathrm{Sym}^2\, V^*)\oplus \mathrm{Sym}^2 \left({\Lambda}^2 V^*\right)
:
\tau S = S \right\}\subset {V^*}^{\otimes 4},\]
where 
$(\tau S)(v_1,v_2,v_3,v_4) := 
S(v_1,v_4,v_3,v_2)$ for all $v_1,\ldots ,v_4\in V$. (For the physical motivation of why we should consider the space of four-tensors with the symmetries described above, see Appendix~\ref{app: tensor M}.) We denote by $R$ and $R'$ the restriction maps
\begin{align*}
R : W \longrightarrow \mathrm{Sym}^2 (\mathrm{Sym}^2\, V^*), \quad   R' : W \longrightarrow \mathrm{Sym}^2 \left( {\Lambda}^2 V^*\right), 
\end{align*}
respectively. We furthermore consider the total symmetrization map
\[ \mathrm{tot} : W\rightarrow \mathrm{Sym}^4\, V^*\]
on $W$ and denote by $K_{W}\subset W$ its kernel. Then we have the following decomposition
\begin{equation*} W = \mathrm{Sym}^4\, V^*\oplus K_W.\end{equation*}

In the following, we will study some properties of the vector space $W$ and the restriction maps $R$ and $R'$. We first note that the restriction map $R$ is in fact an isomorphism.
\begin{prop}\label{prop: properties restriction map R}The restriction map 
\[ R : W \rightarrow \mathrm{Sym}^2 (\mathrm{Sym}^2\, V^*)\]
is an isomorphism, which restricts to the identity map on $\mathrm{Sym}^4\, V^*$.
\end{prop}
\begin{myproof}
It follows from the symmetries of $W$ that $W\cap\Sym^2(\Sym^2 V^*)=\Sym^4 V^*$. Therefore $R$ restricts to the identity on $\Sym^4 V^*$. To prove that $R$ is an isomorphism, we will start showing that $R$ is injective. For this, we need to show that $W\cap\Sym^2(\Lambda^2 V^*)=\{0\}$. Using the symmetries, we find that for every $S\in W\cap\Sym^2(\Lambda^2 V^*)$, we have
\begin{align*}
    S(v_1,v_2,v_3,v_4) &= S(v_1,v_4,v_3,v_2) \\
                        &= -S(v_4,v_1,v_3,v_2)  \\
                        &= -S(v_4,v_2,v_3,v_1)  \\
                        &= S(v_2,v_4,v_3,v_1)  \\
                        &= S(v_2,v_1,v_3,v_4)  \\
                        &= -S(v_1,v_2,v_3,v_4)
\end{align*}
and therefore, $S = 0$. In order to prove that $R$ is surjective, we show that it has a right-inverse. We set
\begin{align*}
    R^{-1}: \Sym^2 \left(\Sym^2 V^*\right) &\longrightarrow W \\
T &\longmapsto T + h(T)
\end{align*}
where 
\begin{align*}
    h: \Sym^2 (\Sym^2 V^*) &\longrightarrow \Sym^2 (\Lambda^2 V^*) \\
    T&\longmapsto h(T)
\end{align*}
is defined by
  \[ h(T)( e_i \wedge e_j, e_k \wedge e_l) = T(e_i \vee e_l, e_j \vee e_k) - T(e_i \vee e_k, e_j \vee e_l). \]
We show that $S:=R^{-1}(T)$ is indeed in $W$ as claimed. All the defining symmetries of $W$ but the symmetry under $\tau$ follow immediately from the definition. It remains to verify the symmetry under $\tau$, i.e.\ $S(v_1,v_2,v_3,v_4) = S(v_1,v_4,v_3,v_2)$: 
\begin{align*}
        S(v_1,v_2,v_3,v_4)
        &= T(v_1 \vee v_2, v_3 \vee v_4) + h(T)(v_1 \wedge v_2, v_3 \wedge v_4)\\
        &= T(v_1 \vee v_2, v_3 \vee v_4) + T(v_1 \vee v_4, v_2 \vee v_3) - T(v_1 \vee v_3, v_2 \vee v_4) \\
        &= T(v_1 \vee v_4, v_3 \vee v_2) +  T(v_1 \vee v_2, v_4 \vee v_3) -  T(v_1 \vee v_3, v_4 \vee v_2)\\
        &=  T(v_1 \vee v_4, v_3 \vee v_2) + h(T)(v_1 \wedge v_4, v_3 \wedge v_2)\\
        &= S(v_1,v_4,v_3,v_2),
\end{align*}
    where we used the definition of $S$ in the last step.
This now shows that $T + h(T) \in W$ for all $T \in  \Sym^2 (\Sym^2 V^*)$ as desired. Since clearly $R(R^{-1}(T)) = T$, surjectivity of $R$ follows.     
\end{myproof}
Next, we will study the kernel and image of $R'$.
\begin{prop}
\label{prop: properties restriction map R'}
The restriction map 
\[ R' : W \longrightarrow \mathrm{Sym}^2 \left( {\Lambda}^2 V^*\right)\]
is an epimorphism onto the space of algebraic curvature tensors 
\[ K(V) := \left\{ S\in \mathrm{Sym}^2 \left( \Lambda^2V^*\right) : \sum_{\mathfrak{S}}S(v_1,v_2,v_3)=0\quad\mbox{for all}\quad
v_1,v_2,v_3\in V\right\},\] 
where $\mathfrak{S}$ indicates the cyclic sum. The kernel of $R'$ is $\mathrm{Sym}^4\, V^*$. 
As a consequence,  
\[ R'|_{K_W} : K_{W}\longrightarrow K(V)\]
is an isomorphism. 
\end{prop}
\begin{myproof}
It follows from $W\cap\Sym^2(\Sym^2 V^*)=\Sym^4 V^*$ that the kernel of $R'$ is the space $\Sym^4 V^*$. To see that the image $R'(K_W)$ is contained in $K(V)$, notice that for $S \in K_W$  we have
\begin{align*}
    0 &= \mathrm{tot}(S)(v_1,v_2,v_3,v_4) \\
    &= \frac{1}{24}\sum_{\sigma \in S_4} (\sigma S)(v_1,v_2,v_3,v_4) \\
    &= \frac{1}{3} \left[ S(v_1,v_2,v_3,v_4) + S(v_2,v_3,v_1,v_4) + S(v_3,v_1,v_2,v_4)\right] \\
    &= \frac{1}{3} \sum_{\mathfrak{S}}S(v_1,v_2,v_3)(v_4),
\end{align*}
where we have used the symmetries of $S$ to reorder the arguments for every permutation $\sigma \in S_4$ from the second line to obtain one of the permutations in the third line. By doing so, we find every permutation eight times since this is the order of the symmetry group of elements in $W$. As a result, we see that the cyclic sum over elements in $K_W$ vanishes. Since the other symmetries are inherited as well, we indeed end up in $K(V)$ after restricting to ${\Lambda}^2 V$. This can be seen explicitly by performing the cyclic sum over $R'(S)(v_1,v_2,v_3,v_4)=(S(v_1,v_2,v_3,v_4)-S(v_2,v_1,v_3,v_4))/2$, which decomposes into two (independently-vanishing) cyclic sums for $S$.

To show that $R'$ maps $W$ surjectively onto $K(V)$, we will provide an explicit formula for the inverse of $R'|_{K_W}$. For $A\in K(V)$ we define a $(0,4)$-tensor 
$\sigma_A$ by 
\[ \sigma_A:= \frac{2}{3}(A + \tau A).\]
We claim that $\sigma_A = (R'|_{K_W})^{-1}(A)$. We first show that $\sigma_A \in W$. By construction $\tau \sigma_A = \sigma_A$. 
To see that  $\sigma_A\in \mathrm{Sym}^2(V^*\otimes V^*)$ it suffices to check 
that $\tau A\in  \mathrm{Sym}^2(V^*\otimes V^*)$. We compute
\[ (\tau A)(v_3,v_4,v_1,v_2)=A(v_3,v_2,v_1,v_4)=A(v_1,v_4,v_3,v_2)=(\tau A)(v_1,v_2,v_3,v_4),\]
proving that indeed $\tau A\in \mathrm{Sym}^2(V^*\otimes V^*)$. 
To prove that $\sigma_A\in W$ it remains only to check that $\sigma_A(x\wedge y, z\vee w)=0$
for all $x,y,z,w\in V$. We compute 
\begin{eqnarray*} \sigma_A(x\wedge y, z\vee w)&=&\frac14 (\sigma_A(x,y,z,w) -\sigma_A(y,x,z,w) + 
\sigma_A(x,y,w,z)-\sigma_A(y,x,w,z))\\
&=&\frac{1}{6}( A(x,y,z,w) -A(y,x,z,w) + 
A(x,y,w,z)-A(y,x,w,z)\\
&&+A(x,w,z,y) -A(y,w,z,x) + 
A(x,z,w,y)-A(y,z,w,x))\\
&=&0,\end{eqnarray*}
where we have used only the symmetries of $A\in \mathrm{Sym}^2 \left( {\Lambda}^2 V^*\right)$.
Using in addition the Bianchi identity (i.e. the vanishing cyclic sum), we 
can easily check that $R'(\sigma_A)=A$: 
\begin{align*}
    \sigma_A(x\wedge y,z\wedge w) =&\,\frac{1}{6}\left(4\,A(x,y,z,w)-2\,A(w,x,z,y)-2\,A(x,z,w,y)\right)\\
    =&\, A(x,y,z,w).
\end{align*}

Finally, we note that indeed $\sigma_A\in K_W$, i.e.~$\mathrm{tot}(\sigma_A)=0$, since $\mathrm{tot}(A) = \mathrm{tot}(\tau A)=0$,
due to the skew-symmetry of $A$ in at least one pair of arguments. 
\end{myproof}
We conclude this section by noting the following useful formula for elements $S\in W$ evaluated on a pair $(u^{\otimes 2},v^{\otimes 2})$ for some $u,v\in V$: 
\begin{lem}
\label{lem: formula S(v2,w2)}
    Let $S\in W$, $u,v\in V$. Then
    \[ S(u^{\otimes 2},v^{\otimes 2})=S(u\vee v,u\vee v)-S(u\wedge v,u\wedge v).\]
\end{lem}
\begin{myproof}
For $S\in W$, $u,v\in V$, we find
\begin{align*}
S(u^{\otimes 2},v^{\otimes 2})=&\,S(u\otimes v,v\otimes u)\\=&\,S(u\vee v+u\wedge v,u\vee v-u\wedge v)\\=&\,S(u\vee v,u\vee v)-S(u\wedge v,u\wedge v),
\end{align*}
where we used the invariance under $\tau$ in the first step and the fact that $\Lambda^2 V$ and $\Sym^2 V$ are orthogonal with respect to $S$ for the last equality.
\end{myproof}
\subsection{The positivity cone and its extremal rays}
\label{sect: ers}
We are interested in the cone 
\[ \mathcal{C}_W:= \{ S\in W : S\ge 0\}\] 
of positive semi-definite elements of $W\subset \mathrm{Sym}^2(V^*\otimes V^*)$. 
We furthermore denote by
\[ \mathcal{C}_{\Sym}:=\left\{ S \in \mathrm{Sym}^2 (\mathrm{Sym}^2\, V^*):S\geq0\right\}  \]
the cone of positive semi-definite symmetric bilinear forms on $\Sym^2V^*$.
The following is clear: 
\begin{lem}
\label{lem: R and R' on CW}
    Let $S\in W$. Then $S\in\mathcal{C}_{W}$ if and only if both $R(S)\geq 0$ and $R'(S)\geq 0$.
\end{lem}
In the following proposition we show that elements in $\mathcal{C}_W$ must always have some non-vanishing totally symmetric part.
\begin{prop} \label{K_W:prop}The subspace $K_W$ has trivial intersection with the cone $\mathcal{C}_W$, i.e.~$K_W\cap \mathcal{C}_W=0$. 
\end{prop}
\begin{myproof}
By the proof of Proposition~\ref{prop: properties restriction map R'}, any element 
of $K_W$ is of the form $\sigma_A$ for some $A\in K(V)$. The tensor $\sigma_A$ belongs to 
$\mathcal{C}_W$ if and only if $R(\sigma_A)\ge 0$ and $R'(\sigma_A)=A\ge 0$. The latter 
condition implies $0\le A(u\wedge v,u\wedge v)=A(u,v,u,v)$ for all $u,v\in V$, while the former
implies 
\begin{eqnarray*} 0&\le& \sigma_A(u^2+v^2,u^2+v^2)=\sigma_A(u,u,v,v) + \sigma_A(v,v,u,u)\\&=&\frac23(A(u,v,v,u)+A(v,u,u,v))
= -\frac43 A(u,v,u,v).\end{eqnarray*}
This proves that $A(u,v,u,v)=0$ for all $u,v\in V$ and hence, $A=0$ by Lemma~\ref{lem: ker of psd bilin form}. 
This proves Proposition~\ref{K_W:prop}. 
\end{myproof}

In what follows we will develop criteria for determining whether a given element of $\mathcal{C}_W$ is extremal, based on simple properties of the element. We begin by studying the elements of $\mathcal{C}_W$ which project to extremal elements in $\mathcal{C}_{\Sym}$. Then we will establish several necessary conditions for an element of $\mathcal{C}_W$ to be extremal, formulated in terms of the dimension of its kernel. We will conclude the section with a full classification of all extremal elements of $\mathcal{C}_W$ in the special case where $\dim V=2$.

\subsubsection{Extremal rays that project to extremal rays}
\label{R-extremal}
We already saw in Lemma~\ref{lem: R and R' on CW} that $R(\mathcal{C}_W)\subset \mathcal{C}_{\Sym}$. We will now consider elements in $\mathcal{C}_W$ projecting onto extremal elements in $\mathcal{C}_{\Sym}$ under $R$. 
\begin{prop} 
\label{prop: er proj to ers}
Let $S\in \mathcal{C}_W$ be such that $R(S)$ is extremal in $\mathcal{C}_{\Sym}$. Then $S$ is extremal in $\mathcal{C}_W$ and there exists $\alpha, \beta \in V^*$ such that 
\begin{equation} \label{R-1:eq}
S=(\alpha \vee \beta)^{\otimes 2} + (\alpha \wedge \beta)^{\otimes 2}=\frac12 (\alpha \otimes \beta \otimes \alpha \otimes \beta+ \beta \otimes \alpha \otimes \beta \otimes \alpha).\end{equation} 
In particular, every element of $\mathcal{C}_W$ that is of the form as in \eqref{R-1:eq} for some $\alpha,\beta\in V^*\setminus\{0\}$ is extremal.
\end{prop} 

To prove Proposition~\ref{prop: er proj to ers}, we will need the following lemmas.
\begin{lem}
\label{lem: expr for phi} 
The map $h := R' \circ R^{-1} : \mathrm{Sym}^2 (\mathrm{Sym}^2\, V^*)\rightarrow \mathrm{Sym}^2 ({\Lambda}^2V^*)$ is given by: 
\[ h (\gamma_1\vee \gamma_2 )(x,y,z,w) = \frac12 (\gamma_1(x,w)\gamma_2(z,y) +
\gamma_2(x,w)\gamma_1(z,y)-\gamma_1(y,w)\gamma_2(z,x)-\gamma_2(y,w)\gamma_1(z,x)),\]
for all $\gamma_1, \gamma_2 \in \mathrm{Sym}^2\, V^*$. 
\end{lem}
\begin{myproof}
It follows from the proof of Proposition~\ref{prop: properties restriction map R} that for every $T\in\Sym^2(\Sym^2V^*)$, we have $R'\circ R^{-1}(T)=h(T)$, where 
\[ h(T)( e_i \wedge e_j, e_k \wedge e_l) = T(e_i \vee e_l, e_j \vee e_k) - T(e_i \vee e_k, e_j \vee e_l). \]
Applying this to $T=\gamma_1\vee \gamma_2$ for $\gamma_1,\gamma_2\in\Sym^2 V^*$, we arrive at the desired formula.
\end{myproof}

\begin{rem} The map $h$ is related to the well-known Kulkarni--Nomizu product by 
 \[ h (\gamma_1\vee \gamma_2 ) = -\frac12 \gamma_1 \owedge \gamma_2.\]
\end{rem}

\begin{lem}
    Let $\gamma \in 
    \mathrm{Sym}^2\, V^*$ and consider $\rho_\gamma :=  h (\gamma^{\otimes 2})$. If $\rho_\gamma\geq 0$, then $\rk \gamma \leq 2$. Moreover, if $\rk \gamma = 2$, $\gamma$ has signature of type $(+,-,0, \dots, 0)$.
\end{lem}
\begin{myproof}
    Let $E \subseteq V$ be a plane with basis $(x,y)$. Then
    \begin{align*}
        \rho_\gamma(x \wedge y, x \wedge y) &= \rho_\gamma(x, y, x, y) \\
        &= \gamma(x,y)^2 - \gamma(x,x) \gamma(y,y) \\
        &= - \det \begin{pmatrix}
            \gamma(x,x) & \gamma(x,y) \\
            \gamma(x,y) & \gamma(y,y)
        \end{pmatrix}
    \end{align*}
This is non-negative if and only if $\gamma$ restricted to $E$ is indefinite or degenerate. Since this must hold for all planes $E$, the claim follows.
\end{myproof}

We are now ready to prove Proposition~\ref{prop: er proj to ers}.
\begin{myproof}[Proof of Proposition~\ref{prop: er proj to ers}]
First, we show that every such element must be of the form as in equation~\eqref{R-1:eq}. To see this, note that the extremal elements in $\mathcal{C}_{\Sym}$ are precisely those of rank one. Suppose $S\in \mathcal{C}_W$ is such that $\rk R(S)=1$. 
Then we can write $R(S)=\gamma^{\otimes2}$ for some $\gamma\in\Sym^2 V^*$. Moreover, since $S \in \mathcal{C}_W$, we must have $R'(S)=\rho_\gamma\geq 0$. 
By the previous lemma, the latter condition is satisfied precisely
if $\gamma$ can be written as $\gamma = \alpha \vee \beta$ for some $\alpha, \beta \in V^*$. It follows that $S=\gamma^{\otimes2}+h(\gamma^{\otimes2})$is of the form as in~\eqref{R-1:eq}, since $h(\gamma^{\otimes 2})=(\alpha \wedge \beta)^2$. 

Next, we show that every element of this form is extremal in $\mathcal{C}_W$, using Corollary \ref{cor: nec extremal}. Let $S\in\mathcal{C}_W$ be of the form~\eqref{R-1:eq} and $S'\in\mathcal{C}_W$ such that $\ker S\subsetneq \ker S'$. Then in particular $\ker R(S)\subset \ker R(S')$. Since $\rk R(S)=1$ this implies that $R(S')=\lambda R(S)$ for some $\lambda\geq 0$. By Proposition~\ref{prop: properties restriction map R} this means that $S'=\lambda\, S$ for some $\lambda\geq 0$ and hence $\ker S\subsetneq \ker S'$ readily implies $S'=0$.
\end{myproof}

\subsubsection{Some necessary conditions for extremality}
We will now study some necessary conditions for extremal elements of $\mathcal{C}_W$. Since the case where $\dim V=1$ is trivial, we will assume that $\dim V\geq 2$ throughout the section. First, we observe that for extremal elements $S\in \mathcal{C}_W$ the kernel of the projection $R(S)$ must be non-trivial.  
\begin{lem}
  \label{lem: R full rank} Let $S\in \mathcal{C}_W$ be extremal. Then $\ker R(S)\neq\{0\}$.    
\end{lem}
\begin{myproof}
 Let $S\in\mathcal{C}_W$. If $R(S)$ is strictly positive, then for every $\alpha\in V^*$, we have \[\ker \alpha^{\otimes 4}\supsetneq \ker S.\] Indeed, $\ker \alpha^{\otimes 4}\supset \Lambda^2 V\supset \ker S$, since $R(S)$ has full rank and we have \[\rk \alpha^{\otimes 4}=1<\dim \Sym^2 V\leq \rk S.\]
 By Corollary~\ref{cor: nec extremal} it follows that $S$ is not extremal.\end{myproof}

A similar statement can be made concerning the kernel of $R'(S)$:
\begin{lem}
    \label{lem: R' full rank}
    Let $S\in\mathcal{C}_W$ such that $R'(S)$ has full rank and such that $R(S)$ is not extremal in $\mathcal{C}_{\Sym}$. Then $S$ is not extremal in $\mathcal{C}_W$. Moreover, if $\dim V>2$, then for extremal $S\in \mathcal{C}_W$, $R'(S)$ has a non-trivial kernel.
\end{lem}
\begin{myproof}
    Let $S\in\mathcal{C}_W$ as above. Then there is some $T\in\mathcal{C}_{\Sym}\setminus \{0\}$ such that $\ker R(S)\subsetneq \ker T$. We will show that $T$ can be chosen such that $T+h(T)\in \mathcal{C}_W$, i.e. $h(T)\geq 0$. Since $R'(S)>0$, we may assume that $R'(S)-h(T)>  0$, possibly after rescaling $T$ by a small positive factor. Furthermore, since $\ker R(S)\subsetneq \ker T$, again up to rescaling $T$ we may assume that $R(S)-T\in\mathcal{C}_{\Sym}$. 

    Let $\Vert \cdot \Vert$ be an auxiliary norm on $\Lambda^2 V$ and let 
    \[ \mathcal{K}_1:= \{ Q\in \Lambda^2 V: \Vert Q\Vert =1\}.\]
    We define the auxiliary functions
    \begin{align*}
        d\colon \Sym^2(\Lambda^2 V^*)&\longrightarrow \R\\
        A&\longmapsto \min_{Q\in \mathcal{K}_1}\{ A(Q,Q)\}
    \end{align*}
    and 
    \begin{align*}
        f\colon [0,1] &\longrightarrow \R\\
        t&\longmapsto d\left( h(T)+ t(R'(S)-h(T))\right).
    \end{align*}
    If $d\left(h(T)\right)\geq 0$, then $h(T)\geq 0$ and therefore $T+h(T)\in\mathcal{C}_W$. Moreover, since by assumption $\ker T\supsetneq \ker R(S)$ and $\ker h(T)\supset \{0\}=\ker R'(S)$, we know that $\ker T+h(T)\supsetneq \ker S$ and it follows from Corollary~\ref{cor: nec extremal} that $S$ is not extremal. If $d\left(h(T)\right)< 0$, then we have $f(0)<0$ and $f(1)>0$, since $R'(S)-h(T)>0$. It follows that there is some $\lambda\in (0,1)$ such that $f(\lambda)=0$. Replacing $T$ by $\Tilde{T}:=T+\lambda \cdot (R(S)-T)\in \mathcal{C}_{\Sym}$, we find $h(\Tilde{T})\geq 0$. Moreover, we know that $\ker\tilde{T}\supset\ker R(S)$ and $\ker h(\tilde{T})\supsetneq \ker R'(S)=\{0\}$, since by construction $d(h(\tilde{T}))=0$. Thus, $\ker\left(\tilde{T}+h(\tilde{T})\right)\supsetneq \ker S$ and by Corollary~\ref{cor: nec extremal} we conclude that $S$ is not extremal. 

    Now suppose $\dim V>2$. By Proposition~\ref{prop: er proj to ers}, elements $S\in\mathcal{C}_W$ such that $R(S)$ is extremal in $\mathcal{C}_{\Sym}$ are of the form
    \[
    S=(\alpha\vee \beta)^{\otimes 2}+ (\alpha\wedge\beta)^{\otimes 2} 
    \]
    for some $\alpha,\beta\in V^*$. Therefore, for such $S$ we have $\rk R'(S)=1$. Since for $\dim V>2$ we have $\dim \Lambda^2 V>1$, in this case $R'(S)$ has a non-trivial kernel.
\end{myproof}

Next, we will establish a lower bound for the kernel of extremal elements.
\begin{prop}
\label{prop: dim ker extremal elements}
    Let $\dim V=n\geq 2$. If $S\in\mathcal{C}_W$ is extremal, then $\dim(\ker S)\geq n$. For odd dimensions $n$, we even have $\dim(\ker S)> n$.
\end{prop}
\begin{myproof}
    By Lemma \ref{lem: R full rank}, we know that for extremal $S\in \mathcal{C}_W$, we have $\ker S\cap \Sym^2 V\neq \{0\}$. If $S\in \mathcal{C}_W\cap \Sym^4 V^*$, then we furthermore know that $\Lambda^2 V\subset \ker S$. Therefore, 
    \[
    \dim \ker S> \dim \Lambda^2 V\geq n-1 \quad \text{for }n\geq 2.
    \]

    Similarly, if $R(S)$ is extremal in $\mathcal{C}_{\Sym}$, then 
    \[
    \dim \ker S \geq \dim \ker R(S)=\frac{n(n+1)}{2}-1\geq n\quad \text{for }n\geq 2.
    \]
   
    Now suppose $S\in\mathcal{C}_W\setminus \Sym^4 V^*$ such that $R(S)$ is not extremal in $\mathcal{C}_{\Sym}$. Let $k=\dim \ker S$ and $\{Q_i\}_{i=1}^k$ a basis for $\ker S$. We may assume that each $Q_i\in \Sym^2 V$ or $Q_i\in \Lambda^2 V$. We pick some $\alpha\in V^*\setminus\{0\}$ and consider the maps 
    \[Q_i(\alpha,\cdot)\colon V^*\longrightarrow \mathbb{R}.\]
    Then for $k<n$, we find
    \[ \bigcap_{i=1}^k \ker Q_i(\alpha,\cdot)\neq \{0\}.\]
It follows that if $k<n$ there is some $\beta\in V^*\setminus\{0\}$ such that
\[ 
Q_i(\alpha,\beta)=\alpha\otimes \beta (Q_i)=0\quad \text{for } i=1,...k.
\]
Since the $Q_i$ are all either symmetric or anti-symmetric, we also have 
\[
\beta\otimes \alpha (Q_i)=0\quad \text{for }i=1,...,k.
\]
It follows that 
\[
(\alpha\otimes \beta)^{\otimes 2}+(\beta\otimes \alpha)^{\otimes 2}|_{\ker S}\equiv 0.
\]
Therefore, by Corollary~\ref{cor: nec extremal}, for extremal $S$, we must have $k\geq n$. 

For odd dimensions $n$, we can furthermore use that $\ker S\cap \Lambda ^2 V\neq \{0\}$ by Lemma \ref{lem: R' full rank}. Therefore, for some $i$ we have $Q_i\in \Lambda^2 V$ and since $V$ is odd-dimensional, this $Q_i$ is degenerate. Choosing $\alpha\in\ker( Q_i\colon V^*\rightarrow V)$ the same argument as above implies that $n>k$.
\end{myproof}

For elements in $\Sym^4 V^*\cap \mathcal{C}_W$, more can be said.

\begin{lem}
\label{lem: rank one in ker of tot sym}
    Let $S\in\Sym^4 V^*\cap\mathcal{C}_W$. Suppose that there is some $x\in V\setminus\{0\}$ such that $x^{\otimes 2}\in\ker S$. Then for every $v\in V$ we have $x\otimes v\in\ker S$ and $v\otimes x\in\ker S$. In particular, $S$ restricts to an element of $\mathcal{C}_{\tilde{W}}$ for the space $\tilde{V}=V/(\R x)$.
\end{lem}
\begin{myproof}
    Let $S\in\Sym^4 V^*\cap \mathcal{C}_W$ and suppose that $x^{\otimes 2}\in \ker S$ for some $x\in V\setminus\{0\}$. Then using the symmetries of $S$ we find
    \begin{align*}
        0=S(x^{\otimes 2},v^{\otimes 2})=S(x\otimes v,x\otimes v)=S(v\otimes x,v\otimes x).
    \end{align*}
    The statement follows now from Lemma~\ref{lem: ker of psd bilin form}.
\end{myproof}

\begin{lem}
\label{lem: indef elements in ker of tot sym}
    Let $S\in \Sym^4V^*\cap\mathcal{C}_{W}$ such that there are no rank one elements in $\ker S$. Then every non-trivial element in the kernel of $R(S)$ is indefinite. Moreover, if $\dim V\geq 2$ and $S\in \Sym^4 V^*\cap\mathcal{C}_W$ is extremal then $\dim\ker R(S)\geq 2$.  
\end{lem}   
\begin{myproof}
    Suppose $S\in\mathcal{C}_W\cap\Sym^4 V^*$ such that there are no rank one elements in the kernel of $S$. Then it follows from the observation
    \begin{align}
\label{eq: formula S(v2,w2) when S tot sym}
    S(v^{\otimes 2},w^{\otimes 2})=S(v\otimes w, v\otimes w)\geq 0
\end{align}
    that every element in the $\ker S\cap\Sym^2 V$ must be indefinite. Indeed, by Sylvester's theorem we know that every element $T\in\Sym^2 V$ can be written in the form $T=\sum_i \varepsilon_i e_i^{\otimes 2}$ for $\varepsilon_i\in\{\pm1\}$ and linearly independent $\{e_i\}\subset V$.  If $\varepsilon_i=1$ for every $i$, it follows from equation~\eqref{eq: formula S(v2,w2) when S tot sym} that
    \begin{align*}
        S(T,T)\geq \sum_i S(e_i^{\otimes 2},e_i^{\otimes 2})>0
    \end{align*}
    for $T\neq 0$. By Lemma~\ref{lem: ker of psd bilin form} we conclude that $T\notin\ker S$.
    
    Now suppose $\dim V\geq 2$ and let $S\in \mathcal{C}_W\cap\Sym^4 V^*$ extremal. Then by Lemma~\ref{lem: R full rank} the kernel of $R(S)$ is at least one-dimensional. Suppose that $\ker S\cap \Sym^2 V=\Span\{T\}$ for some $T\in \Sym^2 V\setminus\{0\}$. By Lemma~\ref{lem: rank one in ker of tot sym} we know that we must have $\rk T>1$ since otherwise $T=x^{\otimes 2}$ and $\ker R(S)\supset x\vee V$ in contrast to the assumption that $\ker R(S)=\R\, T$. But then $T$ must be indefinite and therefore we can find some $\alpha\in V^*\setminus\{0\}$ such that
\begin{align*}
    \alpha^{\otimes 2}(T)=T(\alpha,\alpha)=0.
\end{align*}
It follows that $\ker\alpha^{\otimes 4}\supsetneq\ker S$, which cannot happen for extremal $S$. Thus, for extremal $S$ we must have $\dim (\ker S\cap \Sym^2V)\geq 2$.
\end{myproof}
In the case where $V$ is three-dimensional Lemma~\ref{lem: indef elements in ker of tot sym} has some immediate consequences, which will be useful in the following section.
\begin{cor}
\label{cor: ker S for extremal S in 3d}
    Suppose $\dim V=3$. Then  for extremal $S\in \mathcal{C}_W$, we have 
    \[\dim (\ker S\cap \Sym^2 V)\geq 2. \]
\end{cor}
\begin{myproof}
By Proposition \ref{prop: dim ker extremal elements} we know that $\dim \ker S\geq 4$. If $\dim(\ker S\cap \Lambda^2 V)<3$, this implies $\dim(\ker S\cap \Sym^2 V)\geq 2$. If $\dim(\ker S\cap \Lambda^2 V)=3$, then $S\in \mathcal{C}_W\cap \Sym^4 V^*$ and the assertion follows from Lemma~\ref{lem: indef elements in ker of tot sym}.
\end{myproof}

\begin{cor}
\label{cor: existence element rank <3}
    Suppose $\dim V=3$. If $S\in \mathcal{C}_W$ is extremal, then there is some $Q\in\Sym^2 V\cap\ker S$ such that $\rk Q<3$.    
\end{cor}
\begin{myproof}
    This follows from Corollary~\ref{cor: ker S for extremal S in 3d} and Lemma~\ref{lem: existence of degenerate el in 3d}.
\end{myproof}
The following theorem already appeared in~\cite{Li_et_al21}, but with a different proof.
\begin{thm}
\label{thm: ers in 2d}
    Suppose $\dim V=2$. Then $S\in \mathcal{C}_W$ is extremal if and only if $R(S)$ is extremal in~$\mathcal{C}_{\Sym}$.
\end{thm}
\begin{myproof}
    Suppose $S\in\mathcal{C}_W$ is extremal. If $R(S)$ is not extremal in $\mathcal{C}_{\Sym}$, then since $\dim \Lambda^2 V=1$, we know by Lemma~\ref{lem: R' full rank} that $R'(S)\equiv 0$. Therefore, $S\in\Sym^4 V^*\cap \mathcal{C}_W$. Then by Lemma~\ref{lem: indef elements in ker of tot sym} we know that $\dim\ker R(S)\geq 2$, which implies that $\rk R(S)=1$ and hence that $R(S)$ is extremal in $\mathcal{C}_{\Sym}$, a contradiction.
\end{myproof}

\section{Classification of extremal elements in the case $\dim V=3$}
\label{sect: ers in 3d}
In this section, we will determine the extremal elements of $\mathcal{C}_W$ for three-dimensional $V$. Our goal is to prove the following theorem.
\begin{thm}
\label{thm: ers in 3d}
Suppose $\dim V=3$. Then $S\in \mathcal{C}_W$ is extremal if and only if it is of one of the following types: 
\begin{enumerate}
    \item For some $\alpha\in V^*\setminus \{0\}$, $S=\alpha^4$.
    \item For linearly independent $\alpha_1,\alpha_2\in V^*$, $S=(\alpha_1\vee \alpha_2)^{\otimes 2}+(\alpha_1\wedge \alpha_2)^{\otimes 2}$.
    \item For a basis $\{\alpha_1,\alpha_2,\alpha_3\}$ of $V^*$ and parameters $g,d,h$ such that $g^2>1-d^2+dh$ we have
    \begin{align*}
        S=S_{\mathrm{tot}}+2\,(g^2+d^2-1-dh)\left[(\alpha_2\otimes \alpha_3)^{\otimes 2}+(\alpha_3\otimes \alpha_2)^{\otimes 2}\right],
    \end{align*}
    where
    \begin{align*}
    S_{\mathrm{tot}}=&\,\alpha_1^4+6\,\alpha_1^2\alpha_2^2+6\,\alpha_1^2\alpha_3^2+12d\,\alpha_1\alpha_2^2\alpha_3+12g\,\alpha_1\alpha_2\alpha_3^2+4h\,\alpha_1\alpha_3^3+(1+d^2)\,\alpha_2^4\\&+4dg\,\alpha_2^3\alpha_3+6(1+dh)\,\alpha_2^2\alpha_3^2+4g(d+h)\,\alpha_2\alpha_3^3+(1+g^2+h^2)\,\alpha_3^4.    
    \end{align*}
\end{enumerate}
\end{thm}
It will be useful to distinguish the cases of different rank of $R'(S)$ for $S\in \mathcal{C}_W$. Since $\dim \Lambda^2 V=3$, we have $0\leq \rk R'(S)\leq 3$. By Lemma~\ref{lem: R' full rank}, we know that $S\in \mathcal{C}_W$ is not extremal if $\rk R'(S)=3$. In view of Proposition~\ref{prop: er proj to ers}, we will furthermore restrict our study to $S\in\mathcal{C}_W$ such that $R(S)$ is not extremal in $\mathcal{C}_{\Sym}$, i.e.~we assume $\rk R(S)>1$. The general structure is outlined in the following diagram: 

\begin{center}
\begin{tikzpicture}[sibling distance=10em,
  blocknode/.style = {shape=rectangle, rounded corners,
    draw, align=center},textnode/.style={shape=rectangle, rounded corners,
     align=center},->]

  \node[blocknode] {$S\in \mathcal{C}_W$ s.t.~$\rk R(S)>1$}
    child { node[blocknode] {$\rk R'(S)=0$} child {node[textnode] { \textit{not extremal}\\ \textit{by Theorem \ref{thm: ers in sym4}}}}}
    child { node[blocknode] {$\rk R'(S)=1$}
      child { node[textnode] {\textit{may or may not} \\ \textit{be extremal}\\
      \textit{see Section~\ref{sect: rk R'=1}}}}}
    child { node[blocknode] {$\rk R'(S)=2$}
        child { node[textnode] {\textit{not extremal}\\ \textit{see Section~\ref{sect: rk R'=2}}}}}
        child { node[blocknode] {$\rk R'(S)=3$}
            child { node[textnode] {\textit{not extremal}\\ \textit{by Lemma~\ref{lem: R' full rank}}}}};
\end{tikzpicture}
\end{center}

\subsection{Case $\rk R'(S)=0$}

If $R'(S)$ vanishes for some $S\in W$ then we know that $S\in \Sym^4V^*$. For elements in the intersection of this subspace with $\mathcal{C}_W$ we will show the following:
\begin{thm}
\label{thm: ers in sym4}
    Suppose $\dim V=3$. Then $S\in \Sym^4 V^*\cap \mathcal{C}_W$ is extremal in $\mathcal{C}_W$ if and only if $R(S)$ is extremal in $\mathcal{C}_{\Sym}$.
\end{thm}

As outlined in the following diagram, to prove Theorem~\ref{thm: ers in sym4} we distinguish whether or not an element $S\in\mathcal{C}_W\cap \Sym^4 V^*$ contains an element of rank one.

\begin{center}
\begin{tikzpicture}[sibling distance=15em,
  blocknode/.style = {shape=rectangle, rounded corners,
    draw, align=center},textnode/.style={shape=rectangle, rounded corners,
     align=center},->]

  \node[blocknode] {$S\in \mathcal{C}_W\cap\Sym^4 V^*$ s.t.~$\rk R(S)>1$}
    child { node[blocknode] {$\exists\, x\in V\setminus\{0\}\colon x^{\otimes 2}\in\ker S$} child {node[textnode] { \textit{not extremal}\\ \textit{by Proposition~\ref{prop: tot sym rank one in ker}}}}}
    child { node[blocknode] {$\nexists\,x\in V\setminus\{0\}\colon x^{\otimes 2}\in\ker S$}
      child { node[textnode] {\textit{not extremal} \\
      \textit{by Proposition~\ref{prop: tot sym no rank one in ker}}}}};
\end{tikzpicture}
\end{center}

\begin{prop}
\label{prop: tot sym rank one in ker}
    Suppose $\dim V=3$. Let $S\in \Sym^4 V^*\cap \mathcal{C}_W$ such that $\ker S$ contains an element of rank one. Then $S$ is not extremal unless $R(S)$ is extremal in $\mathcal{C}_{\Sym}$.
\end{prop}
\begin{myproof}
   This follows from Lemma~\ref{lem: rank one in ker of tot sym} and Theorem~\ref{thm: ers in 2d}.
\end{myproof}
In order to show that elements in $\Sym^4 V^*\cap \mathcal{C}_W$ containing no rank one elements in their kernels cannot be extremal, we need the following lemma.
\begin{lem}
\label{lem: explicit tot sym no rank one in ker}
    Suppose $\dim V=3$. Let $S \in \Sym^4 V^*\cap\mathcal{C}_W$ such that $\ker R(S)$ contains no elements of rank one. Then there is a basis $\{x,y,z\}$ of $V$ and parameters $a,b,c,d\in \R$ such that with respect to the basis $\{x^{\otimes2},y^{\otimes 2},z^{\otimes 2},x\vee y,x\vee z,y\vee z\}$ of $\Sym^2 V$, $R(S)$ is represented by the matrix
    \begin{align*}
    R(S) = \left(\begin{matrix}1 & 0 & a & 0 & 0 & 0\\0 & 1 & b & 0 & 0 & 0\\a & b & 1 & 0 & c & d\\0 & 0 & 0 & 0 & 0 & 0\\0 & 0 & c & 0 & a & 0\\0 & 0 & d & 0 & 0 & b\end{matrix}\right).    
    \end{align*}
\end{lem}
\begin{myproof}
    Let $S\in\Sym^4V^*\cap\mathcal{C}_W$ as in Lemma~\ref{lem: explicit tot sym no rank one in ker}. Then we know by Corollary~\ref{cor: existence element rank <3} that $\ker R(S)$ contains an element $T$ of rank two. Moreover, by Lemma~\ref{lem: indef elements in ker of tot sym} we know that $T$ must be indefinite. Therefore, $T=x\vee y\in \ker S$ for some linearly independent $x,y\in V$. Using again that there are no elements of rank one in the kernel of $S$, we may assume that, up to rescaling of $T$, we have
    \begin{align*}
    S(x^{\otimes 2},x^{\otimes 2})=1=S(y^{\otimes 2},y^{\otimes 2}).   
    \end{align*}
    Let furthermore $z\in V\setminus\{0\}$ such that 
    \begin{align*}
        S(x^{\otimes 2},x\vee z)=0=S(y^{\otimes 2},y\vee z).
    \end{align*}
    Such $z$ exists, since $\dim V=3$ and $z$ is constrained by two linear equations. Moreover, we must have $z\in V\setminus \Span\{x,y\}$. Indeed, suppose $z=t_1\,x+t_2\,y$ for some $t_i\in\R$, $i=1,2$. Then it follows that
    \begin{align*}
        0=S(x^{\otimes 2},x\vee z)=t_1,\quad 0=S(y^{\otimes2},y\vee z)=t_2,
    \end{align*}
    in contrast to the assumption that $z\neq 0$. After rescaling $z$ we may assume that $S(z^{\otimes 2},z^{\otimes 2})=1$. 
    
    We furthermore set
    \begin{align*}
        a:=&\,S(x^{\otimes 2},z^{\otimes 2})=S(x\vee z,x\vee z),\\
        b:=&\,S(y^{\otimes 2},z^{\otimes 2})=S(y\vee z,y\vee z),\\
        c:=&\,S(z^{\otimes 2},x\vee z),\\
        d:=&\,S(z^{\otimes 2},y\vee z). 
    \end{align*}
    We observe that 
    \begin{align*}
        S(x^{\otimes 2},y^{\otimes 2})=S(x\vee y,x\vee y)=0,\quad S(x\vee z,y\vee z)=S(z^{\otimes 2},x\vee y)=0
    \end{align*}
    and similarly
    \begin{align*}
        S(x^{\otimes2},y\vee z)=0=S(y^{\otimes 2},x\vee z).
    \end{align*}
    It follows that with respect to the basis $\{x^{\otimes 2},y^{\otimes 2},z^{\otimes 2},x\vee y,x\vee z, y\vee z\}$ the symmetric bilinear form $R(S)$ is represented by the matrix as stated above.
\end{myproof}
\begin{prop}
\label{prop: tot sym no rank one in ker}
    Let $S \in \Sym^4 V^*\cap\mathcal{C}_W$ such that $\ker R(S)$ contains no elements of rank one. Then $S$ is not extremal.
\end{prop}
\begin{myproof}
    We will show that there is some $\eta\in V^*$ such that $\eta^{\otimes 4}$ vanishes on the kernel of $R(S)$. For this, let $a,b,c,d\in \R$ as in Lemma~\ref{lem: explicit tot sym no rank one in ker} and let $\{x,y,z\}$ the basis of $V$ as in Lemma~\ref{lem: explicit tot sym no rank one in ker}. We denote by $\{\alpha,\beta,\gamma\}$ the basis of $V^*$ dual to $\{x,y,z\}$. If $a=0$, we observe that $x^{\otimes 2}$ is perpendicular to $\Span\{y^{\otimes 2},z^{\otimes 2},x\vee y,x\vee z,y\vee z\}$ and since $x^{\otimes 2}\notin \ker R(S)$ it follows that  $\alpha^{\otimes 4}$ vanishes on $\ker R(S)$. Similarly, if $b=0$ we find that $\beta^{\otimes 4}$ vanishes on the kernel of $R(S)$.

    Now suppose that both $a,b\neq 0$. Then we note that the kernel of $R(S)$ is at most two-dimensional, since the image of $\Span\{x^{\otimes 2},y^{\otimes 2},x\vee z,y\vee z\}$ is four-dimensional. We know by Corollary~\ref{cor: ker S for extremal S in 3d} that $S$ cannot be extremal if the kernel of $R(S)$ is one-dimensional. It is two-dimensional if and only if 
    \begin{align*}
        0=\det\left(R(S)|_{V/(\R\,x\vee y)}\right)=-a^3b - ab^3 - bc^2 - ad^2 + ab.
    \end{align*}
    In this case 
    \begin{align*}
        \ker R(S)=\Span\{x\vee y,a^2b\,x^{\otimes 2}+ab^2\,y^{\otimes 2}-ab\,z^{\otimes 2}+bc\,x\vee z+ad\,y\vee z\}
    \end{align*}
    and we find that e.g.~$\eta=2b\,\beta+(d-\sqrt{d^2+4b^3})\,\gamma\in V^*\setminus\{0\}$ satisfies $\eta^{\otimes 4}|_{\ker R(S)}\equiv 0$, where we used that $b\geq 0$ since $R(S)\geq 0$.
\end{myproof}

\subsection{Case $\rk R'(S)=1$}
\label{sect: rk R'=1}

We will now study the case where $\dim V=3$ and $S\in\mathcal{C}_W$ is such that $\rk R'(S)=1$. In this case $R'(S)=(\alpha\wedge\beta)^{\otimes 2}$ for some linearly independent $\alpha,\beta\in V^*$ and $\ker R'(S)=z\wedge U$ for $z\in (\ker\alpha\cap\ker \beta)\setminus\{0\}$. It will be useful to study the cases of different dimensions of the intersection $\ker S\cap (z\vee V)$ separately. The aim of this section is to prove the following:
\begin{thm}
\label{thm: dim V=3 rank R'=1 extremal elements}
    Suppose $\dim V=3$. Let $S\in\mathcal{C}_W$ such that $R'(S)=(\alpha\wedge \beta)^{\otimes 2}\neq 0$ for some $\alpha,\beta\in V^*$ and such that $R(S)$ is not extremal in $\mathcal{C}_{\Sym}$. Let $z\in V$ such that $\ker \alpha\cap \ker \beta =\R\cdot z$. Then $S$ is extremal in $\mathcal{C}_W$ if and only if $S$ is such that $\ker S\cap (z\vee V)=\{0\}$ and $\rk R(S)=3$.    
\end{thm}

The following diagram shows an outline of the structure of this section:

\begin{center}
\scriptsize

\begin{tikzpicture}[sibling distance=9em,
  blocknode/.style = {shape=rectangle, rounded corners,
    draw, align=center},textnode/.style={shape=rectangle, rounded corners,
     align=center},->]

\node[blocknode] {$S\in \mathcal{C}_W$ s.t.~$\rk R(S)>1$, $\rk R'(S)=1$\\ i.e.~$\ker R'(S)=z\wedge V$ for some $z\in V\setminus\{0\}$, $\mathcal{Q}:=\ker S\cap(z\vee V)$}
    child { node[blocknode,yshift=-1.5mm] {$\dim \mathcal{Q}=3$\\ i.e.~$z\vee V\subset \ker S$} 
        child {node[textnode] {reduces to the\\
        2-dimensional case}
            child {node[textnode] {\textit{not extremal}\\ \textit{by Theorem~\ref{thm: ers in 2d}}}}}}
child {node[blocknode, yshift=-2mm] {$\dim\mathcal{Q}=2$\\ i.e.~$\exists\, U\subset V$ s.t.\\ $z\vee U\subset \ker S$}
            child { node[textnode] {\textit{not extremal by}\\ \textit{Proposition~\ref{prop: R' rank 1 sym2u in kernel}}}}}
child {node[blocknode,yshift=-2mm,xshift=2mm] {$\dim \mathcal{Q}=1$,\\ i.e.~$\exists \,x\in V\setminus\{0\}$ s.t.\\ $\R\cdot(x\vee z)\subset \ker S$}
    child {node[textnode,xshift=-3mm] {\textit{not extremal by}\\ \textit{Proposition~\ref{prop: R' rank one case xz in ker S}}}}}
child {node[blocknode, xshift=4em,yshift=-1mm] {$\dim\mathcal{Q}=0$\\ $\implies \rk R(S)\geq 3$}
    child {node[blocknode,teal,xshift=1cm]{$\rk R(S)=3$}
        child{ node[textnode]{\textit{extremal by}\\ \textit{Proposition~\ref{prop: rank R'=1, rank R=3, no xv in kerS}}}}}
    child {node[blocknode,xshift=6.5mm]{$\rk R(S)=4$}
        child {node[textnode] {\textit{not extremal by}\\ \textit{Proposition~\ref{prop: R' rank 1 no xz in kernel}}}}}
    child { node[blocknode,xshift=3mm] {$\rk R(S)>4$}
        child {node[textnode] {\textit{not extremal by}\\ \textit{Proposition~\ref{prop: dim ker extremal elements}}}}}};

\end{tikzpicture}
\end{center}

As a first step, we will exclude the presence of certain elements in $\ker S$ for extremal $S$ with $\rk R'(S)=1$:
\begin{lem}
\label{lem: R' rank 1 case z2 in kernel}
    Suppose $\dim V=3$. Let $S\in\mathcal{C}_W$ such that $R'(S)=(\alpha\wedge \beta)^{\otimes 2}\neq 0$ for some $\alpha,\beta\in V^*$. Let furthermore $z\in V$ such that $\ker\alpha\cap\ker\beta=\R\cdot z$. If $z^{\otimes 2}\in\ker S$, then $S$ is not extremal unless $R(S)$ is extremal in $\mathcal{C}_{\Sym}$.   
\end{lem}
\begin{myproof}
    If $z^{\otimes 2}\in \ker S$, then by Lemma~\ref{lem: formula S(v2,w2)} we find for every $v\in V$
    \[
    0=S(v^{\otimes 2},z^{\otimes 2})=S(v\vee z,v\vee z)-S(v\wedge z,v\wedge z)=S(v\vee z,v\vee z),
    \]
    where we used $z\in\ker\alpha\cap\ker\beta$ in the last step. It follows that $z\vee v\in \ker S$ for every $v\in V$ and hence the problem reduces to the two-dimensional case, which is covered in Theorem~\ref{thm: ers in 2d}.
\end{myproof}

\begin{lem}
\label{lem: R' rank 1 z2+psd in kernel}
Suppose $\dim V=3$. Let $S\in\mathcal{C}_W$ be such that $R'(S)=(\alpha\wedge \beta)^{\otimes 2}\neq 0$ for some $\alpha,\beta\in V^*$. Let furthermore $z\in V$ such that $\ker\alpha\cap\ker\beta=\R z$. If there are some $x,y\in V$, linearly independent of $z$ and such that either $x^{\otimes 2}+z^{\otimes 2}\in\ker S$ or $x^{\otimes 2}+y^{\otimes 2}+z^{\otimes 2}\in\ker S$. Then also $z^{\otimes2}\in\ker S$. In particular, $S$ is not extremal unless $R(S)$ is extremal in $\mathcal{C}_{\Sym}$.    
\end{lem}
\begin{myproof}
    If $x^{\otimes 2}+z^{\otimes 2}\in\ker S$, then 
    \[S(z^{\otimes 2},\cdot)=-S(x^{\otimes 2},\cdot).\]
    Using Lemma~\ref{lem: formula S(v2,w2)}, we deduce
    \[
    0\leq S(z^{\otimes 2},z^{\otimes 2})=-S(x^{\otimes 2},z^{\otimes 2})=-S(x\vee z,x\vee z)+S(x\wedge z,x\wedge z)=-S(x\vee z,x\vee z)\leq 0,
    \]
    where we used $z\in\ker\alpha\cap\ker\beta$. Similarly, if $x^{\otimes 2}+y^{\otimes 2}+z^{\otimes 2}\in\ker S$, we find 
    \[
    0\leq S(z^{\otimes 2},z^{\otimes 2})=-S(x^{\otimes 2},z^{\otimes 2})-S(y^{\otimes 2},z^{\otimes 2})=-S(x\vee z,x\vee z)-S(y\vee z,y\vee z)\leq 0.
    \]
    In both cases, it follows that $z^{\otimes 2}\in\ker S$ and
    the assertion follows from Lemma~\ref{lem: R' rank 1 case z2 in kernel}.
\end{myproof}

As described above, we want to consider the cases of different dimensions of the intersection $\ker S\cap(z\vee V)$ with $S\in \mathcal{C}_W$ and $z\in V$ as above. When this intersection is two-dimensional, by Lemma~\ref{lem: R' rank 1 case z2 in kernel} it is enough to consider the situation where $z\vee U\subset \ker S$ for some $U\subset V$ complementary to the line $\R z$. It will therefore be useful to study the intersection $\ker S\cap \Sym^2 U$ for some $U\subset V$ complementary to $\R z$.

\begin{lem}
    \label{lem: aux rk R' one x vee (x,y)}
    Suppose $\dim V=3$. Let $S\in\mathcal{C}_W$ such that $R'(S)=(\alpha\wedge \beta)^{\otimes 2}\neq 0$ for some $\alpha,\beta\in V^*$, and let $U\subset V$ complementary to $\ker\alpha\cap\ker \beta$. If $\{x,y\}\subset U$ is a basis of $U$, then 
    \[
    \dim\left(\ker S\cap(x\vee \Span\{x,y\})\right)\leq 1.
    \]
    In particular, the intersection $\Sym^2 U\cap \ker S$ is at most two-dimensional. Moreover, if the intersection $\ker S\cap \Sym^2 U$ is two-dimensional, it contains an element of rank 1.
\end{lem}
\begin{myproof}
    Clearly, $\dim\left(\ker S\cap(x\vee \Span\{x,y\})\right)\leq 1$ when $\dim (\Sym^2 U\cap \ker S)\leq1$. Now suppose $\dim (\Sym^2 U\cap \ker S)>1$. By Lemma~\ref{lem: existence u2-v2} we know that there is an element of the form $x\vee y\in\Sym^2 U\cap \ker S$, for some $x,y\in U\setminus\{0\}$.

    We first consider the case where $x$ and $y$ are linearly dependent, i.e. $x^{\otimes 2}\in\ker S$. By Lemma~\ref{lem: formula S(v2,w2)}, we find that for every $w\in U\setminus \{0\}$ linearly independent of $x$, 
    \begin{align}
    \label{eq: lemma sym^2U cap kerS}
    0=S(x^{\otimes 2},w^{\otimes 2})=S(x\vee w,x\vee w)-S(x\wedge w, x\wedge w).    
    \end{align}
    
    By the assumption $R'(S)=(\alpha\wedge \beta)^{\otimes 2}\neq 0$, we know that $\ker R'(S)=z\wedge U$. In particular, we have $S(x\wedge w,x\wedge w)>0$ and it follows from equation \eqref{eq: lemma sym^2U cap kerS} that $x\vee w\notin \ker S$ and therefore $\dim\left(\ker S\cap(x\vee \Span\{x,w\})\right)\leq 1$ for every $w\in U$. 

    Similarly, if $x\vee y\in \ker S$ for linearly independent $x,y\in U$, then Lemma~\ref{lem: formula S(v2,w2)} implies
    \[
    S(x^{\otimes 2},y^{\otimes 2})=S(x\vee y,x\vee y)-S(x\wedge y, x\wedge y)=-S(x\wedge y,x\wedge y)<0. 
    \]
    Therefore, neither $x^{\otimes 2}$ nor $y^{\otimes 2}$ can be in the kernel of $S$ and we conclude that
    \[
    \Sym^2 U\cap \ker S=\Span\{ x\vee y, x^{\otimes2}+\lambda\, y^{\otimes 2}\}
    \]
    for some $\lambda\neq 0$. We claim that $\lambda>0$: 
    
    Indeed, since $x^{\otimes 2}+\lambda\, y^{\otimes 2}\in\ker S$, we have
    \[0=S(x^{\otimes 2}+\lambda\, y^{\otimes 2},x^{\otimes 2})=S(x^{\otimes 2},x^{\otimes 2})+\lambda\, S(x^{\otimes 2},y^{\otimes 2}).\]
    Using again Lemma~\ref{lem: formula S(v2,w2)} and that $x\vee y\in\ker S$, we obtain 
    \[
    0< S(x^{\otimes 2},x^{\otimes 2})=-\lambda\, S(x^{\otimes 2},y^{\otimes 2})=\lambda\, S(x\wedge y,x\wedge y).
    \]
    Since $S(x\wedge y,x\wedge y)>0$, we must have $\lambda>0$. 
    
    We may therefore write
    \[
    \Sym^2 U\cap \ker S=\Span\{ x\vee y, x^{\otimes2}+\lambda\, y^{\otimes 2}\}=\Span\left\{ x\vee y, (x+\sqrt{\lambda}\, y)^{\otimes 2}\right\}.
    \]
This shows that $\Sym^2 U\cap \ker S$ contains an element of rank one.
\end{myproof}

\begin{lem}
\label{lem: R' rank 1 sym2 U}
     Suppose $\dim V=3$. Let $S\in\mathcal{C}_W$ such that $R'(S)=(\alpha\wedge \beta)^{\otimes 2}\neq 0$ for some $\alpha,\beta\in V^*$, and let $U\subset V$ complementary to $\ker\alpha\cap\ker \beta$. Then there exist $\Tilde{\alpha},\Tilde{\beta}\in\Span\{\alpha,\beta\}$ such that $\Tilde{\alpha}\vee\Tilde{\beta}$ vanishes on $\ker S\cap \Sym^2 U$. 
\end{lem}
\begin{myproof}
    If $\Sym^2 U\cap \ker S$ is one-dimensional, then
    \[\dim\left(\mathrm{Ann}(\Sym^2 U\cap \ker S)\cap\Sym^2(\Span\{\alpha,\beta\})\right)\geq2\]
    and by Lemma~\ref{lem: existence u2-v2} there are some $\Tilde{\alpha},\Tilde{\beta}\in \Span\{\alpha,\beta\}$ such that $\Tilde{\alpha}\vee\Tilde{\beta}\in\mathrm{Ann}(\Sym^2 U\cap \ker S)$.

    Now suppose $\dim (\Sym^2 U\cap \ker S)>1$. By Lemma~\ref{lem: aux rk R' one x vee (x,y)} we know that in this case $\dim (\Sym^2 U\cap \ker S)=2$ and there is an element of the form $u^{\otimes 2}\in\Sym^2 U\cap \ker S\setminus\{0\}$. Since $\Sym^2 U\cap\ker S$ is two-dimensional, we have
    \[
    \dim\left(\mathrm{Ann}(\Sym^2 U\cap \ker S)\cap\Sym^2(\Span\{\alpha,\beta\})\right)\geq 1,
    \]
    i.e. $\mathrm{Ann}(\Sym^2 U\cap \ker S)\cap\Sym^2(\Span\{\alpha,\beta\})$ contains a non-trivial element $\gamma$. By Sylvester's theorem, it will be either of the form $\gamma=\Tilde{\alpha}^{\otimes 2}$ for some $\Tilde{\alpha}\in\Span\{\alpha,\beta\}$ or of the form $\gamma=\bar{\alpha}^{\otimes 2}\pm\bar{\beta}^{\otimes 2}$ for linearly independent $\bar{\alpha},\bar{\beta}\in\Span\{\alpha,\beta\}$. Furthermore, since there is a rank one element $u^{\otimes 2}\in\Sym^2 U\cap \ker S$ and $u\notin \ker\alpha\cap\ker\beta$, we may exclude the case $\gamma=\bar{\alpha}^{\otimes 2}+\bar {\beta}^{\otimes 2}$ for linearly independent $\bar{\alpha},\bar{\beta}$. Recalling that 
    \[
    \bar{\alpha}^{\otimes 2}-\bar{\beta}^{\otimes 2}=(\bar{\alpha}+\bar{\beta})\vee (\bar{\alpha}-\bar{\beta}), 
    \]
    it follows that there are $\Tilde{\alpha},\Tilde{\beta}\in \Span\{\alpha,\beta\}$ such that $\Tilde{\alpha}\vee\Tilde{\beta}\in\mathrm{Ann}(\Sym^2 U\cap \ker S)$.  
\end{myproof}

We will now show that $S$ as above with $\rk R(S)>1$ cannot be extremal if for some $U\subset V$ complementary to the line $\R z$ we have $z\vee U\subset \ker S$, where $z\in V$ is as above:

\begin{prop}
\label{prop: R' rank 1 sym2u in kernel}
    Suppose $\dim V=3$. Let $S\in\mathcal{C}_W$ such that $R'(S)=(\alpha\wedge \beta)^{\otimes 2}\neq 0$ for some $\alpha,\beta\in V^*$. Let furthermore $z\in\ker\alpha\cap\ker\beta\setminus\{0\}$. If for some $U\subset V$ complementary to $\R z$ we have $z\vee U\subset \ker S$, then $S$ is not extremal unless $R(S)$ is extremal in $\mathcal{C}_{\Sym}$.
\end{prop}
\begin{myproof}
    If $z\vee v\in \ker S$ for every $v\in U$, observing that for all $v,w\in U$
    \[
    S(v\vee w,z^{\otimes 2})=S(v\vee z,w\vee z)=0,
    \]
    we find that $z^{\otimes 2}\perp_S \Sym^2 U$. By Lemma~\ref{lem: R' rank 1 case z2 in kernel}, we may furthermore assume that $z^{\otimes 2}$ is not in the kernel of $S$. It follows that there is a basis $\{Q_i\}$ of $\ker R(S)$ such that $Q_1,Q_2\in z\vee U$ and $Q_i\in\Sym^2 U$ for $i>2$.
    
    By Lemma~\ref{lem: R' rank 1 sym2 U}, there exist $\Tilde{\alpha},\Tilde{\beta}\in\Span\{\alpha,\beta\}$ such that $\Tilde{\alpha}\vee\Tilde{\beta}\in \mathrm{Ann}(\Sym^2 U\cap \ker S)$. Since $z\in \ker \alpha\cap\ker \beta$, it follows that 
    \[
    \Tilde{\alpha}\vee\Tilde{\beta}(Q_1)= 0=\Tilde{\alpha}\vee\Tilde{\beta}(Q_2)
    \]
    and hence $\Tilde{\alpha}\vee\Tilde{\beta}|_{\ker S}\equiv 0$. Since 
    \[
    h((\Tilde{\alpha}\vee\Tilde{\beta})^{\otimes 2})=(\Tilde{\alpha}\wedge\Tilde{\beta})^{\otimes 2}\in\R(\alpha\wedge\beta)^{\otimes 2},
    \]
     which vanishes on $\ker S$, it follows that
    \[
    \left((\Tilde{\alpha}\otimes \Tilde{\beta})^{\otimes 2}+(\Tilde{\beta}\otimes \Tilde{\alpha})^{\otimes 2}\right)|_{\ker S}\equiv 0.
    \]
    The statement follows now from Corollary~\ref{cor: nec extremal}.
\end{myproof}

We will now improve the statement of Proposition~\ref{prop: R' rank 1 sym2u in kernel} by showing that $S$ as above  with $\rk R(S)>1$ cannot be extremal if $\ker S\cap (z\vee V)\neq \{0\}$:

\begin{prop}
\label{prop: R' rank one case xz in ker S}
Suppose $\dim V=3$. Let $S\in\mathcal{C}_W$ such that $R'(S)=(\alpha\wedge \beta)^{\otimes 2}\neq 0$ for some $\alpha,\beta\in V^*$. Let furthermore $z\in\ker\alpha\cap\ker\beta\setminus\{0\}$. If for some $x\in V\setminus\{0\}$  we have $x\vee z\in \ker S$, then $S$ is not extremal unless $R(S)$ is extremal in $\mathcal{C}_{\Sym}$.
\end{prop}
\begin{myproof}
    By Lemma~\ref{lem: R' rank 1 case z2 in kernel} we may assume that $x$ is linearly independent of $z$. Using Proposition~\ref{prop: R' rank 1 sym2u in kernel} we may furthermore assume that
    \[\ker S\cap (z\vee V)=\R\,x\vee z.\] 
    Since by assumption also $x\wedge z\in\ker S$, it follows that 
    \[x\otimes z=x\vee z +x\wedge z \in\ker S.\]
    Thus, we observe
    \begin{align*}
    S(x\vee v,z^{\otimes 2})=S(x\otimes v,z^{\otimes 2})=S(x\otimes z,z\otimes v)=0, 
    \end{align*}
    where we used that $\Lambda^2 V$ and $\Sym^2 V$ are perpendicular with respect to $S$ for the first equality and the invariance under $\tau$ for the second equality.
    
    Similarly, we find 
    \begin{align*}
    S(x^{\otimes 2},v\vee z)=S(x^{\otimes 2},v\otimes z)=S(x\otimes z, v\otimes x)=0    
    \end{align*}
    for every $v\in V$.
    
    Furthermore, we observe for every $v,w\in V$ that
    \begin{align}
    \label{eq: prop rk R'=1 case xz in ker, aux xv perp wz}
    S(x\vee v, w\vee z)=S(x\otimes v, w\otimes z)=S(x\otimes z, w\otimes v)=0,
    \end{align}
    where we used that $w\wedge z\in\ker S$ for every $w\in V$.

    With these relations in mind, we complete $\{x,z\}\subset V$ to a basis $\{x,y,z\}$ of $V$ such that $S(z,z,z,y)=0$. To see that this is possible, we observe that the linear map
    \[ S(z,z,z,\cdot)\colon V\to \R\]
    has a two-dimensional kernel, since we know $z\notin \ker\left( S(z,z,z,\cdot)\right)$, as we assumed that $z^{\otimes 2}$ is not in the kernel and therefore by Lemma~\ref{lem: ker of psd bilin form}, it follows that $S(z^{\otimes 2},z^{\otimes 2})\neq 0$. We furthermore know that $x\in\ker\left( S(z,z,z,\cdot)\right)$, since by assumption $x\vee z\in\ker S$. Choosing a vector $y\in\ker\left( S(z,z,z,\cdot)\right)$ linearly independent of $x$, we obtain a basis $\{x,y,z\}$ of $V$ such that $S(z,z,z,y)=0$.
    
    Then, by equation \eqref{eq: prop rk R'=1 case xz in ker, aux xv perp wz}, we have 
    \[
    (z\vee V)\perp_S (x\vee \Span\{x,y\}).
    \]
    Furthermore, there is a line $L\subset\Sym^2 V$, which is complementary and orthogonal to $(z\vee V)\oplus (x\vee \Span\{x,y\})$ in $\Sym^2 V$. Let $T\in\Sym^2 V$ such that $L=\R\, T$. We obtained a decomposition
    \[
    \Sym^2 V=(z\vee V)\oplus (x\vee \Span\{x,y\})\oplus \R\, T
    \]
    of $\Sym^2 V$ into pairwise orthogonal subspaces. 
    
    We would now like to study the kernel of $S$ in terms of this decomposition. For the first factor we know by assumption that $\ker S\cap (z\vee V)=\R\, x\vee z$. For the second factor we recall that, by Lemma~\ref{lem: aux rk R' one x vee (x,y)}, the intersection $\ker S \cap x\vee \Span\{x,y\}$ can at most be one-dimensional. 

    If $T\notin \ker S$, it follows that $\ker S\cap \Sym^2 V=\Span\{x\vee z,x\vee u\}$ for some $u\in \Span \{x,y\}$. Therefore, choosing some $\gamma\in V^*\setminus\{0\}$ such that $\gamma(x)=0$, we find that $\gamma^{\otimes 4}|_{\ker S}\equiv 0$ and hence $S$ is not extremal by Corollary~\ref{cor: nec extremal}. 

    Now suppose $T\in \ker S$, i.e. $\ker S\cap\Sym^2 V=\Span\{x\vee z,x\vee u, T\}$ for some $u\in \Span\{x,y\}$. We will study separately the cases, where $T$ is of rank one, two or three.
    
    Firstly, we observe that under the assumptions we made, we must have $\rk T>1$: If $T=w^{\otimes 2}$ for some $w\in V\setminus\{0\}$, then we observe that, since $T$ is complementary to $(z\vee V)\oplus (x\vee \Span\{x,y\})$, $w$ must be linearly independent of $x$ and $z$. Using Lemma~\ref{lem: formula S(v2,w2)} and the fact that $w\wedge z\in\ker S$, we find 
    \[ 
    0=S(w^{\otimes 2},z^{\otimes 2})= S(w\vee z,w\vee z)-S(w\wedge z,w\wedge z)=S(w\vee z,w\vee z),
    \]
    in contrast to the assumption that $\ker S\cap (z\vee V)=\R\, x\vee z$.

    Suppose now that $\rk T=2$. In this case $T$ may either be of the form $v\vee w$ or $v^{\otimes 2}+w^{\otimes 2}$ for some $v,w\in V\setminus\{0\}$. If $T=v\vee w$, then we may choose $\gamma\in V^*\setminus\{0\}$ such that $\gamma(x)=0=\gamma(v)$ and conclude that $\gamma^{\otimes 4}|_{\ker S}\equiv 0$. If $T=v^{\otimes 2}+w^{\otimes 2}$ for linearly independent $v,w\in V\setminus\{0\}$, we show that there exist $\Tilde{\alpha},\Tilde{\beta}\in\Span\{\alpha,\beta\}$ such that $\Tilde{\alpha}\vee \Tilde{\beta}|_{\ker S}\equiv 0$. For this, we observe first that 
    \[
    \Sym^2\left(\Span\{\alpha,\beta\}\right)\subset \ann(x\vee z),    
    \]
    and hence,
    \begin{align*}
    \ann(\ker S\cap\Sym^2 V)\cap\Sym^2\left(\Span\{\alpha,\beta\}\right)=\ann\left(\Span\{x\vee u,\, T\}\right)\cap\Sym^2\left(\Span\{\alpha,\beta\}\right).  
    \end{align*}
    Therefore,
    \[\dim\left(\ann(\ker S\cap\Sym^2 V)\cap\Sym^2\left(\Span\{\alpha,\beta\}\right)\right)\geq1.
    \]
    Since by assumption $T=v^{\otimes 2}+w^{\otimes 2}$ for linearly independent $v,w\in V\setminus\{0\}$, it cannot be annihilated by an element of the form $\Tilde{\alpha}^{\otimes 2}+\Tilde{\beta}^{\otimes 2}$ for linearly independent $\Tilde{\alpha},\Tilde{\beta}\in\Span\{\alpha,\beta\}$. Therefore, there must be some
    \[\Tilde{\alpha}\vee\Tilde{\beta}\in\ann(\ker S\cap\Sym^2 V)\cap\Sym^2\left(\Span\{\alpha,\beta\}\right).
    \]
    Arguing as in the proof of Proposition~\ref{prop: R' rank 1 sym2u in kernel}, it follows that
    \[
    \left((\Tilde{\alpha}\otimes \Tilde{\beta})^{\otimes 2}+(\Tilde{\beta}\otimes \Tilde{\alpha})^{\otimes 2}\right)|_{\ker S}\equiv 0.
    \]
    By Corollary~\ref{cor: nec extremal} it follows that $S$ cannot be extremal in $\mathcal{C}_W$.
    
    Now, suppose that $\rk T=3$. By Lemma~\ref{lem: R' rank 1 z2+psd in kernel} we know that $T$ must be indefinite. Then $T$ determines a Lorentzian metric $g_T$ on $V$, which, up to changing the sign of $T$, we may assume to be of mostly positive signature. The vector $x\in V$ can satisfy $g_T(x,x)>0$, $g_T(x,x)=0$ or $g_T(x,x)<0$. 

    \textit{Case 1:} If $g_T(x,x)>0$, then after rescaling $T$, we may assume that $T=x^{\otimes 2}+v^{\otimes 2}-w^{\otimes 2}$ for some $v,w\in V$, such that $\{x,v,w\}\subset V\setminus\{0\}$ are linearly independent. Choosing $\gamma\in V^*\setminus\{0\}$ such that $\gamma(x)=0$, $\gamma(v)=\gamma(w)$ we find $\gamma^{\otimes 4}|_{\ker S}\equiv 0$ and hence $S$ is not extremal in $\mathcal{C}_W$ by Corollary~\ref{cor: nec extremal}.

    \textit{Case 2:} If $g_T(x,x)=0$, there exist $v,w\in V\setminus\{0\}$ such that $\{x,v,w\}\subset V$ are linearly independent and such that $T=x\vee v+w^{\otimes 2}$. Choosing $\gamma\in V^*\setminus \{0\}$ such that $\gamma(x)=\gamma(w)=0$ we find $\gamma^{\otimes 4}|_{\ker S}\equiv 0$ and conclude that $S$ is not extremal.

    \textit{Case 3:} If $g_T(x,x)<0$, then there exist $v,w\in V\setminus\{0\}$ such that $\{x,v,w\}\subset V$ are linearly independent and such that $T= v^{\otimes 2}+w^{\otimes 2}-x^{\otimes 2}$. We claim that this is not possible under the assumptions of Proposition~\ref{prop: R' rank one case xz in ker S}. Indeed, if $v^{\otimes 2}+w^{\otimes 2}-x^{\otimes 2}\in\ker S$, then using Lemma~\ref{lem: formula S(v2,w2)} and the fact that $z\wedge V\subset \ker S$, yields
    \[
    0=S(x^{\otimes2}, z^{\otimes 2})=S(v^{\otimes 2}+w^{\otimes 2},z^{\otimes 2})= S(v\vee z,v\vee z)+S(w\vee z,w\vee z)>0,
    \]
    as we assumed that $\ker S\cap (z\vee V)=\R\, x\vee z$. 
\end{myproof}

Next, we will focus on $S\in \mathcal{C}_W$ as above such that $\ker S\cap(z\vee V)=\{0\}$. We will first study some properties of elements in the kernel of such $S$:

\begin{lem}
\label{lem: R' rank 1 no xz in ker els in sym2U}
Suppose $\dim V=3$. Let $S\in\mathcal{C}_W$ such that $R'(S)=(\alpha\wedge \beta)^{\otimes 2}\neq 0$ for some $\alpha,\beta\in V^*$. Let furthermore $z\in\ker\alpha\cap\ker\beta\setminus\{0\}$ and suppose that $\ker S\cap (z\vee V)=\{0\}$. Then the following hold:
\begin{enumerate}
    \item For every $U\subset V$ complementary to $\R  z$ the intersection $\ker S\cap \Sym^2 U$ is at most one-dimensional.
    \item Every non-trivial element in $\ker R(S)$ is indefinite.
    In particular, there are no elements of rank one in $\ker R(S)$.
\end{enumerate}
\end{lem}
\begin{myproof}
Let $S\in\mathcal{C}_W$ as in Lemma~\ref{lem: R' rank 1 no xz in ker els in sym2U}. We will start excluding the presence of certain elements in $\ker S$.

We note that if $x^{\otimes 2}\in \ker S$ for some $x\in V$, then using Lemma~\ref{lem: formula S(v2,w2)} and the fact that $z\wedge V\subset \ker S$, we obtain
    \[
    0=S(x^{\otimes 2},z^{\otimes 2})=S(x\vee z,x\vee z),
    \]
    which is impossible by the assumption that $\ker S\cap (z\vee V)=\{0\}$. Similarly, if $x^{\otimes2}+y^{\otimes 2}\in \ker S$ for some $x,y\in V$ linearly independent of $z$, we find
    \[
    0<S(x\vee z, x\vee z)= S(x^{\otimes 2}, z^{\otimes 2})=-S(y^{\otimes 2}, z^{\otimes 2})=-S(y\vee z, y\vee z)<0,
    \]
    a contradiction. 
    
    Thus, we observe that if there is some $T\in\ker S\cap \Sym^2 U\setminus\{0\}$ for some $U\subset V$ complementary to $\ker \alpha\cap\ker\beta$, then it must be indefinite. In fact, any element $T\in \ker R(S)$ must be indefinite: 
        
    If $\rk T=3$, then $T$ must be indefinite by Lemma~\ref{lem: R' rank 1 z2+psd in kernel}. If $T=x^{\otimes 2}+y^{\otimes 2}$ for some $x,y\in V\setminus \{0\}$, then we must have $z\in \Span\{x,y\}$, since otherwise $T\in\Sym^2 U$ for $U:=\Span\{x,y\}$ complementary to $\R\cdot z$, which is in contrast to our above observation. Therefore, we may assume after applying a rotation that $y\in \R\cdot  z$. Again, this cannot happen by Lemma~\ref{lem: R' rank 1 z2+psd in kernel}. Similarly, if $\rk T=1$, then by the above argument we must have $T=\lambda z^{\otimes 2}$ for some $\lambda\in \R\setminus\{0\}$, which is excluded by our assumptions. 
        
    Therefore, $\ker R(S)$ consists of indefinite elements. In particular, there are no rank one elements in the kernel of $R(S)$.
    
    Next, we claim that for every $U\subset V$ complementary to $\ker\alpha\cap\ker\beta$ the intersection $\ker S\cap \Sym^2 U$ can at most be one-dimensional. 
    
    Indeed, suppose that $\Span\{T_1,T_2\}\subset\ker S\cap \Sym^2 U$ for linearly independent $T_1,T_2\in \Sym^2 U$. Since $T_1$ is indefinite, there exist linearly independent $u_1,w_1\in U$ such that $T_1=u_1\vee w_1$. Similarly, we know that also $T_2$ defines a Lorentzian metric $g_{T_2}$ on $U$. If $u_1$ is null for $g_{T_2}$, then there exists some $w$ linearly independent of $u_1$ such that $T_2=u_1\vee w$. If $u_1$ is not null for $g_{T_2}$ then we know that, up to rescaling by a non-zero factor, $T_2=u_1^{\otimes2}-w^{\otimes2}$ for some $w\in U$ linearly independent of $u_1$:

    \textit{Case 1:} If $T_1=u_1\vee w_1$ and $T_2=u_1\vee w$, where $u_1,w\in U$ are linearly independent, we may write $w=\lambda\, u_1+\mu\, w_1$ for some $\lambda,\mu\in\R\setminus\{0\}$. Then, we have
    \[ u_1\vee w=\lambda\, u_1^{\otimes 2}+\mu\, u_1\vee w_1\in\ker S\]
    which implies $u_1^{\otimes 2}\in\ker S$ in contrast to our above observations.

    \textit{Case 2:} If $T_1=u_1\vee w_1$, $T_2=u_1^{\otimes2}-w^{\otimes 2}$, where $u_1,w\in U$ are linearly independent, we again write $w=\lambda\, u_1+\mu\, w_1$ for some $\lambda,\mu\in\R$, $\mu\neq0$. Then,
    \[
    u_1^{\otimes 2}- w^{\otimes 2}=(1-\lambda^2)\, u_1^{\otimes 2}-\mu^2\, w_1^{\otimes 2}-2\lambda\mu\,u_1\vee w_1\in\ker S,
    \]
    and hence, $(1-\lambda^2)\, u_1^{\otimes 2}-\mu^2\, w_1^{\otimes 2}\in\ker S$. By our observation that elements in $\ker S\cap \Sym^2 U$ must be indefinite, we must have $1-\lambda^2>0$. Therefore, after rescaling $w_1$, we conclude that $u_1^{\otimes 2}-w_1^{\otimes 2}\in\ker S$. Then, we apply Lemma~\ref{lem: formula S(v2,w2)} and use $u_1\vee w_1\in \ker S$ to obtain
    \[
    0<S(u_1^{\otimes 2},u_1^{\otimes 2})=S(u_1^{\otimes 2},w_1^{\otimes 2})=-S(u_1\wedge w_1,u_1\wedge w_1)<0,
    \]
    a contradiction.
\end{myproof}   

If $\ker S\cap (z\vee V)=\{0\}$ for $S$ and $z$ as above, then $\rk R(S)\geq 3$. In the following proposition we show that if $\rk R(S)=4$, then $S$ is not extremal.

\begin{prop}
\label{prop: R' rank 1 no xz in kernel}
    Suppose $\dim V=3$. Let $S\in\mathcal{C}_W$ such that $R'(S)=(\alpha\wedge \beta)^{\otimes 2}\neq 0$ for some $\alpha,\beta\in V^*$ and such that $\rk R(S)=4$. Let furthermore $z\in\ker\alpha\cap\ker\beta\setminus\{0\}$ and suppose that $\ker S\cap (z\vee V)=\{0\}$. Then $S$ is not extremal in $\mathcal{C}_W$.
\end{prop}
\begin{myproof}
    By assumption, we have $\dim (\ker R(S))= 2$. By Lemma~\ref{lem: R' rank 1 no xz in ker els in sym2U}, we know that all the elements in $\ker S\cap\Sym^2 V$ are indefinite.

    We write $\ker S\cap \Sym^2 V=\Span\{T_1,T_2\}$. By Lemma~\ref{lem: existence of degenerate el in 3d} we may assume that $\rk T_1<3$ and since $T_1$ must be indefinite, this means that $T_1=u_1\vee v_1$ for some linearly independent $u_1,v_1\in V\setminus\{0\}$. If also $\rk T_2<3$, i.e. $T_2=u_2\vee v_2$ for some linearly independent $u_2,v_2\in V\setminus\{0\}$, then choosing $\gamma\in V^*\setminus\{0\}$ such that $\gamma(u_1)=0=\gamma(u_2)$, we find $\gamma^{\otimes 4}|_{\ker S}\equiv 0$ and therefore, $S$ cannot be extremal by Corollary~\ref{cor: nec extremal}.

    Now suppose $\rk T_2=3$. Then $T_2$ defines a Lorentzian metric $g_{T_2}$ on $V$. Up to changing the sign of $T_2$, we may furthermore assume that $g_{T_2}$ is of mostly positive signature.

    \textit{Case 1:} If $g_{T_2}(u_1,u_1)>0$, then there exist some $v,w\in V\setminus\{0\}$ such that $\{u_1,v,w\}\subset V$ are linearly independent and such that
    \[
    T_2=u_1^{\otimes 2}+v^{\otimes 2}-w^{\otimes 2}.
    \]
    Choosing $\gamma\in V^*\setminus\{0\}$ such that
    \[
    \gamma(u_1)=0,\quad \gamma(v)=\gamma(w)   
    \]
    we have $\gamma^{\otimes 4}|_{\ker S}\equiv 0$, hence $S$ is not extremal in $\mathcal{C}_W$ by Corollary~\ref{cor: nec extremal}.

    \textit{Case 2:} If $g_{T_2}(u_1,u_1)=0$, then there exist some $v,w\in V\setminus\{0\}$ such that $\{u_1,v,w\}\subset V$ are linearly independent and such that
    \[
    T_2=u_1\vee v+w^{\otimes 2}.
    \]
    Then for $\gamma \in V^*\setminus\{0\}$ satisfying $\gamma(u_1)=0=\gamma(w)$, we find $\gamma^{\otimes 4}|_{\ker S}\equiv 0$.

    \textit{Case 3:} Now suppose $g_{T_2}(u_1,u_1)<0$. There exist some $v,w\in V\setminus\{0\}$ such that $\{u_1,v,w\}\subset V$ are linearly independent and such that
    \[
    T_2=v^{\otimes 2}+w^{\otimes 2}-u_1^{\otimes 2}.
    \]
    If $g_{T_2}(v_1,v_1)\geq 0$, we can exchange the roles of $u_1$ and $v_1$ so that we are in case one or two. Therefore, suppose without loss of generality that also $g_{T_2}(v_1,v_1)<0$. After applying a rotation to $v$ and $w$, we may then assume that some non-trivial element in the line $\R v_1$ is related to $u_1$ by a boost in the direction of $v$, i.e. for some $\lambda\in \R\setminus\{0\}$, $\kappa\in \R\setminus\{0\}$, $v'\in V$ we have
\begin{align*}
\begin{pmatrix}
   \lambda\, v_1\\
v'\end{pmatrix} 
= \begin{pmatrix}
    \cosh{\kappa} & \sinh{\kappa}\\
\sinh{\kappa} & \cosh{\kappa}
\end{pmatrix}
\begin{pmatrix}
    u_1\\
v
\end{pmatrix}\end{align*}
Therefore, for some $\lambda\in \R\setminus\{0\}$, $\kappa\in \R\setminus\{0\}$
\begin{align*}
\lambda\, v_1 = \cosh{\kappa}\left(u_1+\tanh{\kappa}\cdot v\right).
\end{align*}
We set $t:=\tanh{\kappa}$. In particular, $0<t^2<1$. Then, after rescaling $v_1$, we find
\begin{align*}
    v_1=u_1+t\, v.
\end{align*}
Solving for $v$, this yields
\[v=\frac{1}{t}\left(v_1-u_1\right)\]
    and hence
    \[T_2=\frac{1}{t^2}\,v_1^{\otimes 2}+w^{\otimes 2}+\left(\frac{1}{t^2}-1\right)\, u_1^{\otimes 2}-\frac{2}{t}\, u_1\vee v_1.\]
    We define 
    \[\tilde{T}_2:= T_2+ \frac{2}{t}\, T_1= \frac{1}{t^2}\,v_1^{\otimes 2}+w^{\otimes 2}+\frac{1-t^2}{t^2}\,u_1^{\otimes 2}\in \ker S.\]
    Since $t^2<1$, $\tilde{T}_2$ is positive definite. But this is in contrast to our observation in Lemma~\ref{lem: R' rank 1 no xz in ker els in sym2U}. Therefore, we cannot have both $g_{T_2}(u_1,u_1)<0$ and $g_{T_2}(v_1,v_1)<0$.
\end{myproof}

Finally, we will show that if for $S$ and $z$ as above $\ker S\cap(z\vee V)=\{0\}$ and if $\rk R(S)=3$, then $S$ is extremal.

\begin{prop}
\label{prop: rank R'=1, rank R=3, no xv in kerS}
Suppose $\dim V=3$. Let $S\in\mathcal{C}_W$ such that $R'(S)=(\alpha\wedge \beta)^{\otimes 2}\neq 0$ for some $\alpha,\beta\in V^*$ and such that $\rk R(S)=3$. Let furthermore $z\in\ker\alpha\cap\ker\beta\setminus\{0\}$ and suppose that $\ker S\cap (z\vee V)=\{0\}$. Then $S$ is extremal in $\mathcal{C}_W$.    
\end{prop}
\begin{myproof}
    Let $S$ as in Proposition~\ref{prop: rank R'=1, rank R=3, no xv in kerS}. We need to show that every element $S'\in\mathcal{C}_W$ such that $\ker S'\supsetneq \ker S$ has to vanish. 
    
    For this, we note that every element $S'\in\mathcal{C}_W$ such that $\ker S'\supsetneq \ker S$ must either have $R'(S')=0$, i.e.~$S'$ is fully symmetric, or its kernel contains an element $u\vee z\neq 0$: Indeed, the proper inclusion $\ker S'\supsetneq \ker S$ implies that either $\ker R'(S')\supsetneq \ker R'(S)$ or $\ker R(S')\supsetneq \ker R(S)$. If $\ker R'(S')\supsetneq \ker R'(S)$, then we must have $R'(S')=0$, since $\rk R'(S)=1$ by assumption. If $\ker R(S')\supsetneq \ker R(S)$ by the assumption $\rk R(S)=3$ we must have $\rk R(S')\leq 2$, i.e.~$\dim (\ker R(S'))\geq 4$ and hence $\ker R(S')\cap (z\vee V)\neq \{0\}$.

We have seen in Theorem~\ref{thm: ers in sym4} and Proposition~\ref{prop: R' rank one case xz in ker S} that in both such cases $S'$ is not extremal unless it projects to an extremal element in $\mathcal{C}_{\Sym}$. It is therefore enough to consider the case where $\rk R(S')=1$, i.e.~$S'$ is of the form $S'=\gamma^{\otimes 4}$ for some $\gamma \in V^*$ or of the form $S'=(\tilde{\alpha}\vee \tilde{\beta})^{\otimes 2}+(\tilde{\alpha}\wedge\tilde{\beta})^{\otimes 2}$ for linearly independent $\tilde{\alpha},\tilde{\beta}\in V^*$ such that $\tilde{\alpha}\wedge\tilde{\beta}|_{\ker R'(S)}\equiv 0$, i.e.~$\tilde{\alpha},\tilde{\beta}\in\Span\{\alpha,\beta\}$. 

    In Proposition~\ref{prop: explicit description extremal R' rank 1}, we show that the kernel of $R(S)$ for an element $S\in\mathcal{C}_W$ as in Proposition \ref{prop: rank R'=1, rank R=3, no xv in kerS} is of one of the following three forms:
    \begin{enumerate}
    \item for parameters $d,g,h\in \R$ such that $d,g\neq 0$ and $g^2>1-d^2+dh$ and some $x,y\in V$ such that $\{x,y,z\}$ is a basis of $V$
    \begin{align*}
    \ker R(S)=\Span\{ &-x^{\otimes 2}+z^{\otimes 2}+d\,y\vee z,\\
    &h\, x^{\otimes 2}-d\, y^{\otimes 2} +(d-h)\, z^{\otimes 2}+dg\,x\vee z,\\
    &-(g^2-dh)\, x^{\otimes 2}-d^2\, y^{\otimes 2}+(d^2+g^2-dh)\, z^{\otimes 2}+dg\,x\vee y\};
    \end{align*} 

\item for parameters $g,h\in\R$ such that $g^2>1$ and some $x,y\in V$ such that $\{x,y,z\}$ is a basis of $V$
\begin{align*}
    \ker R(S)=\Span\left\{ y\vee(x-g\,z),\; -x^{\otimes 2}+ z^{\otimes 2},\;
    g\,x^{\otimes 2}-g\,y^{\otimes 2}+h\, x\vee y+g^2\,x\vee z\right\};
\end{align*}

\item for parameters $d,h\in \R$ such that $d(d-h)>1$ and some $x,y\in V$ such that $\{x,y,z\}$ is a basis of $V$
\begin{align*}
    \ker R(S)=\Span \{ &-x^{\otimes 2}+y^{\otimes 2}+(d-h)\cdot y\vee z,\;\, x\vee( y-d\, z),\\ &h\, x^{\otimes 2}-d\,y^{\otimes 2}+(d-h)\,z^{\otimes 2}\}.
\end{align*}
\end{enumerate}

We will first consider the case where $S'=\gamma^{\otimes 4}$ for some $\gamma\in V^*$. Finding $\gamma\in V^*\setminus\{0\}$ such that $\gamma^{\otimes 2}|_{\ker S}\equiv 0$ would in each case amount to finding a non-trivial real solution $(\gamma_x,\gamma_y,\gamma_z)\in\R^3$ to the respective system of equations: 

        \begin{enumerate}
    \item for parameters $d,g,h\in \R$ such that $d,g\neq 0$ and $g^2>1-d^2+dh$,
    \begin{align*}
    -\gamma_x^{ 2}+\gamma_z^{2}+d\,\gamma_y\gamma_z&=0,\\
    h\, \gamma_x^{2}-d\, \gamma_y^{2} +(d-h)\, \gamma_z^{2}+dg\,\gamma_x \gamma_z&=0,\\
    -(g^2-dh)\, \gamma_x^{2}-d^2\, \gamma_y^{2}+(d^2+g^2-dh)\, \gamma_z^{2}+dg\,\gamma_x \gamma_y&=0;
    \end{align*} 

\item for parameters $g,h\in\R$ such that $g^2>1$, 
\begin{align*}
    \gamma_y(\gamma_x-g\,\gamma_z)&=0,\\
    -\gamma_x^{2}+ \gamma_z^{2}&=0,\\
    g\,\gamma_x^{2}-g\,\gamma_y^{2}+h\, \gamma_x\gamma_y+g^2\,\gamma_x\gamma_z&=0;
\end{align*}

\item for parameters $d,h\in \R$ such that $d(d-h)>1$,
\begin{align*}
    -\gamma_x^{2}+\gamma_y^{2}+(d-h)\cdot \gamma_y\gamma_z&=0,\\ 
    \gamma_x( \gamma_y-d\, \gamma_z)&=0,\\ 
    h\, \gamma_x^{ 2}-d\,\gamma_y^{ 2}+(d-h)\,\gamma_z^{ 2}&=0.
\end{align*}

    \end{enumerate}
However, these systems of equations do not have a non-trivial solution.

Now we consider the case where $S'=(\tilde{\alpha}\vee \tilde{\beta})^{\otimes2}+(\tilde{\alpha}\wedge\tilde{\beta})^{\otimes 2}$ for some $\tilde{\alpha},\tilde{\beta}\in\Span\{\alpha,\beta\}$. We set
\[\tilde{\alpha}(x)=:\tilde{\alpha}_x,\quad \tilde{\alpha}(y)=:\tilde{\alpha}_y,\quad \tilde{\beta}(x)=:\tilde{\beta}_x,\quad \tilde{\beta}(y)=:\tilde{\beta}_y.\]
Then in each case, $(\tilde{\alpha}\vee \tilde{\beta})\mid_{\ker R(S)}\equiv 0$ implies that
\begin{align*}
    \tilde{\alpha}_x\tilde{\beta}_y+\tilde{\beta}_x\tilde{\alpha}_y=&\,0\\
    \tilde{\alpha}_x\tilde{\beta}_x-\tilde{\alpha}_y\tilde{\beta}_y=&\,0.
\end{align*}
This implies that either $\tilde{\alpha}=0$ or $\tilde{\beta}=0$. 

We conclude that every $S'\in\mathcal{C}_W$ such that $\ker S'\supsetneq \ker S$ must vanish. Hence, $S$ is extremal by Corollary~\ref{cor: nec extremal}.
\end{myproof}

We are now prepared to prove Theorem~\ref{thm: dim V=3 rank R'=1 extremal elements}:
\begin{myproof}[Proof of Theorem \ref{thm: dim V=3 rank R'=1 extremal elements}]
Let $S\in\mathcal{C}_W$ such that $R'(S)=(\alpha\wedge \beta)^{\otimes 2}$ for some $\alpha,\beta\in V^*$ and such that $R(S)$ is not extremal in $\mathcal{C}_{\Sym}$. Let furthermore $z\in(\ker\alpha\cap\ker\beta)\setminus\{0\}$. 

If $\dim\left( (z\vee V)\cap\ker S\right)>0$, then $S$ is not extremal by Proposition~\ref{prop: R' rank one case xz in ker S}. 

Now suppose $\dim\left( (z\vee V)\cap\ker S\right)=0$. Then in particular we must have $\dim\ker R(S)\leq 3$. Moreover, by Proposition~\ref{prop: dim ker extremal elements}, we must have $\dim\ker S>3$. Since $\dim \ker R'(S)=2$, this implies that $\dim\ker R(S)>1$ and hence, $\rk R(S)\in\{3,4\}$. If $\rk R(S)=4$, then $S$ is not extremal by Proposition~\ref{prop: R' rank 1 no xz in kernel}, while if $\rk R(S)=3$, then $S$ is extremal by Proposition~\ref{prop: rank R'=1, rank R=3, no xv in kerS}. This proves the assertion.
\end{myproof}

\subsection{Case $\rk R'(S)=2$}
\label{sect: rk R'=2}
We will now consider $S\in\mathcal{C}_W$ such that $\rk R'(S)=2$. In this case, we know by Proposition~\ref{prop: er proj to ers} that $R(S)$ cannot be extremal in $\mathcal{C}_{\Sym}$ and therefore, $\rk R(S)\geq 2$. Moreover, since $V$ is three-dimensional, $\dim\left(\ker R'(S)\right)=1$ and hence $\ker R'(S)=\R\cdot x\wedge y$ for some linearly independent $x,y\in V$ or, equivalently, $\ker R'(S)=\Lambda^2 U$ for some two-dimensional subspace $U\subset V$ of $V$. Our strategy will be to distinguish whether the restriction of the projection map 
\[ \pi\colon \Sym^2 V\longrightarrow \Sym^2 V/\Sym^2 U\]
to the kernel of $R(S)$ is surjective or not. We will see that $S$ as above cannot be extremal if the restriction is not surjective. This will reduce the list of possible candidates for extremal elements $S\in\mathcal{C}_W$ with $\rk R'(S)=2$ to a few cases, in which we are able to give an explicit description of the kernel. Eventually, it will turn out that none of these $S\in\mathcal{C}_W$ is extremal. The structure of the section is as follows:

\begin{center}
\footnotesize
\begin{tikzpicture}[sibling distance=10em,
  blocknode/.style = {shape=rectangle, rounded corners,
    draw, align=center},textnode/.style={shape=rectangle, rounded corners,
     align=center},->]

\node[blocknode] {$S\in \mathcal{C}_W$ s.t.~$\rk R(S)>1$, $\rk R'(S)=2$ i.e.~$\ker R'(S)=\Lambda^2 U$ \\for some 2-dimensional $U\subset V$, $\pi\colon \Sym^2 V\rightarrow \Sym^2 V/ \Sym^2 U$}
    child { node[blocknode,xshift=-1cm] {$\pi|_{\ker R(S)}$ is not surjective} 
        child {node[textnode] {\textit{not extremal by}\\ \textit{Proposition~\ref{prop: R' rank 2 sym2U 2d}}}}}
child {node[blocknode,xshift=1cm] {$\pi|_{\ker R(S)}$ is surjective\\ $\implies \rk R(S)\leq 3$}
            child { node[blocknode,xshift=-5mm] {$\rk R(S)=2$}
                child{ node[textnode] {\textit{not extremal}\\ \textit{by Proposition~\ref{prop: R' and R rank 2 pi surj}}}}}
            child { node[blocknode,xshift=8mm] {$\rk R(S)=3$}
                child{node[blocknode,xshift=3mm] {$\exists \,z\in V\setminus U$ s.t.\\$z\vee U\subset \ker S$}
                    child{node[textnode] {\textit{not extremal by}\\ \textit{Proposition~\ref{prop: R' rank 2 and R rank 3 pi surj zU in ker}}}}}
                child{ node[blocknode]{$\nexists \,z\in V\setminus U$ s.t.\\$z\vee U\subset \ker S$}
                    child{ node[textnode]{\textit{not extremal by}\\ \textit{Proposition~\ref{prop: R' rank 2 and R rank 3 pi surj special kernel}}}}}}}
;

\end{tikzpicture}
\end{center}

We will start proving that $S\in \mathcal{C}_W$ with $\rk R'(S)=2$ cannot be extremal if the restriction of the projection onto $\Sym^2 V/ \Sym^2 U$ to $\ker R(S)$ is not surjective.

\begin{prop}
\label{prop: R' rank 2 sym2U 2d}
    Suppose $\dim V=3$. Let $S\in\mathcal{C}_W$ such that $\rk R'(S)=2$. Let $U\subset V$ such that $\ker R'(S)=\Lambda^2 U$ and consider the projection 
    \[
    \pi\colon \Sym^2 V\longrightarrow \Sym^2 V/\Sym^2 U.
    \]
    We denote the image of $\ker R(S)$ under the projection $\pi$ by $\overline{\ker R(S)}:=\pi(\ker R(S))$.
    If
    \[
    \dim\left(\overline{\ker R(S)}\right)\leq 2,
    \]
    i.e. the restriction of $\pi$ to $\ker R(S)$ is not surjective, then $S$ is not extremal.
\end{prop}
\begin{myproof}
    Let $\gamma\in\ann(U)\setminus \{0\}$. Then $\gamma\vee V^*$ annihilates $\Sym^2 U$. For $S\in\mathcal{C}_W$ as above, let $\{ [T_1],[T_2]\}$ be a generating set of $\overline{\ker R(S)}$, where $[T_i]$ is represented by some $T_i\in\Sym^2 V$ for $i=1,2$. Then
    \[ 
    (\gamma\vee V^*)\cap \ann(T_1)\cap \ann(T_2)\neq \{0\}.
    \]
    Therefore, there is some $\eta \in V^*\setminus\{0\}$ such that $\gamma\vee\eta(T_i)= 0$ for $i=1,2$. Since $\gamma\vee V^*$ annihilates $\Sym^2 U$, $\gamma\vee\eta(\Tilde{T}_i)= 0$ holds for any $\Tilde{T}_i$ representing $[T_i]\in\overline{\ker R(S)}$.
    
    It follows that $\gamma\vee\eta|_{\ker S}\equiv 0$. Since $\gamma\in\ann(U)$, also $\gamma\wedge\eta|_{\ker S}\equiv 0$. We conclude that
    \[
    \left((\gamma\otimes \eta)^{\otimes 2}+(\eta\otimes \gamma)^{\otimes 2}\right)|_{\ker S}\equiv 0.
    \]
    By Corollary~\ref{cor: nec extremal} this shows that $S$ cannot be extremal.
\end{myproof}

Next, we will consider $S\in \mathcal{C}_W$ as above such that $\pi|_{\ker R(S)}$ is surjective. Then we must have $\rk R(S)\leq 3$. If $\rk R(S)=2$, the intersection $\ker S\cap \Sym^2 U$ is one-dimensional. The following lemma shows that in this situation, we have $\ker S\cap \Sym^2U=\R\cdot(u\vee v)$ for linearly independent $u,v\in U$:
\begin{lem}
\label{lem: R' rank 2 sym2 U cap kerS 1d}
    Suppose $\dim V=3$. Let $S\in\mathcal{C}_W$ such that $\rk R'(S)=2$. Let $U\subset V$ such that $\ker R'(S)=\Lambda^2 U$. If $\dim \left( \ker S\cap \Sym^2U\right)=1$, then $\ker S\cap \Sym^2U=\R\cdot(u\vee v)$ for linearly independent $u,v\in U$.
\end{lem}
\begin{myproof}
    For $S$ as above, we know that for $u,v\in U$, we have $u\wedge v\in\ker S$ and, hence, by Lemma~\ref{lem: formula S(v2,w2)},
    \[
    S(u^{\otimes 2},v^{\otimes 2})=S(u\vee v, u\vee v).
    \]
Therefore, if $u^{\otimes 2}\in\ker S$ for some $u\in U$, it follows that $u\vee v\in \ker S$ for every $v\in U$.

    Similarly, if $u^{\otimes 2}+v^{\otimes 2}\in\ker S$ for linearly independent $u,v\in V$. Then
    \[
    0\leq S(u^{\otimes 2},u^{\otimes 2})=-S(u^{\otimes 2},v^{\otimes 2})=-S(u\vee v, u\vee v)\leq 0,
    \]
    and hence, $\Sym^2 U\subset \ker S$.

    Thus, the intersection $\ker S\cap \Sym^2U$ can only be one-dimensional if it is of the form $\ker S\cap \Sym^2U=\R\cdot(u\vee v)$ for linearly independent $u,v\in U$.
\end{myproof}

With the help of Lemma~\ref{lem: R' rank 2 sym2 U cap kerS 1d}, we can deduce further information about $\ker R(S)$, when $S\in\mathcal{C}_W$ is as above with $\pi|_{\ker R(S)}$ surjective and $\rk R(S)=2$:

\begin{lem}
\label{lem: R' rank 2 and R rank 2 pi surj, existence zu in ker}
Suppose $\dim V=3$. Let $S\in\mathcal{C}_W$ such that $\rk R'(S)=2$ and $\rk R(S)=2$. Suppose that $\dim\left(\ker S\cap \Sym^2 U\right)=1$, where $U\subset V$ such that $\ker R'(S)=\Lambda^2 U$. Then there is a $z\in V\setminus U$ such that $z\vee U\subset \ker S$.
\end{lem}
\begin{myproof}
    By Lemma \ref{lem: R' rank 2 sym2 U cap kerS 1d} we know that $\ker S\cap \Sym^2 U=\R\cdot u\vee v$ for some linearly independent $u,v\in U$. Since also $u\wedge v\in \ker S$ this implies $u\otimes v\in \ker S$ and $v\otimes u\in\ker S$. It follows that
\begin{align}
\label{eq: rank R'=2, rank R=2, S(u2,v2)}
S(u^{\otimes 2},v^{\otimes 2})=S(u\otimes v,v\otimes u)=0.
\end{align}
Furthermore, we observe that for every $w\in V$:
\begin{align}
\label{eq: rank R'=2, rank R=2, S(u2,vw)}
S(u^{\otimes 2}, v\vee w)=S(u^{\otimes 2}, v\otimes w)=S(u\otimes w, v\otimes u)=0    
\end{align}
and, similarly,
\begin{align}
\label{eq: rank R'=2, rank R=2, S(v2,uw)}
S(v^{\otimes 2}, u\vee w)=0.
\end{align}

By our assumption that $\dim \left(\ker R(S)\right)=4$ and $\dim (\ker S\cap \Sym^2 U)=1$, we know that the projection
\[
\pi\colon \ker S\longrightarrow \Sym^2 V/\Sym^2 U
\]
is surjective. Hence, for $\bar{z}\in V\setminus U$, there must be $Q_i\in\Sym^2 U$, $i=1,2,3$ such that
\[
\bar{z}\vee u+Q_1,\, \bar{z}\vee v+ Q_2\;\in \ker S
\]
Since $u\vee v\in \ker S$, we may furthermore assume that
\[
Q_i=\lambda_i\, u^{\otimes 2}+\mu_i\, v^{\otimes 2}
\]
for some $\lambda_i,\mu_i\in \R$. 

We claim that $\mu_1=0$: Indeed, by assumption we have
\[
\bar{z}\vee u +\lambda_1 u^{\otimes 2}+\mu_1 v^{\otimes 2}=(\bar{z}+\lambda_1 u)\vee u + \mu_1\, v^{\otimes 2}\in \ker S.
\]
Applying equation \eqref{eq: rank R'=2, rank R=2, S(v2,uw)} for $w=\bar{z}+\lambda_1 u$, we observe that $v^{\otimes 2}$ and $(\bar{z}+\lambda_1 u)\vee u$ are orthogonal with respect to $S$. Since $S$ is semi-definite, this implies that both $\mu_1\cdot v^{\otimes 2}\in\ker S$ and $(\bar{z}+\lambda_1 u)\vee u\in\ker S$. Since by Lemma~\ref{lem: R' rank 2 sym2 U cap kerS 1d}, we know that $v^{\otimes 2}\notin \ker S$, it follows that $\mu_1=0$.

An analogous argument shows that $\lambda_2=0$. Therefore, we obtain
\[
\Span\{(\bar{z}+\lambda_1 u)\vee u,\, (\bar{z}+\mu_2 v)\vee v\}\subset \ker S.
\]

Defining
\[
z:=\bar{z}+\lambda_1\, u+\mu_2\,v\in V\setminus\{0\},
\]
we find
\begin{align*}
    z\vee u-\mu_2 u\vee v\in \ker S\quad &\implies \quad z\vee u\in \ker S\\
    z\vee v-\lambda_1 u\vee v\in \ker S\quad &\implies \quad z\vee v\in \ker S,
\end{align*}
where we used that $u\vee v\in \ker S$. It follows that $z\vee U\subset \ker S$. 
\end{myproof}

Combining Lemma~\ref{lem: R' rank 2 sym2 U cap kerS 1d} and \ref{lem: R' rank 2 and R rank 2 pi surj, existence zu in ker}, we know enough about $\ker R(S)$ for $S\in \mathcal{C}_W$ as above with $\pi|_{\ker R(S)}$ surjective and $\rk R(S)=2$ to conclude that such $S$ cannot be extremal:
\begin{prop}
\label{prop: R' and R rank 2 pi surj}
Suppose $\dim V=3$. Let $S\in\mathcal{C}_W$ such that $\rk R'(S)=2$ and $\rk R(S)=2$. Suppose that $\dim (\ker S\cap \Sym^2 U)=1$, where $U\subset V$ such that $\ker R'(S)=\Lambda^2 U$. Then $S$ is not extremal in $\mathcal{C}_W$.
\end{prop}
\begin{myproof}

    Let $S\in\mathcal{C}_W$ as in Proposition~\ref{prop: R' and R rank 2 pi surj}. Then we know by Lemma~\ref{lem: R' rank 2 sym2 U cap kerS 1d} that for some linearly independent $u,v\in U\setminus \{0\}$, we have
    \[
    \ker S\cap \Sym^2 U=\R \,u\vee v.
    \]
    Furthermore, by Lemma~\ref{lem: R' rank 2 and R rank 2 pi surj, existence zu in ker}, there is some $z\in V\setminus U$ such that $z\vee U\subset \ker S$. 

    Therefore, for some $\lambda,\mu\in \R$,
    \[
    \ker R(S)=\Span\{ u\vee v,\, z\vee u,\, z\vee v,\, z^{\otimes 2}+\lambda\, u^{\otimes 2}+\mu\, v^{\otimes 2}\}.
    \]

   We may assume that 
   \[S(u^{\otimes 2},u^{\otimes 2})=1=S(v^{\otimes 2},v^{\otimes 2}).\]
Using Lemma~\ref{lem: formula S(v2,w2)}, we find
   \begin{align*}
       S(u^{\otimes 2},v^{\otimes 2})=&\,S(u\vee v,u\vee v)-S(u\wedge v,u\wedge v)=0,\\
    S(z\wedge u,z\wedge u)=&\,-S(z^{\otimes 2},u^{\otimes 2})=\lambda\,S(u^{\otimes 2},u^{\otimes 2})+\mu \, S(v^{\otimes 2}, u^{\otimes 2})=\lambda,\\
    S(z\wedge v,z\wedge v)=&\,-S(z^{\otimes 2},v^{\otimes 2})=\lambda\,S(u^{\otimes 2},v^{\otimes 2})+\mu \, S(v^{\otimes 2}, v^{\otimes 2})=\mu.
   \end{align*}
In particular, $\lambda, \mu>0$. 

We set
\[\alpha:=\alpha_1+\sqrt{\lambda}\alpha_3,\quad \beta:=\alpha_1-\sqrt{\lambda}\,\alpha_3,\]
where $\{\alpha_1,\alpha_2,\alpha_3\}$ is the basis of $V^*$ dual to $\{u,v,z\}$. Then $v\in\ker \alpha\cap\ker\beta$ and therefore
\begin{align*}
(\alpha\vee \beta)(u\vee v)=0=(\alpha\vee \beta)(z\vee v),\quad (\alpha\wedge \beta)(u\wedge v)=0.
\end{align*}
Moreover, we compute
\begin{align*}
\left(\alpha\vee \beta\right)(z\vee u)&=0,\\
\left(\alpha\vee \beta\right)(z^{\otimes 2}+\lambda\, u^{\otimes 2}+\mu\, v^{\otimes 2})&=-\lambda+\lambda =0.
\end{align*}
Therefore, 
\[\left.\left((\alpha\vee \beta)^{\otimes 2}+(\alpha\wedge \beta)^{\otimes 2}\right)\right|_{\ker S}\equiv 0.\]
By Corollary \ref{cor: nec extremal} it follows that $S$ is not extremal.
\end{myproof}

Next, we will focus on the situation where $S\in\mathcal{C}_W$ is as above with $\pi|_{\ker R(S)}$ surjective and $\rk R(S)=3$. In this case it is in general no longer true that we can find some $z\in V\setminus U$ such that $z\vee U\subset \ker S$. However, in any case we get a relatively nice description of the kernel of $R(S)$.
\begin{lem}
\label{lem: R' rank 2 and R rank 3 pi surj, existence zu in ker}
 Suppose $\dim V=3$. Let $S\in\mathcal{C}_W$ such that $\rk R'(S)=2$ and $\rk R(S)=3$. Suppose that $\ker S\cap \Sym^2 U=\{0\}$, where $U\subset V$ such that $\ker R'(S)=\Lambda^2 U$. Then either there is some $z\in V\setminus U$ such that $z\vee U\subset \ker S$ or there are $z\in V\setminus U$ and linearly independent $u,v\in U\setminus \{0\}$ such that $\Span\{z\vee u,\, z\vee v +v^{\otimes 2}-u^{\otimes 2}\}\subset \ker S$.
 \end{lem}
 \begin{myproof}
     Let $S\in\mathcal{C}_W$ as in Lemma~\ref{lem: R' rank 2 and R rank 3 pi surj, existence zu in ker}. Let $\bar{z}\in V\setminus U$. By the assumption $\ker S\cap \Sym^2 U=\{0\}$ we know that there is a two-dimensional subspace of $\ker S\cap \Sym^2 V$ which projects surjectively onto $\pi(\bar{z}\vee U)\subset \Sym^2 V/\Sym^2 U$, where $\pi$ is the projection of $\Sym^2 V$ to the quotient space $\Sym^2 V/\Sym^2 U$. In particular by Lemma~\ref{lem: existence of degenerate el in 3d} this subspace contains a non-trivial element $T_1$ of rank $<3$, which must be of the form $T_1=(\bar{z}+v)\vee u$ for some $u,v\in U$, $u\neq 0$. We set $z:=\bar{z}+v\in V\setminus U$. Then $T_1=z\vee u\in \ker S$. 

     Next, we observe that for every $v,w\in U$,
    \begin{align} 
    \label{eq: lem R' rank 2 and R rank 3 pi surj aux}
    S(z\vee v, w\vee u)=S(z\otimes v, w\otimes u)=S(z\otimes u,w\otimes v)=S(z\vee u,w\vee v)=0,\end{align}
    where we used that $\Lambda^2 U\subset \ker S$, the symmetry of $S$ under $\tau$ and $z\vee u\in \ker S$. Using again that $z\vee u\in \ker S$, this shows that for every $v\in U$
    \begin{align}
    \label{eq: R' rank 2 R rank 3 lemma}
    z\vee v\perp_S (u\vee V).  \end{align}

    Moreover, since $\ker R(S)$ projects surjectively onto $\Sym^2 V/\Sym^2 U$, we know that for every $v\in U\setminus\{0\}$ linearly independent of $u$ there should be some $Q\in \Sym^2 U$ such that $T_2:=z\vee v +Q\in \ker S$. 
    
    \emph{Case 1: } If $\rk Q<2$, then $Q=w^{\otimes 2}$ for some $w\in U$. Then
    \begin{align}
    \label{eq: case one in R' rank 2 and R rank 3 pi surj, existence zu in ker}
    0=S(z\vee v+ w^{\otimes 2}, u^{\otimes 2})=S(w^{\otimes 2},u^{\otimes 2})=S(u\vee w, u\vee w),    
    \end{align}
    where we used \eqref{eq: R' rank 2 R rank 3 lemma} and $u\wedge w\in \ker S$. Since by assumption $\ker S\cap\Sym^2 U=\{0\}$, it follows that $w=0$ and hence $Q=0$. It follows that $T_2=z\vee v\in\ker S$. Since also $T_1=z\vee u\in\ker S$ and because we assumed that $u$ and $v$ are linearly independent, we must have $z\vee U\subset \ker S$.

    \emph{Case 2:} Now suppose $\rk Q=2$. By an analogous calculation as in \eqref{eq: case one in R' rank 2 and R rank 3 pi surj, existence zu in ker}, it follows that $Q$ cannot be definite as an element of $\Sym^2 U$. Moreover, it cannot be of the form $Q=u\vee w$ for some $w\in U\setminus\{0\}$: Otherwise, using \eqref{eq: R' rank 2 R rank 3 lemma}, we would find 
    \begin{align*}
    0=S(z\vee v +u\vee w,u\vee w)=S(u\vee w,u\vee w),
    \end{align*}
    in contrast to the assumption that $\ker S\cap \Sym^2 U=\{0\}$. Therefore, for some $w\in U$, we have $\pm Q=w^{\otimes 2}-\lambda\, u^{\otimes 2}$, for some $\lambda >0$. After rescaling we may assume that $\lambda=1$ and after replacing $T_2$ by a linear combination of $
    T_1$ and $T_2$, we may further assume $v=w$. Then, possibly after changing $T_2\rightarrow -T_2$ and $w\rightarrow -w$, we obtain
    \[
    \Span\{ z\vee u, z\vee w+w^{\otimes 2}-u^{\otimes 2}\}\subset \ker S.
    \]
\end{myproof}

\begin{rem}
The two cases described in Lemma~\ref{lem: R' rank 2 and R rank 3 pi surj, existence zu in ker} cannot occur simultaneously: Indeed, if for some $z\in V\setminus U$ we have $z\vee U\subset \ker S$, then any $\bar{z}\in V\setminus U$ such that $\bar{z}\vee v\in\ker S$ for some $v\in U$ must satisfy $\bar{z}\in\R z$, since otherwise, there would be a contradiction with the assumption that $\ker S\cap\Sym^2 U=\{0\}$. Similarly, if $z\vee U\subset \ker S$, there cannot be any linearly independent  $u,w\in U$ such that $z\vee w+w^{\otimes 2}-u^{\otimes 2}$.
\end{rem}

We can now apply Lemma~\ref{lem: R' rank 2 and R rank 3 pi surj, existence zu in ker} to explicitly describe the kernel of $R(S)$ for $S\in \mathcal{C}_W$ as above with $\pi|_{\ker R(S)}$ surjective and $\rk R(S)=3$. We will first consider the case where $z\vee U\subset \ker S$ for some $z\in V\setminus U$, $U\subset V$ as above. We will then conclude that such $S$ is not extremal:
\begin{prop}
\label{prop: R' rank 2 and R rank 3 pi surj zU in ker}
Suppose $\dim V=3$. Let $S\in\mathcal{C}_W$ such that $\rk R'(S)=2$ and $\rk R(S)=3$. Suppose that $\ker S\cap \Sym^2 U=\{0\}$, where $U\subset V$ such that $\ker R'(S)=\Lambda^2 U$. Suppose furthermore that there is some $z\in V\setminus U$ such that $z\vee U\subset \ker S$. Then $S$ is not extremal in $\mathcal{C}_W$. 
\end{prop}
\begin{myproof}

Let $S\in\mathcal{C}_W$, $U\subset V$ and $z\in V\setminus U$ as in Proposition~\ref{prop: R' rank 2 and R rank 3 pi surj zU in ker}. Then for some basis $\{u,v\}$ of $U$ and $\lambda,\,\mu,\,\delta\in \R$ we have
\[
\ker S=\Span \{ u\wedge v,\, z\vee u,\, z\vee v,\, z^{\otimes 2}+\lambda\, u^{\otimes 2}+\mu\,v^{\otimes 2}+\delta\, u\vee v\}.
\]
We may furthermore assume that $\delta=0$, possibly after changing to a different basis of $U$.

\emph{Case 1:} If $\lambda \geq0$, we may proceed as in the proof of Proposition~\ref{prop: R' and R rank 2 pi surj} to find that 
\[\left.\left[\left((\alpha_1+\sqrt{\lambda}\alpha_3)\vee (\alpha_1-\sqrt{\lambda}\,\alpha_3)\right)^{\otimes 2}+\left((\alpha_1+\sqrt{\lambda}\alpha_3)\wedge (\alpha_1-\sqrt{\lambda}\,\alpha_3)\right)^{\otimes 2}\right]\right|_{\ker S}\equiv 0,\]
where $\{\alpha_1,\alpha_2,\alpha_3\}$ is the basis of $V^*$ dual to $\{u,v,z\}$. Hence, $S$ is not extremal by Corollary~\ref{cor: nec extremal}.

\emph{Case 2:} Now suppose that both $\lambda$ and $\mu$ are negative. We will show that this case cannot occur. For this, we assume without loss of generality that $\lambda=-1=\mu$. Then 
\[
S(u^{\otimes 2},\cdot)=S(z^{\otimes 2},\cdot)-S(v^{\otimes 2},\cdot)
\]
and, since we assumed $\ker S\cap\Sym^2 U=\{0\}$, it follows that
\begin{align*}
0<&\,S(u^{\otimes 2},u^{\otimes 2})\\=&\,S(z^{\otimes 2},u^{\otimes 2})-S(v^{\otimes}, u^{\otimes 2})\\
=&\, S(z\vee u,z\vee u)-S(z\wedge u,z\wedge u)-S(u\vee v,u\vee v)+S(u\wedge v,u\wedge v)\\
=&\,-S(z\wedge u,z\wedge u)-S(u\vee v,u\vee v)<0,
\end{align*}
a contradiction. Here, we used Lemma~\ref{lem: formula S(v2,w2)} for the second equality and $z\vee u, u\wedge v\in\ker S$, $z\wedge u, u\vee v\notin \ker S$ and $S\geq 0$ in the last step. 

We conclude that $S$ as in Proposition~\ref{prop: R' rank 2 and R rank 3 pi surj zU in ker} is not extremal in $\mathcal{C}_W$. 
\end{myproof}

Finally, we apply Lemma~\ref{lem: R' rank 2 and R rank 3 pi surj, existence zu in ker} to $S\in\mathcal{C}_W$ as above with $\pi|_{\ker R(S)}$ surjective, $\rk R(S)=3$ and such that there is no $z\in V\setminus U$ such that $z\vee U\subset \ker S$, where $\pi\colon \Sym^2 V\rightarrow \Sym^2 V /\Sym^2 U$ and $U\subset V$ are as above, to show that such $S$ is not extremal:

\begin{prop}
\label{prop: R' rank 2 and R rank 3 pi surj special kernel}
Suppose $\dim V=3$. Let $S\in\mathcal{C}_W$ such that for some basis $\{u,v,z\}$ of $V$ we have
\[\ker S=\Span\{u\wedge v,\,z\vee u,\,z\vee v+v^{\otimes 2}-u^{\otimes 2},\, z^{\otimes 2}+\lambda\,u^{\otimes 2}+\mu\, v^{\otimes 2}+\delta\,u\vee v\}\]
for some $\lambda,\mu,\delta\in\R$. Then $S$ is not extremal in $\mathcal{C}_W$.
\end{prop}
\begin{myproof}
Let $S\in\mathcal{C}_W$ as in Proposition~\ref{prop: R' rank 2 and R rank 3 pi surj special kernel}. Note that the same arguments used in Lemma~\ref{lem: R' rank 2 and R rank 3 pi surj, existence zu in ker} to arrive at \eqref{eq: R' rank 2 R rank 3 lemma} apply in the setting of Proposition~\ref{prop: R' rank 2 and R rank 3 pi surj special kernel}. Thus, we have 
\[ S(z\vee w,u\vee x)=0\]
for every $w\in U$, $x\in V$. Using that $z\vee v+v^{\otimes 2}-u^{\otimes 2}\in\ker S$, we therefore find
\begin{align*}
    0=&\,S(z\vee v, u^{\otimes 2})=S(u^{\otimes 2},u^{\otimes 2})-S(v^{\otimes 2},u^{\otimes 2}),\\
    0=&\,S(z\vee v, u\vee v)=S(u^{\otimes 2},u\vee v)-S(v^{\otimes 2},u\vee v).
\end{align*}
We may furthermore assume that
\[1=S(u^{\otimes 2},u^{\otimes 2})=S(u^{\otimes 2},v^{\otimes 2})=S(u\vee v,u\vee v).\]
We set
\begin{align*}
q_1:=S(v^{\otimes 2},v^{\otimes 2}),\quad
q_2:=S(u^{\otimes 2},u\vee v)=S(v^{\otimes 2},u\vee v).
\end{align*}
Since $R(S)$ is positive semi-definite we find
\begin{align}
\label{eq: case rank r=3, rank r'=2, special kernel, constraint r psd}
q_1\geq 1,\quad -1\leq q_2\leq 1.    
\end{align} 

Since $u\wedge z\notin \ker S$, we must furthermore have
\begin{align}
\label{eq: case rank r=3, rank r'=2, special kernel, constraint 1}
    0<&\,S(u\wedge z,u\wedge z)=-S(u^{\otimes 2},z^{\otimes 2}).
\end{align}

Using that $z^{\otimes 2}+\lambda\,u^{\otimes 2}+\mu\, v^{\otimes 2}+\delta\,u\vee v\in \ker S$, we compute
\begin{align*}
S(u^{\otimes 2},z^{\otimes 2})=S(-\lambda\,u^{\otimes 2}-\mu\,v^{\otimes 2}-\delta\,u\vee v,u^{\otimes 2})=-\lambda-\mu-\delta q_2.
\end{align*}
Hence, equation \eqref{eq: case rank r=3, rank r'=2, special kernel, constraint 1} becomes 
\begin{align}
\label{eq: case rank r=3, rank r'=2, special kernel, constraint 1'}
0<\lambda+\mu+\delta q_2.
\end{align}

We show that there exist $\alpha,\beta \in V^*\setminus\{0\}$ such that 
\begin{align}
\label{eq: prop spec ker eq for alpha beta}
\left.\left((\alpha\wedge \beta )^{\otimes 2}+(\alpha\vee \beta)^{\otimes 2}\right)\right|_{\ker S}\equiv 0.\end{align}

For this, we make the ansatz
\begin{align*}
    \alpha(u)=\beta(u)=1, \quad \alpha(v)=\beta (v)=\varepsilon,\quad \alpha(z)=-\beta(z)
\end{align*}
with 
\begin{align*}
\varepsilon=\begin{cases}
    1 & \text{for }\delta\geq 0\\
    -1 & \text{for }\delta< 0
    \end{cases}. 
\end{align*}

Then we find
\begin{align*}
    (\alpha\wedge \beta)(u\wedge v)=0=
    (\alpha\vee \beta)(z\vee u)=(\alpha\vee \beta)(z\vee v+v^{\otimes 2}-u^{\otimes 2}),\\
    (\alpha\vee \beta)(z^{\otimes 2}+\lambda\,u^{\otimes 2}+\mu\, v^{\otimes 2}+\delta\,u\vee v)=-\alpha(z)^2+\lambda +\mu+|\delta|.
\end{align*}
Therefore, $\alpha$ and $\beta$ satisfy equation \eqref{eq: prop spec ker eq for alpha beta} if and only if $\alpha(z)$
solves
\begin{align*}
-\alpha(z)^2+\lambda +\mu+|\delta|=0.   
\end{align*}
Since by \eqref{eq: case rank r=3, rank r'=2, special kernel, constraint r psd} and \eqref{eq: case rank r=3, rank r'=2, special kernel, constraint 1'} we have
\begin{align*}
\lambda +\mu+|\delta|\geq\lambda+\mu +\delta q_2 >0,    
\end{align*}
we can always find a real solution for $\alpha(z)$. Using Corollary~\ref{cor: nec extremal}, we conclude that $S$ is not extremal.
\end{myproof}

We summarize our results in the following theorem:

\begin{thm}
\label{thm: characterize er in 3d}
Suppose $\dim V=3$. Let $S\in\mathcal{C}_W\setminus\{0\}$ such that $R(S)$ is not extremal in $\mathcal{C}_{\Sym}$. Then $S$ is extremal in $\mathcal{C}_W$ if and only if it is of 
the type as in Proposition~\ref{prop: rank R'=1, rank R=3, no xv in kerS}, i.e. $\rk R'(S)=1$, $\rk R(S)=3$ and $\ker S\cap (z\vee V)=\{0\}$, where $z\in V$ such that $\ker R'(S)=z\wedge V$.
\end{thm}

\subsection{Explicit description of the extremal elements}

We will now derive an explicit form for the extremal elements of the type as in Theorem~\ref{thm: characterize er in 3d}.
\begin{prop}
\label{prop: explicit description extremal R' rank 1}
    Let $\dim V=3$, $S\in\mathcal{C}_W$ such that $\rk R(S)=3$, $\ker R'(S)=z\wedge V$ for some $z\in V\setminus \{0\}$ and such that $\ker S\cap (z\vee V)=\{0\}$. Then for some parameters $d,g,h\in \R$ such that $g^2>1-d^2+dh$, and for some $e_2,e_3\in V$ such that $\{z=:e_1,e_2,e_3\}$ is a basis of $V$, in terms of the basis $\{e_1^{\otimes 2},\,e_2^{\otimes 2},\,e_3^{\otimes 2},\,e_1\vee e_2,\,e_1\vee e_3,\,e_2\vee e_3\}$ of $\Sym^2 V$, we have the representation
    \begin{align*}
    R(S)= \begin{pmatrix}
    1 & 1 & 1 & 0 & 0 & 0 \\
        1&1+d^2 & 1+dh &  dg & 0 & d \\
        1&1+dh & 1+g^2+h^2 & g(d+h) & g & h \\
        0&dg & g(d+h) &  d^2+g^2 & d & g \\
        0&0 & g  & d & 1 & 0 \\
        0&d & h &  g & 0 & 1 
    \end{pmatrix}.
\end{align*}
The kernel of $R(S)$ is given by the following:

If $d,g\neq 0$:
\begin{align*}
\ker R(S)=\Span\{ &e_1^{\otimes 2}-e_2^{\otimes 2}+d\,e_1\vee e_3,\\
&(d-h)\, e_1^{\otimes 2}+h\, e_2^{\otimes 2}-d\, e_3^{\otimes 2} +dg\,e_1\vee e_2,\\
&(d^2+g^2-dh)\, e_1^{\otimes 2}-(g^2-dh)\, e_2^{\otimes 2}-d^2\, e_3^{\otimes 2}+dg\,e_2\vee e_3\}.
\end{align*}
If $d=0$:
\begin{align*}
    \ker R(S)=\Span\left\{ (g\,e_1-e_2)\vee e_3,\; e_1^{\otimes 2}- e_2^{\otimes 2},\;
    g\,e_2^{\otimes 2}-g\,e_3^{\otimes 2}+h\, e_2\vee e_3+g^2\,e_1\vee e_2\right\}.
\end{align*}
If $g=0$:
\begin{align*}
    \ker R(S)=\Span \{ &-e_2^{\otimes 2}+e_3^{\otimes 2}+(d-h)\, e_1\vee e_3,\;\, ( d\, e_1-e_3)\vee e_2,\\ &(d-h)\,e_1^{\otimes 2}+h\, e_2^{\otimes 2}-d\,e_3^{\otimes 2}\}.
\end{align*}

\end{prop}
\begin{myproof}
Since $\ker S\cap (z\vee V)=\{0\}$ and $z\wedge V\subset \ker S$, we deduce that $S(z^{\otimes 2},\cdot)$ defines a positive definite symmetric bilinear form on $V$. Indeed, for any $v\in V$, using Lemma~\ref{lem: formula S(v2,w2)}, we observe
\[
S(z^{\otimes 2}, v^{\otimes 2})=S(z\vee v, z\vee v)-S(z\wedge v,z\wedge v)=S(z\vee v, z\vee v)>0.
\]
We complete $\{z\}$ to an orthonormal base $\{z=:e_1,e_2,e_3\}\subset V$ of $V$ with respect to $S(z^{\otimes 2},\cdot)$. In addition, we require that 
\[S(e_2^{\otimes 2},e_2\vee z)=0. \]
To see that this is possible, let $\{\bar{x},\bar{y}\}$ be a basis of the orthogonal complement of $\R\cdot z$ with respect to $S(z^{\otimes 2},\cdot)$. We may assume that \[S(\bar{y}^{\otimes 2},\bar{y}\vee z)\neq 0\]
Then
\[p_{\bar{x},\bar{y}}(t):=S\left(((\bar{x}+t\bar{y})^{\otimes 2},(\bar{x}+t\bar{y})\vee z)\right)\]
is a cubic polynomial in $t$ and therefore, for some $t_0\in\R$, we have $p_{\bar{x},\bar{y}}(t_0)=0$. Then $e_2:=\bar{x}+t_0\,\bar{y}$ has the desired property.

We set
\begin{align*}
    q:=&\,S(e_2\wedge e_3,e_2\wedge e_3)>0,\\
    a:=&\,S(e_2^{\otimes 2},e_2^{\otimes 2}),\\
    b:=&\,S(e_2^{\otimes 2},e_3^{\otimes 2})=S(e_2\vee e_3,e_2\vee e_3)-S(e_2\wedge e_3,e_2\wedge e_3)=S(e_2\vee e_3,e_2\vee e_3)-q,\\
    c:=&\, S(e_2^{\otimes 2},e_2\vee e_3),\\
    d:=&\, S(e_2^{\otimes 2}, e_1\vee e_3)=S(e_2\vee e_3,e_1\vee e_2),\\
    e:=&\, S(e_3^{\otimes 2}, e_3^{\otimes 2}),\\
    f:=&\, S(e_3^{\otimes 2}, e_2\vee e_3),\\
    g:=&\, S(e_3^{\otimes 2},e_1\vee e_2)=S(e_2\vee e_3,e_1\vee e_3),\\
    h:=&\,S(e_3^{\otimes 2}, e_1\vee e_3).
\end{align*}
With these choices of parameters, we write the quadratic form on $\Sym^2 V$ defined by $R(S)$ in terms of the basis $\{e_1^{\otimes 2},\,e_2^{\otimes 2},\,e_3^{\otimes 2},\,e_1\vee e_2,\,e_1\vee e_3,\,e_2\vee e_3\}$ as
\begin{align*}
       R(S)= \begin{pmatrix}
        1 & 1 & 1 & 0 & 0 & 0 \\
        1&a & b & 0 & d & c \\
        1&b & e & g & h & f \\
        0&0 & g & 1  & 0& d \\
        0&d & h & 0 & 1 & g \\
        0&c & f & d & g & b+q 
    \end{pmatrix}.
\end{align*}
Requiring the fourth principal minors of $R(S)$ to vanish, the first, second and third principal minors to be non-negative and the parameter $q$ to be positive, yields
\begin{align*}
&a=1+d^2, \quad b=1+dh, \quad c=dg, \quad e= 1+g^2+h^2,\quad f=g(d+h)\\ &b+q=d^2+g^2,\quad g^2>1-d^2+dh.    
\end{align*}
Furthermore, with these choices of parameters, we obtain the following basis of the kernel of $R(S)$:

If $d,g\neq 0$:
\begin{align*}
\ker R(S)=\Span\{ &e_1^{\otimes 2}-e_2^{\otimes 2}+d\,e_1\vee e_3,\\
&(d-h)\, e_1^{\otimes 2}+h\, e_2^{\otimes 2}-d\, e_3^{\otimes 2} +dg\,e_1\vee e_2,\\
&(d^2+g^2-dh)\, e_1^{\otimes 2}-(g^2-dh)\, e_2^{\otimes 2}-d^2\, e_3^{\otimes 2}+dg\,e_2\vee e_3\}.
\end{align*}

If $d=0$ (this implies $g^2>1$):
\begin{align*}
    \ker R(S)=\Span\left\{ (g\,e_1-e_2)\vee e_3,\; e_1^{\otimes 2}- e_2^{\otimes 2},\;
    g\,e_2^{\otimes 2}-g\,e_3^{\otimes 2}+h\, e_2\vee e_3+g^2\,e_1\vee e_2\right\}.
\end{align*}

If $g=0$ (this implies $d\neq 0$, $d\neq h$):
\begin{align*}
    \ker R(S)=\Span \{ &-e_2^{\otimes 2}+e_3^{\otimes 2}+(d-h)\cdot e_1\vee e_3,\;\, ( d\, e_1-e_3)\vee e_2,\\ &(d-h)\,e_1^{\otimes 2}+h\, e_2^{\otimes 2}-d\,e_3^{\otimes 2}\}.
\end{align*}

In particular, the kernel of $R(S)$ is three-dimensional and hence $\rk R(S)=3$ as required.
\end{myproof}

\begin{prop}
    Let $\dim V=3$, $S\in \mathcal{C}_W$ such that $\rk R(S)>1$. Then $S$ is extremal if and only if it is of the following form: For some basis $\{\alpha_1,\alpha_2,\alpha_3\}$ of $V^*$ and some parameters $g,h,d\in \R$ such that $g^2>1-d^2+dh$ we have
    \begin{align*}
        S=S_{\mathrm{tot}}+2(g^2+d^2-1-dh)\left[(\alpha_2\otimes \alpha_3)^{\otimes 2}+(\alpha_3\otimes \alpha_2)^{\otimes 2}\right],
    \end{align*}
    where $S_{\mathrm{tot}}\in \Sym^4 V^*$ is given by
    \begin{align*}
    S_{\mathrm{tot}}=&\,\alpha_1^4+6\,\alpha_1^2\alpha_2^2+6\,\alpha_1^2\alpha_3^2+12d\,\alpha_1\alpha_2^2\alpha_3+12g\,\alpha_1\alpha_2\alpha_3^2+4h\,\alpha_1\alpha_3^3+(1+d^2)\,\alpha_2^4\\&+4dg\,\alpha_2^3\alpha_3+6(1+dh)\,\alpha_2^2\alpha_3^2+4g(d+h)\,\alpha_2\alpha_3^3+(1+g^2+h^2)\,\alpha_3^4.    
    \end{align*}
\end{prop}

\begin{myproof}
Let $\{\alpha_1,\alpha_2,\alpha_3\}$ be the basis of $V^*$ dual to the basis $\{e_1,e_2,e_3\}$ in Proposition~\ref{prop: explicit description extremal R' rank 1}. Note that, with our conventions for the symmetric and the anti-symmetric product, we have
\begin{align*}
    2(\alpha_i\vee\alpha_j)(e_i\vee e_j)=1,\quad 2(\alpha_i\wedge \alpha_j)(e_i\wedge e_j)=1\qquad\text{for }i\neq j.
\end{align*}
Using this, it follows from Proposition~\ref{prop: explicit description extremal R' rank 1} that
\begin{align*}
    S=&\,\alpha_1^{\otimes 4}+2\,\alpha_1^{2}\vee \alpha_2^{2}+2\,\alpha_1^{2}\vee \alpha_3^{2}+(1+d^2)\,\alpha_2^{\otimes 4}+2(1+dh)\,\alpha_2^{2}\vee \alpha_3^{2}\\
    & +2d\,\alpha_2^{2}\vee(2\alpha_1\vee \alpha_3)+2dg\,\alpha_2^{2}\vee (2\alpha_2\vee \alpha_3)+(1+g^2+h^2)\,\alpha_3^{\otimes 4}\\
    &+2g\alpha_3^{2}\vee(2\alpha_1\vee \alpha_2)+2h\,\alpha_3^{2}\vee (2\alpha_1\vee \alpha_3)+2g(d+h)\,\alpha_3^{2}\vee(2\alpha_2\vee \alpha_3)\\ &+(2\alpha_1\vee \alpha_2)^{\otimes 2}+2d\,(2\alpha_1\vee \alpha_2)\vee (2\alpha_2\vee \alpha_3)+(2\alpha_1\vee \alpha_3)^{\otimes 2}\\&+2g\,(2\alpha_1\vee\alpha_3)\vee(2\alpha_2\vee \alpha_3)+(d^2+g^2)\,(2\alpha_2\vee \alpha_3)^{\otimes 2}\\&+(d^2+g^2-1-dh)\,(2\alpha_2\wedge\alpha_3)^{\otimes 2}.
\end{align*}
Observe that for any $\beta,\gamma,\delta\in V^*$, we have
\begin{align*}
    4\,\beta^2\vee(\beta\vee \gamma)&=4\,\beta^3\gamma\\
    4\,(\beta\vee \gamma)^{\otimes 2}+2\,\beta^2\vee \gamma^2&=6\,\beta^2\gamma^2\\
    4\,\beta^2\vee(\gamma\vee \delta)+8(\beta\vee\gamma)\vee (\beta\vee \delta)&=12\,\beta^2\gamma\delta.
\end{align*}
After applying these identities and rearranging the terms, we find
\begin{align*}
        S=&\,S_{\mathrm{tot}}+4(g^2+d^2-1-dh)\left[(\alpha_2\vee \alpha_3)^{\otimes 2}+(\alpha_2\wedge \alpha_3)^{\otimes 2}\right]\\
        =&\,S_{\mathrm{tot}}+2(g^2+d^2-1-dh)\left[(\alpha_2\otimes \alpha_3)^{\otimes 2}+(\alpha_3\otimes \alpha_2)^{\otimes 2}\right],
    \end{align*}
    where
    \begin{align*}
    S_{\mathrm{tot}}=&\,\alpha_1^4+6\,\alpha_1^2\alpha_2^2+6\,\alpha_1^2\alpha_3^2+12d\,\alpha_1\alpha_2^2\alpha_3+12g\,\alpha_1\alpha_2\alpha_3^2+4h\,\alpha_1\alpha_3^3+(1+d^2)\,\alpha_2^4\\&+4dg\,\alpha_2^3\alpha_3+6(1+dh)\,\alpha_2^2\alpha_3^2+4g(d+h)\,\alpha_2\alpha_3^3+(1+g^2+h^2)\,\alpha_3^4.   
    \end{align*}
\end{myproof}

In conclusion, we obtain the classification of extremal elements of $\mathcal{C}_W$ for three-dimensional $V$ as stated in Theorem~\ref{thm: ers in 3d}.

\section{Positivity bounds and applications}
\label{sect: pos bounds}
In this section, we will use the extremal rays of the cone $\mathcal{C}_W$ in order to determine a complete set of inequalities describing the dual cone
\begin{align*}
    \mathcal{C}_W^*=\{ M\in W^*:M(S)\geq 0\quad\forall S\in\mathcal{C}_W\}
\end{align*}
of $\mathcal{C}_W$, where $M(S)$ denotes the natural pairing. Note that in order to describe the dual cone $\mathcal{C}_W^*$ it is enough to require $M(S)$ to be non-negative for all extremal elements $S$ of $\mathcal{C}_W$. In this way, every extremal element in the classification in Theorem~\ref{thm: ers in 3d} gives rise to a linear inequality constraining $\mathcal{C}_W^*$. Since $\mathcal{C}_W^*$ is the precisely the cone that was denoted $\mathcal{C}$ in the introduction, these inequalities have the physical interpretation of constraining the two-to-two scattering amplitudes of a relativistic quantum theory that is compatible with the physical principles of unitarity, locality and causality \cite{zhang20,Li_et_al21} (see Appendix~\ref{app:elasticScattering} for details). 

In this section, we will first introduce the notions of \emph{elastic} and \emph{inelastic} bounds. We will then give an alternative description of all elastic bounds and show that it depends on less parameters. After that we will restrict to the spaces of amplitudes satisfying certain additional symmetries and show that in these cases the elastic bounds are sufficient to describe the whole space. Finally, we will discuss the effect of the inelastic bounds in general. If not stated otherwise, in the entire section we will assume that $\dim V=3$.

\subsection{Elastic and inelastic positivity bounds}

In the literature, the bounds coming from the extremal rays of the first and the second type in Theorem~\ref{thm: ers in 3d} are referred to as \emph{elastic (positivity) bounds}. This terminology is justified in Appendix~\ref{app:elasticScattering}. Positivity bounds that are not elastic are called \emph{inelastic}. In the case where $\dim V=3$ these are the bounds corresponding to the extremal ray of the third kind in Theorem~\ref{thm: ers in 3d}.

In the following, we will apply Theorem~\ref{thm: ers in 3d} in order to derive the complete set of positivity bounds (elastic and inelastic) for scattering processes with three flavors of particles, i.e.~$\dim V=3$. 

\subsubsection{The elastic bounds}
Recall that the elastic bounds are the inequalities  corresponding to the extremal rays of the first and second kind. Using the symmetries of $W^*$, they can be summarized as
\begin{align}
\label{eq: elastic constraint general form}
    M(\alpha\otimes \beta,\alpha\otimes \beta)\geq 0,\quad \forall\alpha,\beta\in V^*
\end{align}
for every $M\in (\mathcal{C}_W)^*$. Note that the constraint \eqref{eq: elastic constraint general form} does not change under (non-zero) rescaling of $\alpha$ and $\beta$. Thus, the family of elastic bounds is parameterized by $\R \mathbb{P}^2\times \R \mathbb{P}^2$ and is therefore four-dimensional.

In the following, we will see that the amount of parameters needed to describe all bounds can be reduced further to two. For this, we fix a basis $\{\alpha_1,\alpha_2,\alpha_3\}$ and write $M_{ijkl}=M(\alpha_i,\alpha_j,\alpha_k,\alpha_l)$. Moreover, we use the isomorphism \begin{align}
\label{eq: Phi}
\Phi\colon \R^9\longrightarrow V^*\otimes V^*     
\end{align}
defined by the basis $\{\alpha_i\otimes \alpha_j\}_{i,j=1}^3$ in alphabetic ordering. The following result is well-known:
\begin{lem}
\label{lem: relations for alpha tensor beta}
    A vector $u\in \R^9\cong V^*\otimes V^*$ represents a decomposable tensor $\gamma\in V^*\otimes V^*$, i.e.~$\gamma=\alpha\otimes \beta$ for some $\alpha,\beta\in V^*$, if and only if it satisfies the following relations:
    \begin{align*}
        &u_1 u_5-u_2 u_4=0,\quad u_1 u_6-u_3 u_4=0,\quad u_1 u_8-u_2 u_7=0,\\
        &u_1 u_9-u_3 u_7=0, \quad u_2 u_6-u_3 u_5=0,\quad u_2 u_9-u_3 u_8=0,\\
        &u_4 u_8-u_5 u_7=0,\quad u_4 u_9-u_6 u_7=0,\quad u_5 u_9-u_6 u_8=0.
    \end{align*}
\end{lem}
\begin{myproof}
Given an element $\gamma\in V^*\otimes V^*$ we can think of it as a linear map $\gamma\colon V\rightarrow V^*$. Using the basis $\{\alpha_i\}_{i=1}^3$ we may then represent $\gamma$ by a matrix. The condition that $\gamma=\alpha\otimes \beta$ for some $\alpha,\beta\in V^*$ is then equivalent to the statement that the determinant of all submatrices of order two must vanish. This yields precisely the above conditions on the components $u_i$ of the vector $u\in\R^9\cong V^*\otimes V^*$ representing $\gamma$.
\end{myproof}

\begin{prop}
\label{prop: elastic bounds}
    Let $M\in W^*$. Then $M$ satisfies the elastic bounds if and only if the following hold:
    \begin{align*}
    &M_{1111}\geq0,\quad M_{1212}\geq0,\quad M_{1313}\geq0,\quad M_{2222}\geq0,\quad M_{2323}\geq0,\quad M_{3333}\geq0,\\
    & \qq _1\qq _4-\qq _2^2\geq 0,\quad \qq _1\qq _6-\qq _3^2\geq0,\quad \qq _4\qq _6-\qq _5^2\geq0,\\
    & \qq _1\qq _4\qq _6+2\,\qq _2\qq _3\qq _5-\qq _1\qq _5^2-\qq _2^2\qq _6-\qq _3^2\qq _4\geq0,
    \end{align*}
    where the $\qq _i$ are real quadratic polynomials in two variables given by
    \begin{align*}
        \qq _1(t_1,t_2)&:=M_{1111}+2t_1\,M_{1112}+2t_2\,M_{1113}+t_1^2\,M_{1212}+2t_1t_2\,M_{1213}+t_2^2\,M_{1313},\\
        \qq _2(t_1,t_2)&:=M_{1112}+2t_1\,M_{1122}+2t_2\,M_{1123}+t_1^2\,M_{1222}+2t_1t_2\,M_{1223}+t_2^2\,M_{1323},\\
        \qq _3(t_1,t_2)&:=M_{1113}+2t_1\,M_{1123}+2t_2\,M_{1133}+t_1^2\,M_{1232}+2t_1t_2\,M_{1233}+t_2^2\,M_{1333},\\
        \qq _4(t_1,t_2)&:=M_{1212}+2t_1\,M_{1222}+2t_2\,M_{1232}+t_1^2\,M_{2222}+2t_1t_2\,M_{2223}+t_2^2\,M_{2323},\\
        \qq _5(t_1,t_2)&:=M_{1213}+2t_1\,M_{1223}+2t_2\,M_{1233}+t_1^2\,M_{2223}+2t_1t_2\,M_{2233}+t_2^2\,M_{2333},\\
        \qq _6(t_1,t_2)&:=M_{1313}+2t_1\,M_{1323}+2t_2\,M_{1333}+t_1^2\,M_{2323}+2t_1t_2\,M_{2333}+t_2^2\,M_{3333}.
    \end{align*}
\end{prop}

Before proving Proposition~\ref{prop: elastic bounds}, we will introduce the following notation for better readability:
\begin{align*}
    m_1=&\,M_{1111};\\
    m_2=&\,M_{1112}=M_{1121}=M_{1211}=M_{2111};\\
    m_3=&\,M_{1113}=M_{1131}=M_{1311}=M_{3111};\\
    m_4=&\,M_{1122}=M_{1221}=M_{2112}=M_{2211};\\
    m_5=&\,M_{1212}=M_{2121};\\
    m_6=&\,M_{1123}=M_{1132}=M_{2311}=M_{3211}=M_{1321}=M_{3112}=M_{1231}=M_{2113};\\
    m_7=&\,M_{1213}=M_{1312}=M_{2131}=M_{3121};\\
    m_8=&\,M_{1133}=M_{1331}=M_{3113}=M_{3311};\\
    m_9=&\,M_{1313}=M_{3131};\\
    m_{10}=&\,M_{1222}=M_{2122}=M_{2212}=M_{2221};\\
    m_{11}=&\,M_{1223}=M_{2132}=M_{2213}=M_{1322}=M_{2231}=M_{3122}=M_{3221}=M_{2312};\\
    m_{12}=&\,M_{1232}=M_{3212}=M_{2123}=M_{2321};\\
    m_{13}=&\,M_{1233}=M_{1332}=M_{2133}=M_{3123}=M_{3312}=M_{3213}=M_{3321}=M_{2331};\\
    m_{14}=&\,M_{1323}=M_{2313}=M_{3132}=M_{3231};\\
    m_{15}=&\,M_{1333}=M_{3133}=M_{3313}=M_{3331};\\
    m_{16}=&\,M_{2222};\\
    m_{17}=&\,M_{2223}=M_{2232}=M_{2322}=M_{3222};\\
    m_{18}=&\,M_{2233}=M_{2332}=M_{3223}=M_{3322};\\
    m_{19}=&\,M_{2323}=M_{3232};\\
    m_{20}=&\,M_{2333}=M_{3233}=M_{3323}=M_{3332};\\
    m_{21}=&\,M_{3333}.
\end{align*}
\begin{myproof}[Proof of Proposition~\ref{prop: elastic bounds}]
By Lemma~\ref{lem: relations for alpha tensor beta}, $M\in W^*$ satisfies the elastic bounds if and only if for every $u\in \R^9$ satisfying the constraints in Lemma~\ref{lem: relations for alpha tensor beta} we have 
\begin{align*}
    0\leq &M(\Phi(u),\Phi(u))\\
    =&\,u_1^2\,m_1+2u_1(u_2+u_4)\,m_2+2u_1(u_3+u_7)\,m_3+2(u_2 u_4+u_1u_5)\,m_4+(u_2^2+u_4^2)\,m_5\\
    &+2(u_3u_4+u_1u_6+u_2u_7+u_1u_8)\,m_6+2(u_2u_3+u_4u_7)\,m_7+2(u_3u_7+u_1u_9)\,m_8\\
    &+(u_3^2+u_7^2)\,m_9+2(u_2+u_4)u_5\,m_{10}+2(u_3u_5+u_2u_6+u_5u_7+u_4u_8)\,m_{11}\\
    &+2(u_4u_6+u_2u_8)\,m_{12}+2(u_6u_7+u_3u_8+u_2u_9+u_4u_9)\,m_{13}+2(u_3u_6+u_7u_8)\,m_{14}\\
    &+2(u_3+u_7)u_9\,m_{15}+u_5^2\,m_{16}+2u_5(u_6+u_8)\,m_{17}+2(u_6u_8+u_5u_9)\,m_{18}\\
    &+(u_6^2+u_8^2)\,m_{19}+2(u_6+u_8)u_9\,m_{20}+u_9^2\,m_{21},
\end{align*}
Where $\Phi$ is as in~\eqref{eq: Phi}. Henceforth, we will use the shorter notation $M(u,u)=M(\Phi(u),\Phi(u))$. If all $u_i$ but one vanish, this implies 
\begin{align}
\label{eq: elastic constraints linear}
    m_1\geq 0,\quad m_5\geq0,\quad m_9\geq0,\quad m_{16}\geq 0,\quad m_{19}\geq0,\quad m_{21}\geq0.
\end{align}
Now suppose that $u_i\neq 0$ for precisely two indices $i$. Note that not all choices of pairs of indices are allowed here, since e.g.~the relation $u_1u_5-u_2u_4=0$ prevents us from choosing $u_1,u_5\neq0$ when $u_2=0$. But, for allowed choices such as $u_1,u_2\neq0$, all other $u_i=0$, the relations in Lemma~\ref{lem: relations for alpha tensor beta} are trivially satisfied for all (non-zero) values of $u_1$ and $u_2$. Therefore, in this case the condition $M(u,u)\geq0$ is the same as requiring the restriction of $M$ to the two-dimensional subspace $\Span\{e_1,e_2\}\subset \R^9$, where $e_i$ are the standard orthonormal basis vectors of $\R^9$, to be positive semi-definite. Proceeding like this for all allowed choices of $u$ with only two non-vanishing components we obtain the additional constraints
\begin{equation}
\label{eq: elastic constraints quadratic no params}
\begin{aligned}
    &m_1m_5-m_2^2\geq0,\quad m_1m_9-m_3^2\geq0,\quad m_5m_9-m_7^2\geq0,\\
    &m_5m_{16}-m_{10}^2\geq0,\quad {m_5 m_{19}}-m_{12}^2\geq0,\quad m_9m_{19}-m_{14}^2\geq0,\\ &  m_9m_{21}-m_{15}^2\geq 0,\quad m_{19}m_{16}-m_{17}^2\geq0,\quad m_{19}m_{21}-m_{20}^2\geq0.
    \end{aligned}
\end{equation}
For allowed $u$ with only three $u_i$ non-vanishing the same reasoning yields the additional constraints
\begin{align*}
    &2 m_2 m_3 m_7+ m_1 m_5 m_9-m_3^2 m_5  - m_1 m_7^2 - m_2^2 m_9 
    \geq 0,\\
    & 2 m_{10} m_{12} m_{17}+ m_{16} m_{19} m_5-m_{12}^2 m_{16}  - m_{10}^2 m_{19} - m_{17}^2 m_5 \geq0,\\
    & 2 m_{14} m_{15} m_{20} + m_{19} m_{21} m_9-m_{15}^2 m_{19} - m_{14}^2 m_{21} - m_{20}^2 m_9 \geq0.
\end{align*}
When four components of $u$ are non-vanishing there will always remain a non-trivial constraint. For instance, if all components of $u$ vanish but $u_1$, $u_2$, $u_4$ and $u_5$, we set 
\begin{align*}
    a:=u_1u_5=u_2u_4\quad \implies \quad u_5=a/u_1,\, u_4=a/u_2.
\end{align*}
The elastic constraints become
\begin{align}
\label{eq:elastic with a}
0\leq &\,a^2 \left(u_1^2\,m_5+2u_1u_2\,m_{10}+u_2^2\,m_{16}\right)+2a\left(u_1^3u_2\,m_2+2u_1^2u_2^2\,m_4+u_1u_2^3\,m_{10}\right)\nonumber\\
&+u_1^4u_2^2\,m_1+2u_1^3u_2^3\,m_2+u_1^2u_2^4\,m_5
\end{align}
for all $a, u_1,u_2\neq 0$. Note that by the constraint~\eqref{eq: elastic constraints quadratic no params}, the inequality also holds for $a=0$, as well as
\begin{align*}
    u_1^2\,m_5+2 u_1u_2\,m_{10}+u_2^2\,m_{16}\geq 0.
\end{align*}
Thinking of the right hand-side of~\eqref{eq:elastic with a} as a quadratic form evaluated on $(1,a)$ we find that we get the additional constraint that for all $u_1,u_2\in\R\setminus\{0\}$, we have
\begin{align*}
    0\leq &\left(u_1^2\,m_1 +2u_1u_2\,m_2+u_2^2\,m_5\right)\left(u_1^2\,m_5+2u_1u_2\,m_{10}+u_2^2\,m_{16}\right)\\
    &-\left(u_1^2\,m_2+2u_1u_2\,m_4+u_2^2\,m_{10}\right)^2.
\end{align*}
Using again the previous constraints (or alternatively by continuity), we see that this still holds true for $u_1=0$ or $u_2=0$. We assume without loss of generality that $u_1=1$ and set $u_2:=t$. Then, after introducing the notation
\begin{align*}
    \P_{i,j,k}(t):=m_i+2t\,m_j+t^2\,m_k,
\end{align*}
the constraint becomes
\begin{align*}
    \P_{1,2,5}(t)\,\P_{5,10,16}(t)-\P_{2,4,10}(t)^2\geq0.
\end{align*}
Proceeding similarly for other $u$ with precisely four non-vanishing components, we find moreover
\begin{align*}
    &\P_{1,3,9}(t)\,\P_{5,12,19}(t)-\P_{2,6,14}(t)^2\geq0,\quad\P_{1,2,5}(t)\,\P_{9,14,19}(t)-\P_{3,6,12}(t)^2\geq0,\\ &\P_{1,3,9}(t)\,\P_{9,15,21}(t)-\P_{3,8,15}(t)^2\geq0,\quad\P_{5,7,9}(t)\,\P_{16,17,19}(t)-\P_{10,11,14}(t)^2\geq0,\\ &\P_{5,7,9}(t)\,\P_{19,20,21}(t)-\P_{12,13,15}(t)^2\geq0,\quad\P_{5,10,16}(t)\,\P_{9,14,19}(t)-\P_{7,11,16}(t)^2\geq0,\\ &\P_{5,12,19}(t)\,\P_{9,15,21}(t)-\P_{7,13,20}(t)^2\geq0,\quad\P_{16,17,19}(t)\,\P_{19,20,21}(t)-\P_{17,18,20}(t)^2\geq0.
\end{align*}
Note that these already imply the quadratic constraints~\eqref{eq: elastic constraints quadratic no params} above. 

There are no $u$ satisfying the relations of Lemma~\ref{lem: relations for alpha tensor beta} with precisely 5, 7 or 8 non-vanishing components. 
If there are precisely six non-vanishing components, e.g.~$u_i\neq0$ for $i\leq 6$, $u_i=0$ else, we set 
\begin{align*}
    &a:=u_1u_5=u_2u_4,\quad u_1 u_6-u_3 u_4=0,\quad u_2 u_6-u_3 u_5=0\\ &\implies \quad u_4=a/u_2,\quad u_5=a/u_1,\quad u_6=\frac{u_3}{u_1u_2}a.
\end{align*}
Following the same strategy as above, we find that for all $t_1,t_2\in \R$
\begin{align*}
  P_{1,2,3,5,7,9}(t_1,t_2)\,\P_{5,10,12,16,17,19}(t_1,t_2)-\P_{2,4,6,10,11,14}(t_1,t_2)^2\geq0 , 
\end{align*}
where 
\begin{align*}
    \P_{i,j,k,l,p,q}(t_1,t_2):=m_i+2t_1\,m_j+2t_2\,m_k+t_1^2m_l+2t_1t_2\,m_p+t_2^2\,m_q.
\end{align*}
Similarly, we find that for all $t_1,t_2\in \R$ it holds that
\begin{align*}
    \P_{1,2,3,5,7,9}(t_1,t_2)\, \P_{9,14,15,19,20,21}(t_1,t_2)-\P_{3,6,8,12,13,15}(t_1,t_2)^2&\geq0,\\
    \P_{5,10,12,16,17,19}(t_1,t_2)\,\P_{9,14,15,19,20,21}(t_1,t_2)-\P_{7,11,13,17,18,20}(t_1,t_2)^2&\geq0.    
\end{align*}
Note that when setting either $t_1=0$ or $t_2=0$, these imply the quadratic constraints above.

In the case where none of the components of $u$ vanish we set
\begin{align*}
    &a:=u_1u_5=u_2u_4,\quad b:=u_1u_8=u_2u_7.
\end{align*}
Using all constraints from Lemma~\ref{lem: relations for alpha tensor beta} we find that
\begin{align*}
u_4=\frac{a}{u_2},\quad u_5=\frac{a}{u_1},\quad u_6=\frac{u_3}{u_1u_2}a,\quad u_7=\frac{b}{u_2},\quad u_8=\frac{b}{u_1},\quad u_9=\frac{u_3}{u_1u_2}b.    
\end{align*}

We assume without loss of generality that $u_1=1$ and set $u_2:=t_1$, $u_3:=t_2$. Then the elastic constraints become
\begin{align*}
    0\leq &\,\P_{1,2,3,5,7,9}(t_1,t_2) +2a\, \frac{1}{t_1}\P_{2,4,6,10,11,14}(t_1,t_2) + 2c\, \frac{1}{t_1}\P_{3,6,8,12,13,15}(t_1,t_2)\\
    &+a^2\,\frac{1}{t_1^2}\P_{5,10,12,16,17,19}(t_1,t_2)+2ac\, \frac{1}{t_1^2}\P_{7,11,13,17,18,20}(t_1,t_2)+c^2\,\frac{1}{t_1^2}\P_{9,14,15,19,20,21}(t_1,t_2)
\end{align*}
for all $t_1,t_2\neq 0$. Using the previous constraints, this is equivalent to
\begin{align*}
    0\leq &\,\P_{1,2,3,5,7,9}(t_1,t_2)\,\P_{5,10,12,16,17,19}(t_1,t_2)\,\P_{9,14,15,19,20,21}(t_1,t_2)\\&+2\,\P_{2,4,6,10,11,14}(t_1,t_2)\,\P_{3,6,8,12,13,15}(t_1,t_2)\,\P_{7,11,13,17,18,20}(t_1,t_2)\\
    &-\P_{1,2,3,5,7,9}(t_1,t_2)\,\P_{7,11,13,17,18,20}(t_1,t_2)^2-\P_{2,4,6,10,11,14}(t_1,t_2)^2\,\P_{9,14,15,19,20,21}(t_1,t_2)\\
    &-\P_{3,6,8,12,13,15}(t_1,t_2)^2\,\P_{5,10,12,16,17,19}(t_1,t_2)
\end{align*}
for all $t_1,t_2\in\R\setminus\{0\}$. Moreover, by the previous constraints, the inequality remains true also for $t_1=0$ or $t_2=0$. Furthermore, we note that this constraint implies the cubic constraints above when setting $t_1,t_2=0$, or setting $t_1=0$, $t_2\to\infty$ or $t_2=0$, $t_1\to\infty$. Observing that the $\qq_i$
in the statement agree with the $\P$'s appearing in the constraints, the assertion follows.
\end{myproof}

\subsubsection{The inelastic bounds}
\label{sect: inelastic bounds}
After fixing some basis $\{\alpha_1,\alpha_2,\alpha_3\}$ of $V^*$ we can parameterize the extremal rays of the third kind by the parameters $d,g,h$ as above plus some $A\in\Sl(V)\cong \Sl(3,\R)$ via
\begin{align*}
    S(A,d,g,h)=A^* S(\id,d,g,h),
\end{align*}
where
\begin{align*}
S(\id,d,g,h)=&\,\alpha_1^4+6\,\alpha_1^2\alpha_2^2+6\,\alpha_1^2\alpha_3^2+12d\,\alpha_1\alpha_2^2\alpha_3+12g\,\alpha_1\alpha_2\alpha_3^2+4h\,\alpha_1\alpha_3^3+(1+d^2)\,\alpha_2^4\\&+4dg\,\alpha_2^3\alpha_3+6(1+dh)\,\alpha_2^2\alpha_3^2+4g(d+h)\,\alpha_2\alpha_3^3+(1+g^2+h^2)\,\alpha_3^4\\
    &+2\,(g^2+d^2-1-dh)\left[(\alpha_2\otimes \alpha_3)^{\otimes 2}+(\alpha_3\otimes \alpha_2)^{\otimes 2}\right].
\end{align*}
Then the inelastic bounds become
\begin{align*}
    0\leq S(A,g,d,h)(M)=A^*S(\id,d,g,h)(M)=S(\id,d,g,h)(A\cdot M),
\end{align*}
for all $A\in\Sl(V)$ and $d,g,h\in\R$ such that $g^2+d^2-1-dh>0$. Here, $A\cdot M$ denotes the induced action of $A$ on $M$.

In the following section we will consider amplitudes that are invariant under additional symmetries. In these cases we will see that the number of parameters describing the bounds can be reduced.

\subsection{Positivity bounds for amplitudes with symmetries}
We will now study the spaces of amplitudes that are invariant under the action of some symmetry group. More precisely, for some group $G$ acting on the space $W^*$, we would like to determine the intersection of the space
\begin{align*}
(\mathcal{C}_W^*)^{G}=\{ M\in \mathcal{C}_W^* : A\cdot M=M\;\forall A\in G\} =\mathcal{C}_{W}^*\cap (W^*)^G 
\end{align*}
of $G$-invariant elements in $W^*$ with the cone $\mathcal{C}_W^*$. Note that $(\mathcal{C}_W^*)^{G}=(W^G\cap\mathcal{C}_W)^*$, where $W^G$ denotes the space of elements in $W$ that are invariant under the action of $G$ induced by the action on $W^*$.

In this section we will study the cases where $G=\O(3)$, $G=\Z_2^{\,3}$ and $G=\SO(2)$ and such that the action on $W^*$ is induced by some action of $G$ on $V$. In each of these cases we will see that the elastic bounds are sufficient in order to describe the cone $(\mathcal{C}_W^*)^{G}$.

\subsubsection{$\O(3)$-invariant amplitudes and chiral perturbation theory}
\label{sect: O3 inv}
We will now consider amplitudes that are invariant under some $\O(3)$-symmetry. 

Fixing a basis $\{e_i\}_{i=1}^3$ of $V$ we will now identify $V\cong \R^3$ and consider the group $\O(3)=\O(V)$. The action of $\O(3)$ on $V$ induces an action on $W^*$. In the following, we will consider the cone $(\mathcal{C}_W^*)^{\O(3)}$ of invariant elements in $\mathcal{C}_W^*$.

\begin{thm}
\label{thm: O(3) invariant is elastic}
    Let $M\in (W^*)^{\O(3)}$. Then $M\in \mathcal{C}_W^*$ if and only if it satisfies the elastic positivity bounds.
\end{thm}
\begin{myproof}
Every four-tensor $M\in {V^*}^{\otimes4} $ that is invariant under the action of $\O(3)$ is of the form \cite{jeffrey}
\begin{align*}
    M=&\,a_1\left(e_1^2+e_2^2+e_3^2\right)^{\otimes2}+a_2\left(e_1^4+e_2^4+e_3^4+2(e_1\vee e_2)^{\otimes2}+2(e_1\vee e_3)^{\otimes2}+2(e_2\vee e_3)^{\otimes2}\right)\\&+a_3\left((e_1\wedge e_2)^{\otimes2}+(e_1\wedge e_3)^{\otimes2} +(e_2\wedge e_3)^{\otimes2}\right)\\
    =&\,(a_1+a_2)\left(e_1^4+e_2^4+e_3^4\right)\\&+2a_1\left(e_1^2\vee e_2^2+2(e_1\vee e_2)^{\otimes2}+e_1^2\vee e_3^2+2(e_1\vee e_3)^{\otimes2}+e_2^2\vee e_3^2+2(e_2\vee e_3)^{\otimes2}\right)\\
    &+(2a_2-4a_1)\left((e_1\vee e_2)^{\otimes2}+(e_1\vee e_3)^{\otimes2}+(e_2\vee e_3)^{\otimes2}\right)\\&+a_3\left((e_1\wedge e_2)^{\otimes2}+(e_1\wedge e_3)^{\otimes2} +(e_2\wedge e_3)^{\otimes2}\right).
\end{align*}
Note that the terms with the prefactors $(a_1+a_2)$ and $2a_1$ in the last equality are fully symmetric tensors. This shows that $M$ is invariant under $\tau$ (and therefore contained in $W^*$) if and only if $a_3=2a_2-4a_1$.

From now on, we will assume that $a_3=2a_2-4a_1$. Then, using Proposition~\ref{prop: elastic bounds}, it follows that the elastic bounds for $M$ are equivalent to $a_2\geq|a_1|$. 

Next, we would like to derive the inelastic bounds. Note that since $M$ is invariant under $\O(3)$, using the polar decomposition it is enough to consider $A\in\Sym^2(V)$, $A>0$ in the parameterization given in Section~\ref{sect: inelastic bounds}. We may furthermore assume without loss of generality that the orthonormal basis vectors $\{e_i\}_{i=1}^3$ are eigenvectors of $A$ with eigenvalues $\lambda_i>0$. Then we find

\begin{align*}
    A\cdot M=&\,(a_1+a_2)\sum_i \lambda_i^4 e_i^4+ 2a_1\sum_{i<j}\lambda_i^2\lambda_j^2\,e_i^2\vee e_j^2+ 2a_2\sum_{i<j}\lambda_i^2\lambda_j^2\,(e_i\vee e_j)^{\otimes2}\\&+(2a_2-4a_1)\sum_{i<j}\lambda_i^2\lambda_j^2\,(e_i\wedge e_j)^{\otimes2}.
\end{align*}

Hence, the inelastic bounds become
\begin{align*}
0\leq &\,S(\id,d,g,h)(A\cdot M)\\
=&\,(a_1+a_2)\,\left\{\lambda_1^4+2\lambda_1^2\lambda_2^2+2\lambda_1^2\lambda_3^2+(1+d^2)\,\lambda_2^4+2(1+dh)\,\lambda_2^2\lambda_3^2+(1+g^2+h^2)\,\lambda_3^4\right\}\\
&+4(a_2-a_1)(g^2+d^2-1-dh)\,\lambda_2^2\lambda_3^2
\end{align*}
for all $g,d,h\in\R$ such that $g^2+d^2-1-dh>0$ and $\lambda_i>0$. However, these inequalities already follow from the elastic bounds: If $a_2\geq |a_1|$, we find that
\begin{align*}
S(\id,d,g,h)(A\cdot M)\geq &\,(a_1+a_2)\left\{d^2\,\lambda_2^4+2dh\,\lambda_2^2\lambda_3^2+h^2\,\lambda_3^4\right\}\\
    \geq &\, (a_1+a_2)\left\{d^2\,\lambda_2^4-2|dh|\,\lambda_2^2\lambda_3^2+h^2\,\lambda_3^4\right\}\\
    =&\,(a_1+a_2)(|d|\,\lambda_2^2-|h|\,\lambda_3^2)^2\geq0.
\end{align*}   
\end{myproof}

\begin{app}[Chiral perturbation theory]
\label{appl: ChPT}
Chiral perturbation theory is an effective field theory for quantum chromodynamics\, i.e.~the theory of strong interactions between quarks and gluons. It captures the dynamics of quantum chromodynamics below its confinement scale, in terms of the lightest mesons and baryons. In particular, the field theory which describes the three lightest mesons, namely \textit{pions}, is built out of three scalar fields, adding up to three flavors. Moreover, the pion fields are constrained to transform as a triplet under the $\operatorname{SU}(2)$ flavor symmetry of quantum chromodynamics, that exchanges up and down quarks. As a result, the amplitude for pion two-to-two scattering at low energies is $\O(3)$-symmetric, which can be directly checked from the associated four-tensor, 
\begin{align*}
    M=&\,(2\ell_1+3\ell_2)\left\{e_1^4+e_2^4+e_3^4+2(e_1\vee e_2)^{\otimes 2}+2(e_1\vee e_3)^{\otimes 2}+2(e_2\vee e_3)^{\otimes 2}\right\}\\
    &+(2\ell_1+\ell_2)(e_1^2+e_2^2+e_3^2)^{\otimes2}+2(\ell_2-2\ell_1)\left\{(e_1\wedge e_2)^{\otimes 2}+(e_1\wedge e_3)^{\otimes 2}+(e_2\wedge e_3)^{\otimes 2}\right\},
\end{align*}
where $\ell_1$ and $\ell_2$ are free parameters (see Appendix~\ref{app: ChPT} for how to obtain $M$ and the physical meaning of $\ell_1$ and $\ell_2$). By Theorem~\ref{thm: O(3) invariant is elastic} it follows that the elastic positivity bounds $\ell_2\geq -\ell_1$, $\ell_2\geq 0$ are the strongest positivity bounds that can be derived from two-to-two forward scattering of pions. They agree with the bounds previously derived in the literature, see e.g. Ref.~\cite{Wang:2020jxr}.
\end{app}

\subsubsection{$\Z_2^{\,3}$-invariant amplitudes}
\label{sect: Z2 inv ampl}
We will now study the case where the amplitude is invariant under the action of $\Z_2^{\,3}$ on $V$ generated by $Q_i\in \End(V)$, $i\in\{1,2,3\}$, which are represented by the matrices
\begin{align*}
Q_1=\begin{pmatrix}-1&0&0\\0&1&0\\0&0&1\end{pmatrix},\quad Q_2=\begin{pmatrix}1&0&0\\0&-1&0\\0&0&1\end{pmatrix},\quad Q_3=\begin{pmatrix}1&0&0\\0&1&0\\0&0&-1\end{pmatrix}
\end{align*}
with respect to some fixed basis $\{e_i\}_{i=1}^3$ of $V$. Introducing the notation
\begin{align*}
    U_0:= \Span\{e_1^2,e_2^2,e_3^2\},\quad U_{ij}:= \Span\{e_i\otimes e_j,e_j\otimes e_i\}, \; i<j,
\end{align*}
we have
\begin{align*}
    \left(\Sym^2(V^{\otimes 2})\right)^{\Z_2^{\,3}}=\Sym^2 U_0\oplus \Sym^2 U_{12}\oplus \Sym^2 U_{13} \oplus \Sym^2 U_{23},    
\end{align*}
and $M\in (W^*)^{\Z_2^{\,3}}$ if and only if $M\in \left(\Sym^2(V^{\otimes 2})\right)^{\Z_2^{\,3}}$ and for all $i,j$ we have
\begin{align*}
M(\alpha_i^{\otimes 2},\alpha_j^{\otimes 2})=M(\alpha_i\otimes \alpha_j,\alpha_j\otimes \alpha_i),\quad M(\alpha_i\otimes \alpha_j,\alpha_i\otimes \alpha_j)=M(\alpha_j\otimes \alpha_i,\alpha_j\otimes \alpha_i),    
\end{align*}
where $\{\alpha_i\}_{i=1}^3$ denotes the dual basis of $\{e_i\}_{i=1}^3$. 

As in the case of the $\O(3)$-symmetric amplitudes, it will turn out that the elastic bounds are sufficient to describe the whole cone $(\mathcal{C}_W^*)^{\Z_2^{\,3}}$. Since $\Z_2^{\,3}$ is a subgroup of $\O(3)$ this is a strict relaxation of Theorem~\ref{thm: O(3) invariant is elastic}. For the proof, we will however follow a different approach. First, we prove the following lemma.
\begin{lem}
\label{lem: rank proj S}
    Let $p\colon W\rightarrow W^{\Z_2^{\,3}}$ be the orthogonal projection. Then for every $S\in \mathcal{C}_W$, we have $\rk p(S)\geq \rk S$, where $\rk p(S)$ denotes the rank of $p(S)$ as an element in $\Sym^2(V^*\otimes V^*)$. Similarly, we have $\rk R(p(S))\geq \rk R(S)$ and $\rk R'(p(S))\geq \rk R'(S)$, where $R$ and $R'$ are as defined in Secrion~\ref{sect: pos cone}.
\end{lem}
\begin{myproof}
    Let $S\in \mathcal{C}_W$. Observe that the projection $p(S)$ is given by removing the entries of $S$ away from the blocks $\Sym^2 U_0$ and $\Sym^2 U_{ij}$ in the above notation. Thus, $p(S)$ is block-diagonal of the form 
\begin{align*}
    p(S)=S|_{U_0}+S|_{U_{12}}+S|_{U_{13}}+S|_{U_{23}}.
\end{align*}
    Suppose that $\rk S=k$. Since by assumption $S$ is positive semi-definite this means that there must be at least one principal submatrix $A$ of order $k$ which is positive definite (see e.g.~\cite{horn13}). The corresponding principal submatrix of $p(S)$ will be block-diagonal and its determinant will be a product of principal minors of $A$. Hence, if it was zero, this would imply that at least one of the principal minors of $A$ must vanish, in contrast to the assumption that $A$ is positive definite. Therefore, $p(S)$ has a non-vanishing principal minor of order $k$ and hence $\rk p(S)\geq k=\rk S$. The statement about $\rk R(p(S))$ and $\rk R'(p(S))$ follows analogously, after decomposing the spaces $U_{ij}$ into the symmetric and antisymmetric part, which are orthogonal with respect to $p(S)$ since $S\in W$.
\end{myproof}
\begin{thm}
\label{thm: Z2 cubed invariant is elastic}
Let $M\in (W^*)^{\Z_2^{\,3}}$. Then $M\in \mathcal{C}_W^*$ if and only if it satisfies the elastic bounds.
\end{thm}
\begin{myproof}
First, we observe that $(\mathcal{C}_W^*)^{\Z_2^{\,3}}=\left(\mathcal{C}_W^{\;\Z_2^{\,3}}\right)^*$. We prove that the projection of an extremal element of the third kind in Theorem~\ref{thm: ers in 3d} is never extremal. Since by Proposition~\ref{prop: er proj to er general} every extremal element in $\mathcal{C}_W^{\;\Z_2^{\,3}}$ is the projection of some extremal element in $\mathcal{C}_W$, this will show that every extremal element in $\mathcal{C}_W^{\;\Z_2^{\,3}}$, and therefore every necessary inequality describing $(\mathcal{C}_W^*)^{\Z_2^{\,3}}$, comes from an elastic extremal element in $\mathcal{C}_W$.

Let $S\in\mathcal{C}_W$  be some extremal element in $\mathcal{C}_W$ of the third kind in Theorem~\ref{thm: ers in 3d}. Then $\rk S =4$, $\rk R(S)=3$, $\rk R'(S)=1$ (cf.\ Theorem~\ref{thm: characterize er in 3d}), and it follows from Lemma~\ref{lem: rank proj S} that $\rk p(S)\geq \rk S =4$, $\rk R(p(S))\geq 3$ and $\rk R'(p(S))\geq 1$. We will consider separately the cases of different rank of the restrictions $p(S)|_{U_0}$ and $p(S)|_{U_{ij}}$.

Suppose first that $p(S)|_{U_0}>0$ or $p(S)|_{U_{ij}}>0$  for some indices $i,j$ with $i< j$. Then, there are some $i,j\in\{1,2,3\}$ such that
\begin{align*}
 \ker \left[(\alpha_i\otimes \alpha_j)^{\otimes2}+(\alpha_j\otimes \alpha_i)^{\otimes2}\right]\supsetneq \ker p(S), \end{align*}
 where $\{\alpha_i\}$ is the basis of $V^*$ dual to $\{e_i\}$. By Corollary~\ref{cor: nec extremal}
 this shows that $p(S)$ is not extremal in $p(\mathcal{C}_W)$. We will from now on assume that $\rk p(S)|_{U_0}\leq 2$ and $\rk p(S)|_{U_{ij}}\leq 1$ for all $i<j$.

 The case where $\rk p(S)|_{U_0}=2$ and $\rk p(S)|_{U_{ij}}\leq 1$ for all $i<j$ will be treated in Proposition~\ref{prop: rk S on U0 is 2} below, where we show that such $p(S)$ is not extremal.
 
Finally, suppose that $\rk p(S)|_{U_0}=1$ and $\rk p(S)|_{U_{ij}}\leq 1$. Since $\rk p(S)\geq 4$, this implies $\rk p(S)|_{U_{ij}}= 1$, for all $i,j$. We show that this is impossible for $S$ as above. To see this, we observe that for $p(S)\in\mathcal{C}_W^{\; \Z_2^{\,3}}$ satisfying $\rk p(S)|_{U_0}=1=\rk p(S)|_{U_{ij}}$ for all $i<j$, there must be $\lambda_i\neq 0$, $i=1,2,3$ such that
 \begin{align*}
     p(S)=\left(\sum_{i=1}^3\lambda_i\,\alpha_i^{\otimes 2}\right)^{\otimes 2}+\sum_{i<j}|\lambda_i\lambda_j|\,(\alpha_i\otimes \alpha_j+\operatorname{sgn}(\lambda_i\lambda_j)\, \alpha_j\otimes \alpha_i)^{\otimes 2}.
 \end{align*}
 If $\lambda_i\lambda_j>0$ for all $i,j\in\{1,2,3\}$, then we observe that $R'(p(S))=0$, which is impossible as we must have $\rk R'(p(S))\geq 1$. If there is some pair $i<j$ such that $\lambda_i\lambda_j<0$ there must be precisely two such pairs and we find $\rk R(p(S))=2$ (as $\rk p(S)=4$), in contrast to $\rk R(p(S))\geq 3$. This completes the proof.
\end{myproof}

\begin{prop}
\label{prop: rk S on U0 is 2}
    Let $S\in \mathcal{C}_W$ be an extremal element of the third kind in Theorem~\ref{thm: ers in 3d}. Suppose furthermore that $\rk p(S)|_{U_0}=2$ and $\rk p(S)|_{U_{ij}}\leq 1$ for all $i<j$. Then $p(S)$ is not extremal in $p(\mathcal{C}_W)$.
\end{prop}
\begin{myproof}
We will first show that we must have $\rk p(S)|_{U_{ij}}= 1$ for every pair $i<j$. Indeed, recall that by Theorem~\ref{thm: characterize er in 3d} every inelastic extremal element $S$ in $\mathcal{C}_W$ has  $\rk R(S)=3$, $\ker R'(S)=z\wedge V$ for some $z\in V\setminus\{0\}$ and $z\vee V\cap \ker S=\{0\}$. If there was some pair $i<j$ such that $p(S)|_{U_{ij}}=0$ (there can be at most one such pair since $\rk p(S)\geq 4$), then we would have $\{e_i\vee e_j,e_i\wedge e_j\}\subset \ker p(S)$ and hence also $\{e_i\vee e_j,e_i\wedge e_j\}\subset \ker S$. Moreover, by Lemma~\ref{lem: rank proj S} and the properties of $S$ we know that $\rk R(p(S))\geq 3$ and $\rk R'(p(S))\geq 1$, hence $\rk R(p(S))= 3$ and $\rk R'(p(S))= 1$, as we assumed that $\rk p(S)=4$. In particular, for the unique $k\in\{1,2,3\}$ such that $i\neq k\neq j$, we must have $e_i\wedge e_k\in\ker p(S)$ or $e_j\wedge e_k\in\ker p(S)$ 
(here we have used that the kernel of $p(S)$ is the sum of the kernels of 
its restrictions to $U_0$, $U_{12}$, $U_{13}$ and $U_{23}$). 
Hence, either $\{e_i\wedge V,e_i\vee e_j\}\subset \ker S$ or $\{e_j\wedge V,e_i\vee e_j\}\subset \ker S$, a contradiction.

Next, we observe that in the case where $\rk p(S)|_{U_0}=2$, $\rk p(S)|_{U_{ij}}= 1$ for all $i,j$, there are some linearly independent vectors $(\lambda_i),(\mu_i)\in\R^3\setminus\{0\}$ such that 
 \begin{align*}
     p(S)=&\,\left(\sum_{i=1}^3 \lambda_i\,\alpha_i^{\otimes 2}\right)^{\otimes 2}+\left(\sum_{i=1}^3 \mu_i\,\alpha_i^{\otimes 2}\right)^{\otimes 2}\\&+\sum_{i<j}|\lambda_i\lambda_j+\mu_i\mu_j|\,\left(\alpha_i\otimes \alpha_j+\sgn(\lambda_i\lambda_j+\mu_i\mu_j)\,\alpha_j\otimes \alpha_i\right)^{\otimes 2}.
 \end{align*}
Note that since $\rk p(S)|_{U_{ij}}=1$ for all pairs $i<j$, we must have $\lambda_i\lambda_j+\mu_i\mu_j\neq 0$.

\emph{Case 1:} If $\sgn(\lambda_i\lambda_j)=\sgn(\lambda_i\lambda_j+\mu_i\mu_j)$ for all $i,j$, we note that 
 \begin{align*}
T:=\left(\sum_{i=1}^3\lambda_i\,\alpha_i^{\otimes 2}\right)^{\otimes 2}+\sum_{i<j}|\lambda_i\lambda_j|\,(\alpha_i\otimes \alpha_j+\operatorname{sgn}(\lambda_i\lambda_j)\, \alpha_j\otimes \alpha_i)^{\otimes 2}
\end{align*}
satisfies $\ker T\supsetneq \ker p(S)$. Since $T\in p(\mathcal{C}_W)$ this shows that $p(S)$ is not extremal in $p(\mathcal{C}_W)$. In the case where $\sgn(\mu_i\mu_j)=\sgn(\lambda_i\lambda_j+\mu_i\mu_j)$ for all $i,j$ it follows similarly that $p(S)$ is not extremal. 
 
\emph{Case 2:} Now suppose that $\lambda_i\neq 0$ and $\mu_i\neq 0$ for all $i$ and that there are $i\neq j,k\neq l\in\{1,2,3\}$ such that $\sgn(\lambda_i\lambda_j+\mu_i\mu_j)\neq \sgn(\lambda_i\lambda_j)$ and $\sgn(\mu_k\mu_l)\neq\sgn(\lambda_k\lambda_l+\mu_k\mu_l)$. Up to a 
permutation we will assume without loss of generality that these two pairs are $(1,2)$ and $(1,3)$. Since we assumed that $\lambda_i,\mu_i\neq 0$ for all $i$, we must have $\sgn(\lambda_1\lambda_2+\mu_1\mu_2)= \sgn(\mu_1\mu_2)$ and $\sgn(\lambda_1\lambda_3)=\sgn(\lambda_1\lambda_3+\mu_1\mu_3)$. Then there is some $t>1$, such that for some pair $k<l$ we have $\lambda_k\lambda_l+t\,\mu_k\mu_l=0$ and, for the other pairs $i,j$ we have $\lambda_i\lambda_j+t\,\mu_i\mu_j=0$ or $\sgn (\lambda_i\lambda_j+t\,\mu_i\mu_j)=\sgn(\lambda_i\lambda_j+\mu_i\mu_j)$
(note that the pair $(k,l)$ can be only $(1,3)$ or $(2,3)$). In particular, we find that
 \begin{align*}
     T_t:=&\, \left(\sum_{i=1}^3\lambda_i\,\alpha_i^{\otimes 2}\right)^{\otimes 2}+\left(\sum_{i=1}^3\sqrt{t}\mu_i\,\alpha_i^{\otimes 2}\right)^{\otimes 2}\\&+\sum_{i<j}|\lambda_i\lambda_j+t\mu_i\mu_j|\,(\alpha_i\otimes \alpha_j+\operatorname{sgn}(\lambda_i\lambda_j+t\mu_i\mu_j)\, \alpha_j\otimes \alpha_i)^{\otimes 2}
 \end{align*}
 satisfies $\ker T_t\supsetneq \ker p(S)$ and hence $p(S)$ cannot be extremal in $p(\mathcal{C}_W)$.

\emph{Case 3:} It remains to consider the case where there is some $i$ such that $\lambda_i=0$ or $\mu_i=0$. If only one of the $\lambda_i$ is non-vanishing, we may assume after relabeling that $\lambda_1\neq 0$, $\lambda_2=0=\lambda_3$. But then it follows that $\ker\alpha_1^{\otimes4}\supsetneq \ker p(S)$ and thus $p(S)$ is not extremal. If only one of the $\lambda_i$ is vanishing, we assume after relabeling that it is $\lambda_3$. Then we know that $\sgn(\lambda_1\lambda_3+\mu_1\mu_3)=\sgn(\mu_1\mu_3)$ and $\sgn(\lambda_2\lambda_3+\mu_2\mu_3)=\sgn(\mu_2\mu_3)$. If also $\sgn(\mu_1\mu_2)=\sgn(\lambda_1\lambda_2+\mu_1\mu_2)$, we are in case 1 above and we already know that $p(S)$ is not extremal. If $\sgn(\mu_1\mu_2)\neq\sgn(\lambda_1\lambda_2+\mu_1\mu_2)$ then since $\lambda_1\lambda_2+\mu_1\mu_2\neq 0$, we must have $\sgn(\lambda_1\lambda_2)=\sgn(\lambda_1\lambda_2+\mu_1\mu_2)$. But then it follows that 
\begin{align*}
T:=\left(\sum_{i=1}^3\lambda_i\,\alpha_i^{\otimes 2}\right)^{\otimes 2}+\sum_{i<j}|\lambda_i\lambda_j|\,(\alpha_i\otimes \alpha_j+\operatorname{sgn}(\lambda_i\lambda_j)\, \alpha_j\otimes \alpha_i)^{\otimes 2}
\end{align*}
satisfies $\ker T\supsetneq \ker p(S)$ and hence $S$ is not extremal. The case where $\mu_i=0$ for some $i\in \{1,2,3\}$ works analogously.
\end{myproof}
\begin{rem}
In the appendix of~\cite{Li_et_al21}, a classification of all extremal rays of the cone of $\Z_2^{\,3}$-invariant amplitudes is given. Theorem~\ref{thm: Z2 cubed invariant is elastic} can then alternatively be proved providing explicit elastic extremal elements in $\mathcal{C}_W$ projecting to the extremal elements in the classification result of~\cite{Li_et_al21}. More precisely, in~\cite{Li_et_al21}, it is shown that if $T\in \mathcal{C}_W^{\Z_2^{\,3}}$ is extremal in $\mathcal{C}_W^{\Z_2^{\,3}}$ then either there is some $\mu\in \R$, $i\neq j$ such that
\begin{align*}
    T=\mu\,\left[(\alpha_i\otimes \alpha_j)^{\otimes 2}+(\alpha_j\otimes \alpha_i)^{\otimes 2}\right]
\end{align*}
or for some $(\lambda_i)\in\R^3\setminus\{0\}$ we have
\begin{align*}
T=\left(\sum_{i=1}^3\lambda_i\,\alpha_i^{\otimes 2}\right)^{\otimes 2}+\sum_{i<j}|\lambda_i\lambda_j|\,(\alpha_i\otimes \alpha_j+\operatorname{sgn}(\lambda_i\lambda_j)\, \alpha_j\otimes \alpha_i)^{\otimes 2}.
\end{align*}

In the first case $T$ is obviously the projection of an elastic extremal element in $\mathcal{C}_W$. Moreover, if in the second case only one of the $\lambda_i$ is non-vanishing, then clearly $T=\mu\,\alpha_i^{\otimes 4}$ for some $i\in \{1,2,3\}$, which again means that it is the projection of an elastic extremal element in $\mathcal{C}_W$. 

Now suppose that there is precisely one $i\in\{1,2,3\}$ such that $\lambda_i=0$. We may assume without loss of generality that $\lambda_3=0$, $\lambda_1\lambda_2\neq 0$. If $\lambda_1\lambda_2>0$ then we observe that
\begin{align*}
    T=p\left((\sqrt{|\lambda_1|}\,\alpha_1+\sqrt{|\lambda_2|}\,\alpha_2)^{\otimes 4}\right).
\end{align*}
If $\lambda_1\lambda_2<0$, then we set
\begin{align*}
\alpha:=\sqrt{|\lambda_1|}\,\alpha_1+\sqrt{|\lambda_2|}\,\alpha_2,\quad \beta:=\sqrt{|\lambda_1|}\,\alpha_1-\sqrt{|\lambda_2|}\,\alpha_2
\end{align*}
and observe that
\begin{align*}
    T=p\left( (\alpha\vee \beta)^{\otimes 2}+(\alpha\wedge \beta)^{\otimes 2}\right).
\end{align*}

Finally we suppose that $\lambda_i\neq 0$ for all $i\in\{1,2,3\}$. Then either $\lambda_i\lambda_j>0$ for all $i,j\in\{1,2,3\}$ and in this case we find 
 \begin{align*}
     T=p\left( (\sqrt{|\lambda_1|}\alpha_1+\sqrt{|\lambda_2|}\alpha_2+\sqrt{|\lambda_3|}\alpha_3)^{\otimes 4}\right)
 \end{align*}
 or $\lambda_i\lambda_j<0$ for precisely two pairs $i\neq j$. In the latter case we may assume that $\lambda_1,\lambda_2>0$ and $\lambda_3<0$. Then for 
 \begin{align*}
    \alpha:=\sqrt{\lambda_1}\,\alpha_1+\sqrt{\lambda_2}\,\alpha_2+\sqrt{-\lambda_3}\,\alpha_3,\quad \beta:=\sqrt{\lambda_1}\,\alpha_1+\sqrt{\lambda_2}\,\alpha_2-\sqrt{-\lambda_3}\,\alpha_3
 \end{align*}
 we find
 \begin{align*}
     p(S)=p\left( (\alpha\vee \beta)^{\otimes 2}+(\alpha\wedge \beta)^{\otimes 2} \right).
 \end{align*}
\end{rem}

\subsubsection{$\SO(2)$-invariant amplitudes}

Given some basis $\{e_1,e_2,e_3\}$ of $V$, we will consider the subgroup $\SO(2)\subset \O(3)$ given by rotations with axis $e_3$ and its action on $W^*$. In the following, we will show that the cone $(\mathcal{C}_W^*)^{\SO(2)}$ is completely described by the elastic bounds.

\begin{thm}
    Let $M\in (W^*)^{\SO(2)}$. Then $M\in \mathcal{C}_W^*$ if and only if it satisfies the elastic bounds.
\end{thm}
\begin{myproof}
We show that every element in $(W^*)^{\SO(2)}$ is contained in $(W^*)^{\Z_2^{\,3}}$. First, we observe that since every four-tensor is invariant under the action of $-1\in \O(3)$ and since every $M\in (W^*)^{\SO(2)}$ is in particular invariant under the action of
$\left(\begin{smallmatrix}
-1 & 0 & 0\\
    0 & -1 & 0\\
    0 & 0 & 1
\end{smallmatrix}\right)$, we find that every $M\in(W^*)^{\SO(2)} $ is invariant under the action of $Q_3$ in Section~\ref{sect: Z2 inv ampl}.

In order to see that $M\in (W^*)^{\SO(2)}$ is also invariant under the action of $Q_1$ and $Q_2$, we observe that the condition that $M$ is invariant under the infinitesimal rotation $\epsilon_{ij3}\in\mathfrak{so}(2)\subset \mathfrak{so}(3)$ around $e_3$ translates to the condition
\begin{align}
\label{eq: inv under rotation 3}
    0=\sum_{p=1}^3 M_{pjkl}\,\epsilon_{ip3}+M_{ipkl}\,\epsilon_{jp3}+M_{ijpl}\,\epsilon_{kp3}+M_{ijkp}\,\epsilon_{lp3}
\end{align}
for every $i,j,k,l\in \{1,2,3\}$. Writing out equation~\eqref{eq: inv under rotation 3} for $i=j=k=l=1$, we find 
\begin{align*}
    0=M_{2111}+M_{1211}+M_{1121}+M_{1112}=4\, M_{1112},
\end{align*}
where we used the symmetries of $W^*$ in the last step. Similarly, applying \eqref{eq: inv under rotation 3} for $i=j=k=l=2$, we find that $M_{1222}=0$. Moreover, for $i=j=1$, $k=l=3$, equation~\eqref{eq: inv under rotation 3} yields
\begin{align*}
    0=M_{2133}+M_{1233}=2\, M_{1233}
\end{align*}
and, similarly, choosing $i=k=1$ and $j=l=3$ in \eqref{eq: inv under rotation 3} yields $M_{1323}=0$. Since we already know that $M$ is invariant under $Q_3$, we conclude that $M_{ijkl}$ must vanish, if the index 1 or 2 appears an odd amount of times. Thus, $M$ is also invariant under the action of $Q_1$ and $Q_2$ and therefore $M\in(W^*)^{\Z_2^{\,3}}$. The statement follows now from Theorem~\ref{thm: Z2 cubed invariant is elastic}.
\end{myproof}

\subsection{Elastic vs. inelastic bounds}
In the previous section we have seen several examples of restrictions of the cone to amplitudes with additional symmetries. In each of the cases we considered, the elastic bounds were sufficient to describe the dual cone. 

However, in general, the extremal rays of the third kind in Theorem~\ref{thm: ers in 3d} give rise to new constraints. A way to see this is as follows: We denote by $\mathcal{C}_{\mathrm{el}}$ the closed convex cone generated by the directions of the elastic extremal rays of $\mathcal{C}_W$. Then $\mathcal{C}_{\mathrm{el}}\subsetneq \mathcal{C}_W$, since the extremal rays of $\mathcal{C}_W$ of the third in Theorem~\ref{thm: ers in 3d} are not contained in $\mathcal{C}_{\mathrm{el}}$. Suppose now that the inelastic extremal rays in $\mathcal{C}_W$ do not give rise to any new constraint on the dual cone $\mathcal{C}_W^*$, i.e.~$\mathcal{C}_W^*=\mathcal{C}_{\mathrm{el}}^*$. Then we have $\mathcal{C}_W^{**}=\mathcal{C}_{\mathrm{el}}^{**}$. Since both $\mathcal{C}_W$ and $\mathcal{C}_{\mathrm{el}}$ are closed convex cones, it follows that 
\begin{align*}
    \mathcal{C}_W=\mathcal{C}_W^{**}=\mathcal{C}_{\mathrm{el}}^{**}=\mathcal{C}_{\mathrm{el}},
\end{align*}
a contradiction.

\begin{figure}[H]
\centering
\includegraphics[width=0.7\textwidth]{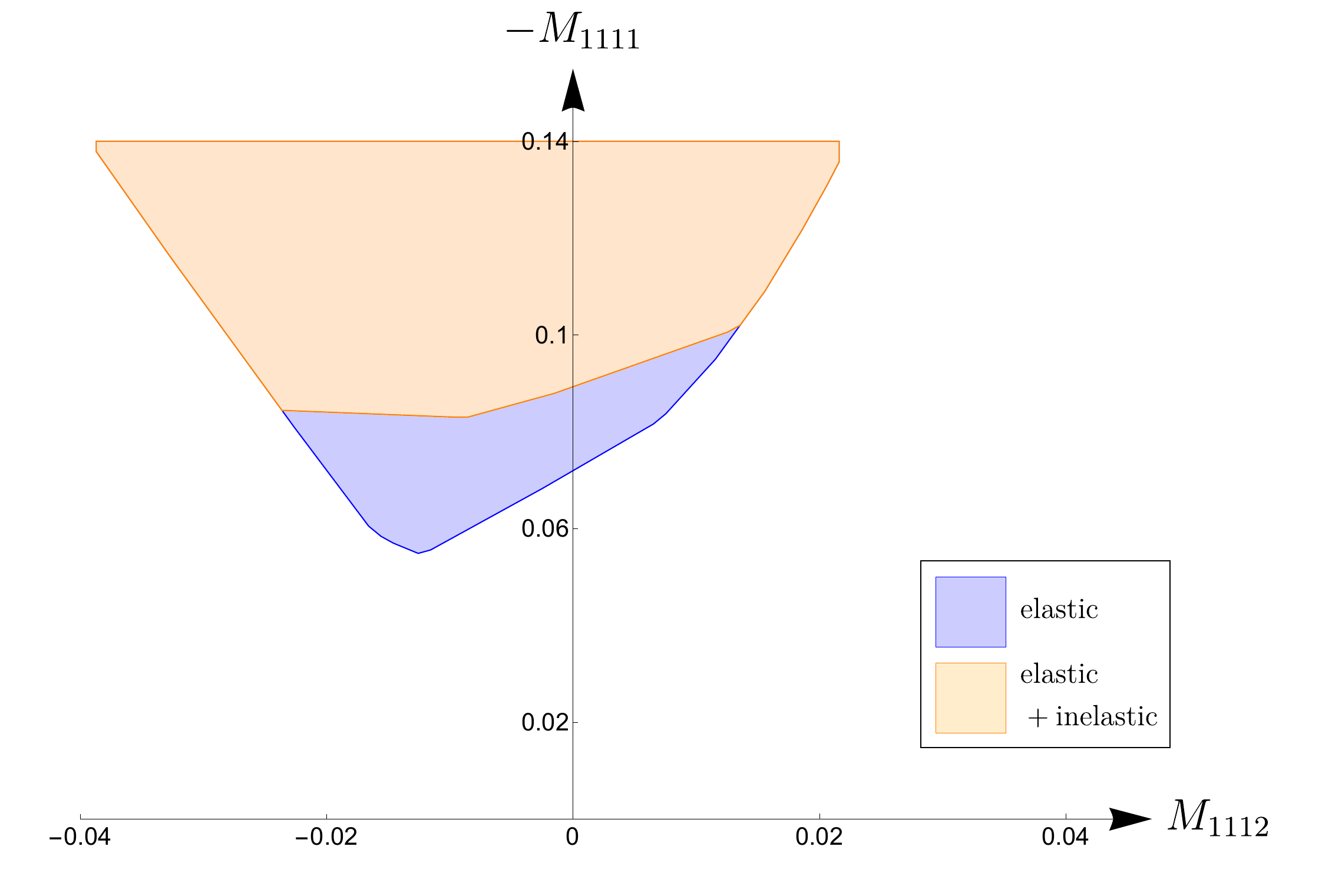}
\caption{ 
Regions of parameters consistent with positivity bounds. The union of the blue and orange regions corresponds to points which abide by the elastic bounds and the orange region corresponds to imposing all (elastic and inelastic) bounds. The 21-dimensional parameter space has been restricted to $M_{1111}$ and $M_{1112}$, more precisely to the perturbations in their direction around the following reference point : $M_{1111}=0.332,
M_{1112}=0.022,
M_{1113}=0.12,
M_{1122}=0.056,
M_{1123}=0.086,
M_{1133}=-0.2,
M_{1212}=0.428,
M_{1213}=-0.252,
M_{1222}=0.12,
M_{1223}=-0.184,
M_{1232}=-0.172,
M_{1233}=-0.04,
M_{1313}=0.356,
M_{1323}=-0.05,
M_{1333}=-0.124,
M_{2222}=0.476,
M_{2223}=0.12,
M_{2233}=-0.14,
M_{2323}=0.626,
M_{2333}=-0.028,
M_{3333}=0.332$. The exact regions have been approximated by numerically sampling extremal rays, with a focus on the neighborhood of the extremal rays which vanish when evaluated on the reference point. The choice of reference point, as well as that of the directions of the perturbations around it, were heavily inspired from Ref.~\cite{Li_et_al21}. Consequently, the Figure above is qualitatively similar to Figure 3 in the supplemental material of Ref.~\cite{Li_et_al21} (where it described positivity bounds for four-fermion EFT operators with three flavors of fermions). We have observed that uninformed choices of reference point and/or directions of perturbations around it are very unlikely to yield an observable difference between elastic and full positivity bounds. A systematic exploration of these aspects is left for future work.}
\label{fig: plot of elastic vs inelastic}
\end{figure}

Figure~\ref{fig: plot of elastic vs inelastic} illustrates the difference between $\mathcal{C}_W^*$ and $\mathcal{C}_{\mathrm{el}}^*$.
\subsubsection*{Acknowledgements} 
We are grateful to Mateo Galdeano for valuable comments on the paper. This work was funded by the Deutsche Forschungsgemeinschaft under Germany’s Excellence Strategy -- EXC 2121 “Quantum Universe” -- 390833306. PP and VC are also funded by the Deutsche Forschungsgemeinschaft under – SFB-Gesch\"aftszeichen 1624 -- Projektnummer 506632645. This project also has received funding from the European Union’s Horizon Europe research and innovation programme under the Marie Sk\l odowska-Curie Staff Exchange grant agreement No 101086085 -- ASYMMETRY. And C.G. thanks the Berkeley Center for Theoretical Physics for hospitality while this work was finalised. 

\appendix
\section{Bounds on physical theories}

In this appendix, we discuss the connection between the various spaces described above and positivity bounds in the form under which they appear in the physics literature. 

\subsection{Positivity bounds from dispersion relations on scattering amplitudes}
\label{app: tensor M}
Positivity bounds are associated with the properties of specific scatterings in relativistic quantum theories. The simplest such scattering, namely the two-to-two scattering of identical particles, is described by a quantum amplitude $\cal A$, from which it is possible to extract a positive quantity $M$~\cite{Adams06}. Following~\cite{zhang20,Li_et_al21}, we now recall the technical steps allowing on how this is done and how to show that $M$ belongs to ${\cal C}_W^*$. Our notation applies straightforwardly to the scattering in scalar theories, but the logic can be extended to spinning particles~\cite{Bellazzini:2016xrt,deRham:2017zjm,Li_et_al21}.

When several flavors (or species) of otherwise-identical particles are present, there are several relevant scatterings. The scattered states are one-particle quantum states, 
that, after fixing their momentum, are described by elements of a finite-dimensional complex vector space $V_c$, dubbed flavor space. We construct a distinguished basis of $V_c$ as follows: we will start with some basis $\{|\psi_i\rangle\}_{i=1}^n$ such that for every basis vector $|\psi_i\rangle$ there is some $j\in\{1,...,n\}$ such that the basis vector $|\psi_j\rangle$ is the anti-particle of $|\psi_i\rangle$ (with flipped spin). Then we consider all the pairs $(|\psi_i\rangle,\,|\psi_j\rangle)$  of particles and antiparticles (with flipped spin) in the basis satisfying $i\neq j$ and replace them by $\frac{1}{\sqrt{2}}\,(|\psi_i\rangle +|\psi_j\rangle)$ and $\frac{1}{\sqrt{2}i}\,(|\psi_i\rangle -|\psi_j\rangle)$. We will from now on denote this basis by $\{|i\rangle\}_{i=1}^n$. Moreover, we denote by $\bar{\cdot}$ the complex conjugate defined by the basis $\{|i\rangle\}_{i=1}^n$. With respect to that conjugation, our basis vectors are self-conjugate.

Now, we consider the forward scattering $|i\rangle\otimes|j\rangle\to |k\rangle\otimes|l\rangle$, where the kinematics (momenta and spin) of particles $i$ and $k$, as well as those of particles $j$ and $l$, are identical. It is described by the quantum amplitude ${\cal A}_{ijkl}(s)$, where $s$ denotes the leftover Mandelstam variable in the forward limit. (See Ref.~\cite{Bellazzini:2016xrt} for a careful definition of the forward limit in the presence of spinning particles and for the demonstration that the amplitude only depends on $s$.) Then we define\footnote{Here we are ignoring IR (i.e., small $s$) singularities due to light particles, which can be explicitly computed and subtracted~\cite{Bellazzini:2016xrt,deRham:2017imi}, so that $\mathcal{A}_{ijkl}$ is analytic in $s=m^2/2$ and therefore $M_{ijkl}$ is well-defined.}
\begin{align*}
    M_{ijkl}:= \frac{\mathrm{d}^2}{\mathrm{d}s^2}\,\mathcal{A}_{ijkl}(s=m^2/2) +\text{c.c.},
\end{align*}
where $m^2$ is the sum of the four squared masses of the scattered particles, and c.c.~stands for the complex conjugate, so that the $M_{ijkl}$ are real numbers. Note that since the basis $|i\rangle$ consists of self-conjugate vectors, using the crossing symmetry\footnote{For the sake of illustration, let us derive the first relation. Given our choice of a self-conjugate basis $|i\rangle$, $1 \leftrightarrow 3$ crossing symmetry implies that $\mathcal{A}_{ijkl}(s)=\mathcal{A}_{ilkj}(u=m^2-s)$. Moreover, $s+t+u=m^2$ implies that, in the forward limit, the points $u=m^2/2$ and $s=m^2/2$, which are mapped by $1 \leftrightarrow 3$ crossing symmetry, are identical.} and CPT-invariance of the amplitude $\mathcal{A}$, we find
\begin{align*}
    M_{ijkl}=M_{ilkj}=M_{kjil}=M_{klij} \ .
\end{align*}
Moreover, we assume the additional symmetry $M_{ijkl}=M_{jilk}$, which follows from Lorentz and Bose symmetry for scalars, but only holds for spinning particles if parity is a good symmetry (or if one averages the amplitude and its parity-transform). See Ref.~\cite{Zhang:2021eeo} for more details. 

Denoting by $V$ the real subspace of self-conjugate vectors, i.e.~the real vector space spanned by the self-conjugate basis vectors $|i\rangle$, we can think of $M_{ijkl}$ as the components of a real four-tensor on $V$, which by the symmetries above lies in fact in $W^*$. It is worth noticing that, despite the restriction to $V$, the $M_{ijkl}$ contain information about two-to-two scatterings of arbitrary states (also of those which are represented by vectors that are not self-conjugate).

Moreover, using the dispersion relation and applying the generalized optical theorem (see \cite{zhang20,Li_et_al21}) one can show that in fact $M\in \mathcal{C}_W^*$. Let us sketch how this arises. Ignoring IR (i.e., small $s$) singularities due to light particles, which can be explicitly computed and subtracted~\cite{Bellazzini:2016xrt,deRham:2017imi}, one can rely on the analytic properties of the amplitude to express $M_{ijkl}$ as a contour integral over complex and small $s$,
\begin{align*}
    M_{ijkl}= \frac{1}{\pi i}\oint ds\,\frac{\mathcal{A}_{ijkl}(s)}{(s-m^2/2)^3}+\text{c.c.} \ .
\end{align*}
Using analyticity of the amplitude, one can deform the integration contour to larger values of $s$. Then, the Froissart-Martin bound~\cite{Froissart:1961ux,Martin:1962rt} allows one to drop the integral over the circle at infinity, leaving behind the integration along the real $s$-axis, which hosts the physical singularities of the amplitude. The discontinuities of the amplitude over these singularities can be related to other amplitudes via the optical theorem, while the integral over negative values of $s$ can be mapped to the the integral over positive values of the $1\leftrightarrow 3$-crossed amplitude. Altogether, one finds
\begin{align*}
    M_{ijkl}= \frac{1}{\pi}\int_{\Lambda_\text{UV}}^\infty ds\,\frac{\sum_X\mathcal{A}_{ij\to X}\mathcal{A}_{kl\to X}^*}{(s-m^2/2)^3}+(j\leftrightarrow l)+\text{c.c.} \ ,
\end{align*}
where $\Lambda_\text{UV}$ is the scale above which one did not subtract singularities (either known IR or unknown UV ones), which we take larger than $m^2/2$, and $X$ refers to any state.

Expanding $\mathcal{A}_{ij\to X}$ into real and imaginary parts, one finds that $M_{ijkl}$ is a sum of terms of the form $m_{ij}m_{kl}+(j\leftrightarrow l)$, for unknown real tensors $m$. Thus, $M$ belongs to the cone generated by elements of the form $m_{ij}m_{kl}+(j\leftrightarrow l)$. The dual cone is made out of four-tensors $S$ so that $S\cdot M\geq 0$ for all $M$ with the aforementioned symmetries and such that $M_{ijkl}=\sum_{a}m^a_{ij}m^a_{kl}+m^a_{il}m^a_{kj}$ for some real tensors $m^a$. In particular, we may assume that $S$ has the same symmetries as $M$, i.e. we may consider the duality within the linear space of tensors with such symmetries. Then, treating $S$ as a bilinear form on two-tensors, the condition $S\cdot M\geq 0$ for all $M$ as above becomes $S(m,m)\geq 0$ for all $m\in V^{\otimes 2}$. Due to the symmetries of $S$, one immediately notices that the symmetric and anti-symmetric parts of $m$ are orthogonal with respect to $S$, and one recovers the characterization of $\mathcal{C}_W$.

\subsection{Elastic bounds and extremal rays}\label{app:elasticScattering}

As said in the main text, positivity bounds can be further separated into elastic and inelastic ones. The former can be deduced by only considering elastic scattering processes, i.e.~processes where the initial state of the scattering agrees with its final state, of a specific kind (namely, the scattered state is a tensor product of two arbitrary quantum superpositions of one-particle states, see below). Other scatterings yield inelastic bounds (and are associated with elastic scattering of states describing entangled particles~\cite{zhang20,Li_et_al21} -- making our wording "inelastic" slightly misleading).

Finally, let us establish that elastic bounds, as defined above, are the inequalities  corresponding to the extremal rays of the first and second kind in Theorem~\ref{thm: ers in 3d}. Indeed, they are associated to the elastic scattering of quantum states of the form $|\psi_1\rangle\otimes|\psi_2\rangle$ with $|\psi_1\rangle=\sum_i \alpha_i|i\rangle$ and $|\psi_2\rangle=\sum_i \beta_i|i\rangle$. $\alpha,\beta\in V^*$ define the state. Following the aforementioned steps that allowed one to extract $M_{ijkl}\in{\cal C}_W^*$, one obtains, due to the multi-linearity of amplitudes,
\begin{align*}
\label{eq: elastic constraint general form}
\sum_{i,j,k,l}\alpha_i\beta_j\alpha_k\beta_lM_{ijkl}=M(\alpha\otimes \beta,\alpha\otimes \beta)=((\alpha\vee \beta)^{\otimes 2}+(\alpha\wedge \beta)^{\otimes 2})(M) \geq 0 , \quad \forall\alpha,\beta\in V^*,
\end{align*}
where the last equality follows from the symmetry of $M\in W$, yielding the second kind of extremal ray (as well as the first when $\alpha=\beta$).

\subsection{Chiral perturbation theory}
\label{app: ChPT}

Chiral perturbation theory ($\chi$PT) describes the low-energy dynamics of mesons, which are bound states of the theory of strong interactions in the presence of light quarks. The $\chi$PT description is appropriate below the scale at which strong interactions break the (approximate) chiral symmetry of the theory, with respect to which mesons are interpreted as (pseudo-)Nambu-Goldstone bosons. Focussing on the three lightest mesons, namely pions, the chiral symmetry corresponds to the spontaneously-broken approximate $SU(2)_L\times SU(2)_R\times U(1)_V$ symmetry of strong interactions with two flavors of light quarks. In that case, the field space of $\chi$PT is built upon the coset $SU(2)_L\times SU(2)_R\times U(1)_V/U(2)_V$, where $V$ refers to a simultaneous action on all helicity components of a given quark, unlike $L,R$ which refer to a definite helicity. The coordinates of the coset then correspond to the pion fields. $\chi$PT interactions respect the full $SU(2)\times SU(2)\times U(1)$ invariance, non-linearly realized by the action of the group on the coset (and thereby on the pions)\footnote{There is also a spurion associated to the non-zero quark masses which make the $SU(2)\times SU(2)\times U(1)$ symmetry only approximate, but it is irrelevant for the positivity bounds as discussed in this paper.}. That action is conveniently encoded via a $2\times 2$ unitary matrix $U$,
\begin{align*}
    U:=\sqrt{1-\frac{1}{\pi^a\pi^a}}\,\mathbbm{1}+i\frac{\pi^a\tau^a}{f_{\pi}} \ ,
\end{align*}
where $a$ runs from 1 to 3, the summation over repeated indices is implied, $\tau^a$ denote the Pauli matrices and the pion fields $\{\pi^a\}_{a=1}^3$ map the coset. With respect to the unbroken $SU(2)\times U(1)$ subgroup, they form a neutral triplet (that is automatically self-conjugate). The parameter $f_{\pi}$ is the pion decay constant. The aforementioned action of $SU(2)_L\times SU(2)_R\times U(1)_V$ on $U$ reads
\begin{equation*}
U\to R U L^\dagger \ ,
\end{equation*}
where $L,R$ refer to unitary matrices belonging to the corresponding group factor. $U(1)_V$ acts trivially, while the unbroken $U(2)_V=SU(2)_V\times U(1)_V$ acts via the identification $L=R$. The vacuum state, defined so that $\langle\pi^a\rangle=0$, i.e. $\langle U\rangle=1$, leaves $U(2)_V$ unbroken.

We furthermore denote by $u$ the square root of $U$ and set
\begin{align*}
    u_{\mu}:=i u^{\dagger}\del_{\mu}U u^{\dagger}=-\frac{\partial_\mu \pi^a \tau^a}{F_\pi}+... \ ,
\end{align*}
where the ellipses refer to terms with more than one pion field, which are irrelevant for our purposes. (As a consequence, the simplest positivity bounds do not differentiate between pions, associated with a rich coset structure, and any triplet of scalars.) From the perspective of positivity bounds, which constrain four-field four-derivative interactions, the relevant parts of the Lagrangian are
\begin{align*}
    \mathcal{L}_{\mathrm{ChPT},4}\supset&\, \frac{\ell_1}{4}\tr^2(u_{\mu}u^{\mu})+\frac{\ell_2}{4}\tr(u_{\mu}u_{\nu})\tr(u^{\mu}u^{\nu})\\
    =&\,\frac{\ell_1}{f_{\pi}^4} \del_{\mu}\pi^a\del^{\mu}\pi^a\del_{\nu}\pi^b\del^{\nu}\pi^b+\frac{\ell_2}{ f_{\pi}^4}\del_{\mu}\pi^a\del_{\nu}\pi^a\del^{\mu}\pi^b\del^{\nu}\pi^b \ .
\end{align*}

Computing the forward scattering amplitude of $|\pi^a\rangle\otimes |\pi^b\rangle\to |\pi^c\rangle\otimes |\pi^d\rangle$ and taking the second derivative by $s$ at $2m_{\pi}^2$ yields
\begin{align*}
    M_{abcd}=\frac{\mathrm{d}^2}{\mathrm{d}s^2}\,\mathcal{A}_{abcd}(s=2m_{\pi}^2)=\frac{2}{f_{\pi}^4}\left[(2\ell_1+\ell_2)\,\delta_{ab}\delta_{cd}+2\ell_2\,\delta_{ac}\delta_{bd}+(2\ell_1+\ell_2)\,\delta_{ad}\delta_{bc}\right].
\end{align*}
We observe that for some dual bases $\{e_i\}_{i=1}^3$ and $\{\alpha_i\}_{i=1}^3$ of $\R^3$ and ${\R^3}^*$, we have
\begin{align*}
    \delta_{ab}\delta_{cd}&=\sum_{i,j=1}^3e_i^{\otimes2}\otimes e_j^{\otimes2}(\alpha_a,\alpha_b,\alpha_c\alpha_d)=\left(\sum_{i=1}^3e_i^{\otimes 2}\right)^{\otimes 2}(\alpha_a,\alpha_b,\alpha_c,\alpha_d),\\
    \delta_{ac}\delta_{bd}&=\sum_{i,j=1}^3e_i\otimes e_j\otimes e_i\otimes e_j(\alpha_a,\alpha_b,\alpha_c,\alpha_d)\\&=\left[\sum_{i=1}^3 e_i^{\otimes 4} +2\sum_{i<j}(e_i\vee e_j)^{\otimes 2}+2\sum_{i<j}(e_i\wedge e_j)^{\otimes 2}\right](\alpha_a,\alpha_b,\alpha_c,\alpha_d),\\
    \delta_{ad}\delta_{bc}&=\sum_{i,j=1}^3e_i\otimes e_j\otimes e_j\otimes e_i(\alpha_a,\alpha_b,\alpha_c,\alpha_d)\\
    &=\left[\sum_{i=1}^3 e_i^{\otimes 4} +2\sum_{i<j}(e_i\vee e_j)^{\otimes 2}-2\sum_{i<j}(e_i\wedge e_j)^{\otimes 2}\right](\alpha_a,\alpha_b,\alpha_c,\alpha_d).
\end{align*}
Putting everything together and rescaling by $f_{\pi}^4/2$ we arrive at the expression as in Application~\ref{appl: ChPT}.

\bibliographystyle{jhep}
\bibliography{ref}
\end{document}

%% file: math_part/settings.tex
\usepackage{comment}
\usepackage[automark]{scrlayer-scrpage}								
\usepackage{amsfonts}												
\usepackage{amsmath}												
\usepackage{mathtools}												
\usepackage{stmaryrd}												
\usepackage{amssymb}												
\usepackage{mathrsfs}												
\usepackage{accents}												
\usepackage[T1]{fontenc}											
\usepackage[utf8]{inputenc}											
\usepackage[top=1in,bottom=1in,left=1.25in,right=1.25in]{geometry}	
\usepackage{enumerate} 												
\usepackage{authblk}												
\usepackage{abstract}												
\usepackage{etoolbox}												
\usepackage[normalem]{ulem}
\usepackage{csquotes}                                               

\usepackage{tikz-cd}	\usepackage{tikz}											

\usepackage{amsthm}													

\usepackage{cite}

\usepackage[hidelinks,colorlinks=false]{hyperref}				
\usepackage[noabbrev]{cleveref}										
\usepackage{float} 
\usepackage{bbm}
\makeatletter
\g@addto@macro\bfseries{\boldmath}
\makeatother

\newcommand{\Z}{\mathbb{Z}}

\newcommand{\R}{\mathbb{R}}

\renewcommand{\P}{\mathrm{P}}
\newcommand{\qq}{\mathrm{q}}
\newcommand{\del}{\partial}

\DeclareMathOperator{\tr}{tr}
\DeclareMathOperator{\Span}{span}
\renewcommand{\O}{\operatorname{O}}
\DeclareMathOperator{\SO}{SO}

\DeclareMathOperator{\End}{End}

\DeclareMathOperator{\Sym}{Sym}
\DeclareMathOperator{\Sl}{SL}

\DeclareMathOperator{\id}{id}

\DeclareMathOperator{\rk}{rank}
\DeclareMathOperator{\ann}{Ann}
\DeclareMathOperator{\ri}{relint}
\DeclareMathOperator{\rb}{\partial_{\mathrm{rel}}}
\DeclareMathOperator{\cl}{cl}

\DeclareMathOperator{\aff}{aff}
\DeclareMathOperator{\conv}{conv}

\DeclareMathOperator{\sgn}{sgn}

\newtheoremstyle{mythm}
{}
{}
{\slshape}
{}
{\bfseries\sffamily}
{.}
{ }
{}
\newtheoremstyle{mydef}
{}
{}
{}
{}
{\bfseries\sffamily}
{.}
{ }
{}

\theoremstyle{mythm}
\newtheorem*{thm*}{Theorem}
\newtheorem{thm}{Theorem}[section]

\newtheorem{prop}[thm]{Proposition}
\newtheorem{cor}[thm]{Corollary}
\newtheorem{lem}[thm]{Lemma}
\theoremstyle{mydef}
\newtheorem{mydef}[thm]{Definition}
\newtheorem{rem}[thm]{Remark}

\newtheorem{app}[thm]{Application}
\newenvironment{myproof}[1][\proofname]{
	\proof[\sffamily\upshape#1]
}{\endproof}

\clearscrheadfoot
\ihead[]{\headmark}
\ohead[]{\pagemark}
\deffootnote[1em]{0em}{1em}{%
	\textsuperscript{\thefootnotemark}%
}
\setfootnoterule{3em}

\apptocmd{\sloppy}{\hbadness 10000\relax}{}{}